\documentclass[preprint,flushrt]{aastex}
\usepackage{graphicx,natbib,morefloats}
\newcommand{\WG}{{\bf W}$_G$}
\newcommand{\PKS}{PKS~2155-304}
\newcommand{\ESO}{ESO~141-G55}  

\newcommand{\z}{\ensuremath{z}}
\newcommand{\bb}{\ensuremath{b}}
\newcommand{\ang}{~\mbox{\AA}}
\newcommand{\persecond}{\ {\rm s}\ensuremath{^{-1}}~}

\newcommand{\percmtwo}{\ {\rm cm}\ensuremath{^{-2}}~}
\newcommand{\percmtwono}{\ {\rm cm}\ensuremath{^{-2}}}
\newcommand{\perhz}{\ {\rm Hz}\ensuremath{^{-1}}~}
\newcommand{\perhzno}{\ {\rm Hz}\ensuremath{^{-1}}}

\newcommand{\perMpcno}{~{\rm Mpc}\ensuremath{^{-1}}}

\newcommand{\nomang}{{\rm m}\mbox{\AA}}
\newcommand{\noang}{\mbox{\AA}}
\newcommand{\mang}{~{\rm m}\mbox{\AA}}

\newcommand{\Ang}{\ang\ }
\newcommand{\Mang}{\mang\ }
\newcommand{\Vr}{{\rm V}\ensuremath{_r}}
\newcommand{\AVr}{\ensuremath{\mid}\Vr\ensuremath{\mid}}
\newcommand{\Nno}{\ensuremath{\mathcal{N}}}
\newcommand{\Dperp}{{\rm D}\ensuremath{_{\perp}}}

\newcommand{\logNh}{\ensuremath{\log{ \Nhno }}}
\newcommand{\Wno}{\ensuremath{\mathcal{W}}}
\newcommand{\W}{\Wno~}
\newcommand{\Ws}{EWs}

\newcommand{\etno}{et~al.}
\newcommand{\et}{\etno\ }
\newcommand{\eti}{\etno}
\newcommand{\etl}{\et}

\newcommand{\hseventy}{\ensuremath{h_{70}}\ }
\newcommand{\hsfi}{\ensuremath{h^{-1}_{70}}\ }

\newcommand{\hone}{\ion{H}{1}\ }

\newcommand{\Xtwo}{\ensuremath{\chi^2}\ }

\newcommand{\sig}{\ensuremath{\sigma}\ }
\newcommand{\signo}{\ensuremath{\sigma}}
\newcommand{\about}{\ensuremath{\sim}}
\newcommand{\nokmsno}{{\rm km~s}\ensuremath{^{-1}}}
\newcommand{\kmsno}{~\nokmsno}

\newcommand{\kms}{\kmsno\ }

\newcommand{\lya}{Ly\ensuremath{\alpha} }
\newcommand{\lyano}{Ly\ensuremath{\alpha}}
\newcommand{\lyb}{Ly\ensuremath{\beta} }
\newcommand{\lybno}{Ly\ensuremath{\beta}}
\newcommand{\lyg}{Ly\ensuremath{\gamma} }
\newcommand{\lyd}{Ly\ensuremath{\delta} }
\newcommand{\lye}{Ly\ensuremath{\epsilon} }
\newcommand{\Nhno}{\ensuremath{N_{\rm HI}}}
\newcommand{\Nh}{\Nhno\ }

\newcommand{\nh}{\Nh}

\newcommand{\bobs}{\ensuremath{b_{\rm obs}}}

\newcommand{\bmsd}{\ensuremath{b}}

\newcommand{\expanded}{\ensuremath{SL \ge 3\signo}}
\newcommand{\tent}{\ensuremath{3\signo \le SL<4\signo}}
\newcommand{\real}{\ensuremath{SL \ge 4\signo}}
\newcommand{\Dld}{\ensuremath{\Delta\lambda_{\rm D}}}

\newcommand{\Dv}{\ensuremath{\Delta v}}
\newcommand{\dz}{{\rm d}\z}

\newcommand{\gt}{\ensuremath{>}}
\newcommand{\lt}{\ensuremath{<}}
\newcommand{\Wlya}{\Wno$_{\lyano}$}

\newcommand{\wrange}{1218--1300\ang}
\newcommand{\zrange}{\ensuremath{0.002 < \z < 0.069}}
\newcommand{\zem}{\ensuremath{z_{\rm em}}}
\newcommand{\prox}{c\zem -- 1,200\kms}
\newcommand{\proxno}{c\zem -- 1,200\kmsno}
\newcommand{\lowz}{low-\z\ }
\newcommand{\lowzno}{low-\z}
\newcommand{\lowzya}{low-\z\ \lya}
\newcommand{\dn}{{\rm d}\Nno}

\newcommand{\dndzno}{\dn/\dz}

\newcommand{\cz}{\ensuremath{cz}}

\newcommand{\lognh}{\logNh}

\newcommand{\Exp}[2]{\ensuremath{#1\times10^{#2}}}
\newcommand{\bvalue}{\bb-value\ }
\newcommand{\bvalues}{\bb-values\ }
\newcommand{\bvalueno}{\bb-value}
\newcommand{\bvaluesno}{\bb-values}
\newcommand{\lam}{\ensuremath{\lambda}}

\newcommand{\NV}[1]{\ion{N}{5}~\lam#1}
\newcommand{\MgII}[1]{\ion{Mg}{2}~\lam#1}
\newcommand{\CI}[1]{\ion{C}{1}~\lam#1}
\newcommand{\CIII}[1]{\ion{C}{3}~\lam#1}

\newcommand{\FeII}[1]{\ion{Fe}{2}~\lam#1}
\newcommand{\SII}[1]{\ion{S}{2}~\lam#1}
\newcommand{\SiII}[1]{\ion{Si}{2}~\lam#1}
\newcommand{\Sithree}{\ion{Si}{3}~\lam1206.5}
\newcommand{\OVI}[1]{O~VI~\lam#1}
\newcommand{\SiIII}[1]{\ion{Si}{3}~\lam#1}
\newcommand{\SFsixty}{\SiII{1260.4}~+\ \FeII{1260.5}}
\newcommand{\NVdoublet}{\ion{N}{5}\ensuremath{~\lambda\lambda}1238, 1242}
\newcommand{\OVIdoublet}{\ion{O}{6}\ensuremath{~\lambda\lambda}1032, 1038}
\newcommand{\CIVdoublet}{\ion{C}{4}\ensuremath{~\lambda\lambda}1548, 1550}
\newcommand{\SIVdoublet}{\ion{S}{4}\ensuremath{~\lambda\lambda}1393, 1402}
\newcommand{\MGdoublet}{\ion{Mg}{2}\ensuremath{~\lambda\lambda}1239.9, 1240.4}
\newcommand{\SIItriplet}{\ion{S}{2}\ensuremath{~\lambda\lambda}1250, 1253, 1259}
\newcommand{\SIIdoublet}{\ion{S}{2}\ensuremath{~\lambda\lambda}1250, 1253}

\newcommand{\vlsrno}{\ensuremath{V_{\rm lsr}}}
\newcommand{\vlsr}{\vlsrno~}
\newcommand{\vobsno}{\ensuremath{V_{\rm obs}}}

\newcommand{\Nabs}{81}
\newcommand{\Npos}{30}

\begin{document}
\title{The Local \lya Forest. I. Observations with \\
   the GHRS/G160M on the Hubble Space Telescope
\footnote{Based on observations with the NASA/ESA Hubble Space Telescope, obtained at the Space
Telescope Science Institute, which is operated by the Association of Universities for Research in
Astronomy, Inc. under NASA contract No. NAS5-26555.}}
\author{Steven V. Penton, John T. Stocke, and J. Michael Shull 
\footnote{Also at JILA, University of Colorado
and National Institute of Standards and Technology.}}
\affil{Center for Astrophysics and Space Astronomy, 
Department of Astrophysical and Planetary Sciences,
University of Colorado, Boulder CO, 80309}
\email{spenton@casa.colorado.edu, stocke@casa.colorado.edu, 
mshull@casa.colorado.edu}
\shorttitle{HST/GHRS observations of the \lowya Forest}
\shortauthors{Penton, Stocke, \& Shull}
\begin{abstract}

We present the target selection, observations, and data reduction and
analysis process for a program aimed at discovering numerous, weak
(equivalent width $\leq 100$\mang) \lya absorption lines in the 
local Universe ($0.003 \leq z \leq 0.069$). The purpose of this program
is to study the physical conditions of the local intergalactic medium,
including absorber distributions in Doppler width  and \hone column density,
redshift evolution of absorber number density, line-of-sight
two-point correlation function, and the baryonic content
and metallicity. By making use of large-angle, nearby
galaxy redshift surveys, we will investigate the
relationship (if any) between these \lya absorbers and galaxies,
superclusters and voids.
In Paper~I, we present high resolution ($\sim 19$\kmsno) spectroscopic
observations of 15 very bright ($V\leq 14.5$) AGN targets made with the
Goddard High Resolution Spectrograph (GHRS) aboard the Hubble Space
Telescope (HST). We find \Nabs\ definite ($\ge 4\sigma$) and \Npos\ possible
($3-4\sigma$) \lya absorption lines in these
spectra, which probe a total pathlength of 116,000\kms 
($\Delta z \sim 0.4$) at very low redshift ($z \leq 0.069$) and column density
($12.5\leq \logNh \leq 14.5$). We found numerous metal lines arising in the
Milky Way halo, including absorption from high velocity
clouds in 10 of 15 sightlines and numerous absorptions intrinsic to the target AGN. 
Here, we describe the details of the
target selection, HST observations, and spectral reduction and analysis.
 We present reduced spectra, absorption line lists, ``pie
diagrams'' showing the known galaxy distributions in the direction of
each target, and nearest galaxy tables for each absorber. In Papers~II and III, we use the data presented here to
determine the basic physical characteristics of the \lowzya forest
and to investigate the relationship of the 
absorbers to the local galaxy distribution.

\end{abstract}
\section{Introduction}\label{sec:obs}
Since the discovery of the high-redshift \lya forest over
25 years ago, these abundant absorption features in the spectra
of QSOs have been used as evolutionary probes of the intergalactic
medium (IGM), galactic halos, large-scale structure, and chemical 
evolution. These absorbers were originally divided into ``metal-bearing"
systems, which occur primarily at higher \hone column densities, and
\lyano-only systems, which occur primarily at lower \hone column 
densities \citep{Sargent87}. Recent observations of  \ion{C}{4} 
lines in \lya absorbers with column densities 
$\nh \geq 10^{14.5}$ cm$^{-2}$ \citep{Tytler95,Cowie95}
have blurred the distinction between these two population of absorbers.
However, the clustering properties of the high-redshift C IV systems are more
like those of galaxies than the \lya clouds \citep{Cristiani97}.
Certainly many of the higher column density absorbers, particularly the
Lyman limit systems \citep{Steidel95} and damped-\lya systems \citep{Wolfe95},
appear to be associated with the gaseous halos of
individual galaxies.  Thus, the distinction between these two 
populations may still be an important one.  

In the past few years, these discrete \lya lines have been 
interpreted theoretically
by N-body hydrodynamical models \citep{Cen94,Hernquist96,Zhang97,Dave99} as arising from baryon density
fluctuations associated with gravitational instability during the epoch
of structure formation. However, these models do not contain the 
detailed physics of star formation and supernova ejecta, which can
move material from the disks of galaxies into the halo or beyond 
in a ``galactic fountain" \citep{Shapiro93,Slavin93}.  Therefore, these simulations cannot exclude 
the possibility that at least some \lya absorbers are ``recycled''
gas from sites of active star formation in galaxies. Regardless, the 
lowest column density absorbers ($\nh \leq 10^{14.5}$ cm$^{-2}$) 
are the most likely systems to be unassociated with
galaxies and primordial in their chemical abundances.

At high redshift, ground-based observations indicate 
 rapid evolution in the distribution of 
\lya absorption lines per unit redshift, 
$d{\cal N}/dz \propto (1+z)^{\gamma}$, where $\gamma \approx 2.5$ 
for $z \geq 1.5$. This evolution is consistent with a
picture of these features as highly ionized ``clouds'' whose numbers
and sizes were controlled by the evolution of the IGM pressure, the
metagalactic ionizing radiation field, and galaxy formation.
If the evolution in cloud numbers at $z > 1.6$  continued into the present epoch, few 
\lya clouds would remain by $z\sim0$. However, one of the delightful 
spectroscopic surprises from the first year of the {\it Hubble 
Space Telescope} (HST) was the discovery of many \lya 
absorption lines toward the quasar 3C~273 at $z_{\rm em}=0.158$ by 
both the Faint Object Spectrograph \citep[FOS,][]{Bahcall91b} and 
the Goddard High Resolution Spectrograph \citep[GHRS,][]{Morris91,Morris93}.  
More recent data from HST support the slowing of the evolution of
\lya systems at $z\leq1.5$ \citep{Weymann98}. 
This evolutionary shift is probably a result of the collapse and 
assembly of baryonic structures in the IGM \citep{Dave99} 
together with the decline in the intensity of the 
ionizing radiation field \citep{Haardt96,Shull99b}. 
Detailed results of the \lya forest evolution in the redshift 
interval $0 < z < 1.5$ are described in the FOS Key Project papers:  
the three catalog papers \citep{Bahcall93,Bahcall96,Jannuzi98} 
and the evolutionary analysis \citep{Weymann98}. 

Because the HST/FOS studies were primarily detections of strong \lya 
lines, with equivalent widths (EWs) greater than 0.24\ang, they most likely
probe only the higher column density population of absorbers. 
A great deal more information about the \lowz IGM can be gained
from studies of the more plentiful weak absorbers, including the
evolution of the lower column density population, which may be
physically distinct. Realizing the 
importance of spectral resolution in detecting weak \lya 
absorbers, the Colorado group has engaged in a long-term program with
the HST/GHRS, using the G160M grating at 19\kms resolution
to study the very low-redshift ($z < 0.07$) \lya forest.
Earlier results from our study have appeared in various  papers 
\citep{Stocke95,Shull96,pks} and reviews \citep{Shull97,Shull99a}.
While other groups have
used the GHRS to probe these lower column density clouds 
\citep[e.g.,][]{Tripp98a,Tripp98b,Impey99},
 their observations used the
lower-resolution GHRS/G140L grating, which failed to detect the
weakest \lya lines and did not resolve the stronger ones.   
The program described here detects lines as weak as EW $\cong 13$\Mang 
($N_{\rm HI} \sim 10^{12.5}$ cm$^{-2}$), comparable to the detection 
limit reached with the Keck Observatory plus HIRES spectrograph 
at high redshift \citep{Songaila95}.  Because the HST/G160M 
resolves most of the \lya absorption lines, it provides,
for the first time, the range of observed Doppler parameters 
(\bvaluesno) and the two-point correlation function of the 
\lowzya forest at velocity separations 
$ \leq 100$\kmsno. 

In order to attain the required resolution and sensitivity, we made
 two important compromises in the observing strategy. 
First, only the brightest UV targets could be used for the program; 
the rare Seyferts, QSOs, and BL Lacs with V\about14.5 are our primary candidates.
 Second, because the
HST/G160M grating yields only \about35\Ang of useful coverage for \lya
forest lines,  little pathlength is surveyed per
target compared to either the FOS Key Project survey observations or
the ground-based spectra at higher redshift. However, in many ways the \about8,000\kms pathlength surveyed by these spectra are the most
important piece of the \lya forest, being the nearest \about100
Mpc to us.  The distances to \lya absorbers discovered in the
present survey allow an immediate comparison with galaxy locations derived 
from redshift surveys that sample approximately the same range in distances
\citep[e.g., the CfA redshift surveys, ][]{CFA}. One of the first
discoveries of our program was that some \lya absorbers are found
in nearby galaxy ``voids", providing the first detection of gas in those
voids \citep{Stocke95,Shull96}. Additionally, the
small distances (on the cosmic scale) to these absorbers allow deeper
galaxy surveys in the optical and 21~cm emission line, 
from which the most sensitive limits have been set on the non-detection 
of galaxies nearby to \lya clouds \citep{VG93,VG96,Morris93,Rauch96,pks,Impey99}.
Observations of the latter type led us to investigate strong \lya
systems toward \PKS\ \citep{pks} with very low
metal abundance, which may be among the first detected primordial \hone 
clouds in the local Universe. 

The first set of four papers discuss
our full GHRS dataset of low column density, \lowzya absorption
systems. In Paper~I (this article)
we describe our HST/GHRS observations and catalog of \lya
absorbers. In Paper~II \citep{PaperII}
we describe the primary scientific results from our program, including
information on the physical parameters mentioned above and the nature of the 
low-redshift \lya forest. In Paper~III \citep{PaperIII} we discuss the relationship between
the \lya absorbers discovered in our GHRS program and the nearby
galaxy distribution from available galaxy redshift surveys. In Paper~IV \citep{PaperIV}, we will
combine these GHRS observations with results from 13 STIS sightlines to discuss the two-point correlation function
of the \lowzya forest.
These \lowzya absorption lines, combined with information
about the distribution of nearest galaxies, can
probe large-scale baryonic structures (filaments) in the IGM, some of which may be remnants of physical conditions
set up during the epoch of galaxy formation. 

In this first paper, we describe data from the HST/GHRS
 of very bright AGNs, which we use to detect low column density (\lognh~$= 12.5-14.5$) \lya
absorption clouds at very low redshift ($\z < 0.07$). The GHRS with the G160M grating delivers spectral resolution R\about15,000. Subsections of this paper describe the target selection and  basic observational setup
(\S~\ref{sec:setup}), calibration (\S~\ref{sec:cal}), spectral processing (\S~\ref{sec:sp}), and a thorough description
of each sightline (\S~\ref{sec:each}). This paper concludes with a tabulation of all quantities directly measured from
the spectra.
\section{HST Observations, Calibration, and Spectral Processing}\label{sec:setup}
\subsection{HST Observations}
To study the local \lya forest, we used all available HST GHRS/G160M 
spectra that meet the following simple criteria:
	(1) Available: public domain GHRS/G160M spectra or our own HST Guest Observer observations, and (2)
Appropriate: spectra blueward of the target's \lya emission and in the wavelength
range  \wrange\ (\zrange\ for \lyano).
The majority of these observations were obtained from our HST Guest Observer projects 
3584, 5715, and 6593. We selected our targets based upon two factors: brightness and location in the sky.
To determine the brightest targets in the UV, we used the IUEAGN database
 \citep{IUEAGN}
to coadd all IUE spectra  of AGN to determine the brightest UV
targets with appropriate redshifts. We limited our selection to targets with \zem$ < 0.18$, to
avoid having
\lyb appear in our wavelength band of interest (\wrange), but we required \zem$ > 0.025$ to
maximize pathlength beyond the region absorbed by the Milky Way interstellar \ion{H}{1}. 
In project 3584 (which carried over into project 5715) we 
observed 4 objects behind well-studied galaxy voids in the foreground of the ``Great Wall''
 \citep{Stocke95,Shull96}.  In the hope of determining
the relationship between galaxies and \lya absorbers in the \lowz universe, we selected this region of the sky where the CfA
galaxy redshift surveys were most complete \citep{CFA,Giovanelli84}.  Owing
to the lack of sufficiently bright targets in the direction of the first CfA slice, in project 6593 we observed 7 additional
targets in other directions.  Our sample is rounded out by 5 archival spectra and a few additional exposures on targets that we
had already observed.  One of our archival targets is at higher
\z\ than the above limit (\z=0.297;\ H~1821+632) but is also included.  
In this sightline we take special care to differentiate
\lya lines from other lines, such as \lyb for \z~\gt~0.19 absorbers.  
Recent FUSE observations of several of these sightlines \citep{Shull00} detect \lyb corresponding
to the stronger \lya absorbers, with EWs~\gt~200\mang. Our survey of \lowzya clouds continues with
HST/STIS observations of an additional 13 targets, which will be reported elsewhere.

 Table~\ref{observations} summarizes the HST observations and observational parameters of the forty-one GHRS/G160M exposures
towards the 15 targets included in our dataset. The first column in Table~\ref{observations}  identifies the 
observed target, while the second column lists the HST exposure or dataset identifier. Additional columns
provide: (3) the exposure calendar date, (4) whether the COSTAR (Corrective Optics Space Telescope Axial
Replacement) optics were in place at the time of the exposure, (5) the exposure time (in seconds), (6 \& 7) the wavelength
extent of the spectrum, (8) the HST/GHRS wavelength calibration (SPYBAL) exposure used in the calibration of the spectrum, and
 (9) the HST proposal identification number (PID) and Principal Investigator (PI) of the observation.

Table~\ref{objects} summarizes the J2000 positions (columns 2-5) and redshifts (\z; column 6)
of our HST targets. All redshifts are the optical, narrow-line, emission redshifts as reported from 
the NASA/IPAC Extragalactic Database (NED).\footnote{The NASA/IPAC Extragalactic Database (NED) is operated by the Jet
Propulsion Laboratory, California Institute of Technology, under contract with NASA.}  Also included in Table~\ref{objects}, but discussed in detail later, are, by column: (7) the Galactic
\hone velocity centroid relative to the Local Standard of Rest (LSR) along this sightline as determined from 
the Leiden/Dwingeloo or Parkes multibeam survey, (8)
the adjustment required to convert the observed wavelength scale to the LSR, where 
$\Delta \vlsr = \vlsr - \vobsno$, 
as determined from the location of the \ion{S}{2} Galactic absorption lines (1250, 1253, and 1259\ang, see \S~\ref{sec:LSR})
(9) the order ({\bf O}) of the polynomial
used to normalize the continuum of each target, (10) the dominant
\hone\ 21~cm velocity of the Galactic High Velocity Clouds (HVCs) along each sightline as reported by \citet{Wakker91}, and
 (11) the mean signal-to-noise ratio (S/N) per resolution element (RE) of each spectrum. The Galactic \hone HVC velocities are taken from
\citet{Wakker91}, and the Galactic
\hone LSR velocity measurements are taken from the Leiden/Dwingeloo survey \citep{Dwing}.
Note that \objectname[]{Fairall 9},
Mrk~335 and  \PKS\ are located near the Magellanic Stream \citep{GibsonMS}, while Mrk~279, Mrk~290, Mrk~501,
Mrk~817, and H~1821+643, are close to the HVC complex C \citep{GibsonHVCC}. The sightline to Mrk~421 lies between the Galactic HVC
MI and MII clouds, and the Mrk~509 sightline passes  near the Galactic center negative velocity HVC feature.
These objects are therefore more likely to contain Galactic absorption features at velocities other than the LSR.
\setlength{\tabcolsep}{2mm}
\begin{center}
\begin{table}[htbp] \caption{HST GHRS/G160M Observations}
\scriptsize
\label{observations}
\begin{tabular}{lllcrcccl}
\ \\
\tableline\tableline
Target &  Dataset & Date & COSTAR &  ET~~~
& $\lambda_{\rm start}$& $\lambda_{\rm end}$ & Spybal &  PID/PI\tablenotemark{a} \\
       &          &      & Deployed ? & (sec)~~ & (\noang) & (\noang)  & & \\
\tableline
\ \ (1)&\ \ (2)&\ (3)&(4)&(5)\ \ &(6)&(7)&(8)&\ \ \ \ \ (9)\\
\tableline
3C273  &  z0gu010mn  &  02/23/91  &  No &  1824.8  &  1265.430  &  1301.460  &  z0gu010nm  &  1140/WEYMANN \\
3C273  &  z0gu010om  &  02/23/91  &  No &  2737.1  &  1234.680  &  1270.790  &  z0gu010pm  &  1140/WEYMANN \\
3C273  &  z1760105t  &  11/27/94  & Yes &  1115.1  &  1214.110  &  1250.270  &  z1760103t  &  3951/WEYMANN \\
3C273  &  z1760107t  &  11/27/94  & Yes &  1115.1  &  1214.130  &  1250.290  &  z1760106t  &  3951/WEYMANN \\
3C273  &  z1760108t  &  11/27/94  & Yes &  1115.1  &  1214.130  &  1250.290  &  z1760106t  &  3951/WEYMANN \\
AKN120  &  z3e70604t  &  12/26/96  & Yes &  5702.4  &  1222.542  &  1258.674  &  z3e70603t  &  6593/STOCKE\\
AKN120  &  z3e70606t  &  12/26/96  & Yes &  5702.4  &  1222.587  &  1258.707  &  z3e70605t  &  6593/STOCKE\\
AKN120  &  z3e70608m  &  12/27/96  & Yes &  5702.4  &  1222.609  &  1258.726  &  z3e70607t  &  6593/STOCKE\\
AKN120  &  z3e7060at  &  12/27/96  & Yes &  2433.0  &  1222.638  &  1258.751  &  z3e70609t  &  6593/STOCKE\\
ESO141-G55  &  z3e70204t  &  08/15/96  & Yes &  7248.4  &  1222.511  &  1258.638  &  z3e70203t  &  6593/STOCKE \\
ESO141-G55  &  z3e70206t  &  08/15/96  & Yes &  7248.4  &  1222.542  &  1258.662  &  z3e70205t  &  6593/STOCKE \\
ESO141-G55  &  z3i70105t  &  10/09/96  & Yes &  9085.0\tablenotemark{b}  &  1228.710  &  1265.920  &  z3i7010ct  &  6451/SAVAGE \\
FAIRALL9  &  z26o0208n  &  04/09/94  & Yes &  7096.3  &  1231.660  &  1267.750  &  z26o020at  &  5300/SAVAGE \\
FAIRALL9  &  z3e70404m  &  08/01/96  & Yes &  13407.0  &  1219.702  &  1255.843  &  z3e70403t  &  6593/STOCKE \\
FAIRALL9  &  z3e70406t  &  08/01/96  & Yes &  6412.0  &  1240.312  &  1276.384  &  z3e70405t  &  6593/STOCKE \\
H1821+643  &  z15f0208m  &  04/17/93  & Yes &  6386.7  &  1231.660  &  1267.750  &  z15f0207t  &  4094/SAVAGE \\
H1821+643  &  z27n0108m  &  04/05/94  & Yes &  9123.8  &  1231.660  &  1267.750  &  z27n0107t  &  5299/LU \\
IZW1  &  z1a60404n  &  09/09/93  &  No &  24178.1  &  1221.401  &  1257.544  &  z1a60403t  &  3584/STOCKE \\
IZW1  &  z2ia0204n  &  02/07/95  & Yes &  21897.2  &  1221.299  &  1257.437  &  z2ia0203t  &  5715/STOCKE \\
MARK279  &  z3e70304t  &  01/17/97  & Yes &  5702.4  &  1222.537  &  1258.673  &  z3e70303t  &  6593/STOCKE \\
MARK279  &  z3e70306t  &  01/17/97  & Yes &  5702.4  &  1222.576  &  1258.703  &  z3e70305t  &  6593/STOCKE \\
MARK279  &  z3e70308t  &  01/17/97  & Yes &  5702.4  &  1222.615  &  1258.737  &  z3e70307t  &  6593/STOCKE \\
MARK279  &  z3e7030at  &  01/17/97  & Yes &  1419.3  &  1222.643  &  1258.762  &  z3e70309t  &  6593/STOCKE \\
MARK290  &  z3kh0104t  &  01/10/97  & Yes &  1622.0  &  1231.760  &  1268.950  &  z3kh0103t  &  6590/WAKKER\\
MARK290  &  z3kh0105t  &  01/10/97  & Yes &  2433.0  &  1231.780  &  1268.970  &  z3kh0103t  &  6590/WAKKER\\
MARK290  &  z3kh0107t  &  01/10/97  & Yes &  2230.3  &  1231.810  &  1268.990  &  z3kh0106t  &  6590/WAKKER\\
MARK335  &  z1a60304n  &  09/03/93  &  No &  13787.1  &  1221.408  &  1257.550  &  z1a60303t  &  3584/STOCKE \\
MARK421  &  z2ia0104t  &  02/02/95  & Yes &  9085.0\tablenotemark{c}  &  1221.426  &  1257.569  &  z2ia0103t  &  5715/STOCKE \\
MARK501  &  z1a65204m  &  02/28/93  &  No &  29196.3  &  1221.418  &  1257.561  &  z1a65203t  &  3584/STOCKE \\
MARK509  &  z1790208m  &  05/23/93  &  No &  6082.6  &  1231.730  &  1267.840  &  z1790207t  &  3463/SAVAGE \\
MARK509  &  z3e70704t  &  10/19/96  & Yes &  4257.8  &  1219.468  &  1255.610  &  z3e70703t  &  6593/STOCKE \\
MARK817  &  z3e70104t  &  01/12/97  & Yes &  8870.4  &  1222.568  &  1258.704  &  z3e70103t  &  6593/STOCKE \\
MARK817  &  z3e70106t  &  01/12/97  & Yes &  8236.8  &  1222.616  &  1258.741  &  z3e70105t  &  6593/STOCKE \\
MARK817  &  z3e70108t  &  01/12/97  & Yes &  7907.3  &  1222.645  &  1258.767  &  z3e70107t  &  6593/STOCKE \\
PKS2155-304  &  z1aw0106t  &  05/11/93  &  No &  1558.7  &  1222.580  &  1258.700  &  z1aw0105t  &  3965/BOGGESS \\
PKS2155-304  &  z1aw0107t  &  05/11/93  &  No &  1558.7  &  1222.590  &  1258.710  &  z1aw0105t  &  3965/BOGGESS \\
PKS2155-304  &  z1aw0108m  &  05/11/93  &  No &  1520.6  &  1222.600  &  1258.720  &  z1aw0105t  &  3965/BOGGESS \\
PKS2155-304  &  z3e70504t  &  10/06/96  & Yes &  3801.6  &  1257.580  &  1293.628  &  z3e70503t  &  6593/STOCKE \\
PKS2155-304  &  z3e70505t  &  10/06/96  & Yes &  3168.0  &  1257.634  &  1293.669  &  z3e70503t  &  6593/STOCKE \\
Q1230+0115  &  z3cj0105t  &  07/11/96  & Yes &  5474.3  &  1216.960  &  1254.200  &  z3cj0103t  &  6410/RAUCH \\
Q1230+0115  &  z3cj0108t  &  07/11/96  & Yes &  5474.3  &  1217.000  &  1254.230  &  z3cj0106t  &  6410/RAUCH \\
\tableline
\end{tabular}
\footnotesize
\tablenotetext{a}{HST Proposal Id number (PID) / HST Principal Investigator (PI).}
\tablenotetext{b}{An HST anomaly caused the final exposure to be cut short by 38.8 seconds.}
\tablenotetext{c}{Subexposures 5 through 15 were unusable due to an HST pointing error. The
originally scheduled observation was 14598.1 seconds in length.}
\end{table}
\end{center}
\normalsize 
\setlength{\tabcolsep}{2mm}
\setlength{\tabcolsep}{2mm}
\begin{center}
\begin{table}[htbp]
\caption{HST GHRS/G160M Targets}
\scriptsize
\label{objects}
\begin{tabular}{lcccccrrcrr}
\ \\
\tableline\tableline
Target&RA (J2000) &DEC (J2000)&$l$&$b$&\z&
\vlsrno\tablenotemark{a}~(H~I)&
$\Delta \vlsrno$\tablenotemark{b}&O\tablenotemark{c}&V$_{\rm hvc}$\tablenotemark{d}&S/N\\
      &(hh:mm:ss)&($\pm$dd:mm:ss)&($^{\circ}$)&($^{\circ}$)&&(km/s)&(km/s)&&(km/s)&\\
\tableline
(1)&(2)&(3)&(4)&(5)&(6)&(7)&(8)&(9)&(10)&(11)\\
\tableline
3C273&12 29 06.8&+02 03 07.8& -70.05&  64.36&  0.1583&          -5&  10.5&          13&         115&27.2\\
AKN120&05 16 11.4&-00 08 59.4& -158.31& -21.13&  0.0331&           5&  -2.7&          14*&        -116&22.2\\
ESO141-G55&19 21 14.3&-58 40 14.9& -21.82& -26.71&  0.0371&           0&  -2.6&          20*&         200&25.8\\
FAIRALL9&01 23 45.9&-58 48 20.9&   -64.93& -57.83&  0.0461&          -6&  -8.8&          20*&         190&32.7\\
H1821+643&18 21 57.1&+64 20 36.7&  94.00&  27.42&  0.2968&          -1&  46.5&           6&        -180&20.0\\
IZW1&00 53 34.9&+12 41 36.3&   123.75& -50.17&  0.0607&          -5&  33.3&          14&        -100&13.2\\
MARK279&13 53 03.4&+69 18 29.6&  115.04&  46.86&  0.0306&           5&  50.0&          25*&        -135&30.0\\
MARK290&15 35 52.4&+57 54 09.6&   91.49&  47.95&  0.0296&          -5&  27.9&          9*&        -120&17.1\\
MARK335&00 06 19.5&+20 12 10.3&  108.76& -41.42&  0.0256&          -3&  30.5&          25*&        -400&33.3\\
MARK421&11 04 27.4&+38 12 30.8&  179.83&  65.03&  0.0300&          -8&  38.2&           5&        -110&23.9\\
MARK501&16 53 52.2&+39 45 36.6&   63.60&  38.86&  0.0337&           0&  55.4&          9&        -115&15.3\\
MARK509&20 44 09.8&-10 43 24.5&   35.97& -29.86&  0.0344&           4&  15.1&          23*&        -240&28.4\\
MARK817&14 36 22.1&+58 47 39.4&  100.30&  53.48&  0.0313&          -2&  19.6&          27*&        -115&42.7\\
PKS2155-304\tablenotemark{e}&21 58 52.0&-30 13 32.3&  17.73& -52.25&  0.1165&          -4&  17.5&           4&        -200&20.3\\
PKS2155-304\tablenotemark{f}&21 58 52.0&-30 13 32.2&  17.73& -52.25&  0.1165&          -4&   3.0&           3&        -200&34.2\\
Q1230+0115&12 30 50.0&+01 15 21.7& -68.74&  63.66&  0.1170&          -5&   1.1&           7&         115& 9.2\\
\tableline
\end{tabular}
\footnotesize
\tablenotetext{a}{The velocity centroid (in \kmsno), relative to the LSR, of the Galactic H~I along this sightline as determined
from the Leiden/Dwingeloo or Parkes multibeam survey.}
\tablenotetext{b}{$\Delta \vlsr$ = $\vlsr - \vobsno$, as determined from the location of the S~II Galactic lines (1250, 1253, and 1259\ang).}
\tablenotetext{c}{The order of the polynomial used to normalize the spectrum. An asterisk (*) indicates that a broad 
Gaussian component was included to model intrinsic \lya emission, present in the observed band.}
\tablenotetext{d}{LSR velocity in km s$^{-1}$ of the dominant HI-HVC in this direction as reported by \citet{Wakker91}.}
\tablenotetext{e}{Values for Pre-COSTAR exposure covering  1222--1258\ang.}
\tablenotetext{f}{Values for Post-COSTAR exposure covering 1257--1294\ang.}
\end{table}
\end{center}
\normalsize
\subsection{GHRS Calibration}\label{sec:cal}
	All of our HST/G160M spectra were taken with the Large Science 
Aperture (LSA) and are a mixture of pre- and post-COSTAR observations as noted in
Table~\ref{observations}.
All data were re-calibrated with the final GHRS recommendations 
\citep{Sherbert97}, using the familiar standard IRAF/STSDAS/CALHRS environment,
with the following exceptions:
\begin{enumerate}
\item {\it Background Subtraction}: We find that the null order polynomial 
(identified in CALHRS nomenclature as {\bf PLY\_CORR}) background subtraction gives
superior results to the  median ({\bf MDF\_CORR}) and mean ({\bf MNF\_CORR}) background
options, mainly  due to the inappropriately small size of the default filter box.

\item {\it Wavelength Calibration}: As listed in Table~\ref{observations}, the indicated
 GHRS spectrum Y-balances (SPYBAL's) were used during the recalibration process to correct
the  initial wavelength zero point offsets.

\item {\it Blemishes and Degradation Accounting}: To avoid any possible 
detector artifacts from appearing as weak spectral features, 
photocathode blemishes, dead, noisy or ``flaky'' diodes, pixels with 
Reed-Solomon decoding errors (which indicate possible data transmission errors) are removed from the spectra. For this 
reason, some spectra have incomplete wavelength coverage, while 
others have varying S/N ratios across the spectrum.
\end{enumerate}

All subsequent data reductions were performed using the Interactive Data
Language (IDL) of Research Systems, Inc (\url{http://www.rsinc.com}).
Because of the low S/N of the individual data files, the subexposure 
coadditions were performed in wavelength space weighted by exposure 
time, and not by using the IRAF default STSDAS/CALHRS routines {\bf poffsets} and {\bf specalign}. To achieve
this, an exposure time vector was created that defines the actual on-target 
exposure time for each pixel. The exposure time for each sub-exposure is calculated from the
information in the FITS header.  Specifically, for each
exposure, the following procedure was followed:
\begin{enumerate}
	\item If the spectra were taken in FP-SPLIT mode, coadd all spectra at identical 
carrousel positions by exposure time. 
	\item Check the quality of the sub-exposures for anomalies.
	\item Recalculate the exposure time from the dataset headers and 
compare this to the STDSAS/FITS {\bf EXPTIME} header keyword. 
	\item If the spectral subexposures were not quarter-stepped, resample the spectra 
to quarter-stepped (quarter GHRS diode) pixels, conserving flux and adjusting the error 
vectors accordingly. 
	\item Smooth the spectra to the resolution appropriate for the line-spread function (LSF)
of the detector at the time of observation. The pre-COSTAR large science aperture (LSA)
line spread function (LSF) can be represented by a core of Full Width Half Maximum (FWHM)=1.2 diodes 
(\about20\kms at 1240\ang) and broad wings \citep{Gilliland93}. The
post-COSTAR GHRS LSF is well-modeled \citep{Gilliland92} by a  single sharp Gaussian with FWHM=1.1
diodes or \about19\kms in  our wavelength range. The smoothing is performed by convolving the LSF
of the GHRS/LSA at the time of observation with the flux and error 
vectors.
	\end{enumerate}

Once all the subexposures are in place, we used the following procedure to determine the final spectrum:

\begin{enumerate}
	\item Use the first subexposure of the first dataset to define the 
wavelength scale.
	\item Populate the final flux, error and exposure time vectors with the first 
subexposure values, except where any anomaly appears in the 
calibration flags. Possible anomalies include photocathode blemishes, faulty diodes, and decoding errors as
previously described.
	\item Coadd the next exposure, weighting by the exposure time in the 
flux vector.
Add the exposure times to the exposure vector.  Assume Poisson 
statistics and combine, in quadrature, the exposure's 1$\sigma$ error vector into the merged error
vector. The  coaddition is
performed pixel by pixel into the closest wavelength bin of the first 
exposure. If no wavelength
bin is detected within one-half of a quarter-stepped pixel width, a new wavelength bin is created and populated. This most 
commonly occurs at the edges of the spectra, as the coverage is
extended by multiple observations.
	\item Repeat until all exposures of this dataset are merged.
	\item If the flux level of the target is constant, or if the 
temporal separation between observations is small, combine all 
subexposures of all appropriate datasets to the existing merged 
exposure. If the flux levels between exposures at different epochs are not constant, then we
scale the flux level of the subsequent exposures to the first spectrum for each target listed in
Table~\ref{observations}. The error vectors of these exposures are also scaled to preserve the
proper S/N weighting. The scaling is determined by
the average fluxes in the overlap region of the exposures. For \ESO\ (HST~exposure~z3i70105t),
Markarian~509 (HST~exposure z3e70704t), and Fairall 9 (HST~exposures z3e70404m and
z3e70406t), we allow a scaling that varies linearly across the overlap region to compensate for any
``Baldwin Effect''.\footnote{The ``Baldwin Effect'', first noted in \citet{Baldwin77}, pertains to the
observed systematic decrease of the equivalent width of AGN line emission (e.g., \lyano) with an increase in
continuum luminosity. This changes the continuum slope in the vicinity of the \lya
emission in the indicated spectra.}  For \PKS, the pre- and post-COSTAR observations covered
different wavelength ranges and are analyzed separately.
\end{enumerate}
\subsection{Spectral Processing}\label{sec:sp}
\subsubsection{Line Identification: Local Continuum Method and \bvalues}\label{sec:local}
	To begin the spectral processing, we treat all spectral fluctuations with~\gt~1$\sigma$ negative departures from the
continuum as our  initial absorption line list. Linear continuum and Gaussian features 
(with observed Doppler velocities constrained in the range of $12 < \bobs < 100$\kmsno) are fitted to 
a 4\Ang region centered on prominent features. 
This wide range of \bvalues allows us to
include all possible absorption lines in these spectra. Gaussian features are fitted to
the functional form, $\exp\left[-(\lam-\lam_c)^2 / 2 \Dld^2 \right]$, where $\lam_c$ is the line center and
the observed Doppler widths (\bobs) are related to the Gaussian widths by 
\WG~=~\bobs/$\sqrt{2}$, and \Dld$ = \left(\lambda/c\right)$\WG. The \lya \bvalue is related to the physical properties
of the absorbing gas by :
\begin{equation}{\label{bval} \bb = \sqrt{{2kT \over m_p} + V_{turb}^2} = \sqrt{V_{therm}^2 + V_{turb}^2} ~~~~,}\end{equation} 
where $T$ is the temperature of the cloud, $m_p$ is the proton mass, $k$ is the Boltzmann constant,
and $V_{turb}$\ represents non-thermal turbulent motion of the absorbing gas. 
We selected Gaussian components because, at the spectral resolution and S/N of these observations, 
we are unable to distinguish between Gaussian and Voigt line profiles at high confidence level, and
at the column densities that we are sampling, Gaussian and Lorentzian profiles are virtually indistinguishable.
In addition it is unclear whether we are resolving discrete individual
absorption clouds or nonÐdiscrete  ensembles of neutral hydrogen spread in velocity space.
Up to 10 Gaussian
components are fitted simultaneously to each region of interest, with  features taken from the 
initial line list. If more
than 10 Gaussian  components are required, or if the continuum is non-linear, we decrease the 
wavelength region accordingly until the fitting can 
proceed. When we encounter obvious asymmetries, features broader than \bobs=100\kmsno, or distinct
absorption minima, we count them as additional
features and add them to the line list.

We repeat this procedure of adding Gaussian components and refitting all features when we encounter
an absorption feature broader than 100\kmsno, until all
modeled features have \bobs~\lt~100\kmsno. Occasionally, we relax this limit to allow
strong intrinsic and Galactic features to have \bobs~=~100\kmsno. 
This minimizes the number of components in
exceptionally  broad features that are almost certainly not intervening absorptions. 
Although we allow absorption features  with \bobs$ \leq 12$\kms  to 
remain in the initial absorption line list, they are removed after we perform a continuum normalization, 
or  by our selection of a $\ge 3\sigma$ significance cutoff. 		
In addition, we observe broad low-contrast absorption troughs  (\bobs~\gt~100\kmsno) 
in some of the low S/N spectra. We interpret these as continuum features, detector artifacts, or possible
nondiscrete absorption features.  In these cases, we fit the troughs with Gaussians to maintain the  continuum integrity, but
we do not count them as \lya absorption  features.

We express the strength of an absorption feature in terms of its observed equivalent width (EW) in\mang.
An accurate determination of the EW  depends on an accurate modelling of the continuum level in the absence of any
absorption. 
For the majority of this paper, we are interested in the rest-frame equivalent widths, \W~=~EW/(1+\z).
We quote \W in all Tables. However we often use the abbreviation EW in contexts where the difference
between observed and rest-frame values are unimportant. 
We calculate the significance level (SL) of each potential absorber by integrating the S/N per 19\kms 
FWHM resolution element (RE) of the best-fit Gaussian. 
This SL is different (usually larger) than that indicated by the uncertainty in \Wno, which is derived from the uncertainties of the best-fit
Gaussian components.
Features with \real\ are termed ``definite'' absorbers, while features with \tent\ are termed ``possible'' absorbers.
The continuum and absorption feature fits are performed in IDL using a 
Levenberg-Marquardt \Xtwo minimization implementation based on MINPACK-1 \citep{More93,Markwardt99}.

The measured \bvalues are independent of the EW measurements and are important determinations
in their own right. These are the first \bvalue measurements published at \lowzno, and are
essential for converting \W to column density in higher column density absorbers.
To obtain \bvaluesno, the measured Gaussian widths (\WG) are deconvolved using the {\bf smoothed} LSFs of the
pre- and post-COSTAR data (as appropriate) since the data themselves have been smoothed as
described in the previous section. In the limit of very high S/N data, this smoothing could decrease
our ability to determine accurate \bvalues for very narrow features. But, for the modest S/N present
in the current data, two effects ``trade-off'' against each other to make small \bvalues equally accurate in smoothed and 
unsmoothed data: The error inherent in fitting the data with a Gaussian and the error
in determining whether that Gaussian is distinctly broader than the LSF. We see no evidence that smoothing
these data have made it more difficult to measure very narrow \bvaluesno; indeed a couple
of \bvalues at $\leq$ 15\kms are found. As \citet{Lu96} have pointed out on the basis of simulations using
moderate S/N ratio spectra, these very \lowz \bvalues have large errors and are underestimates
of the true line widths due to the presence of noise. 
Specifically, since the data are inherently smeared by the instrumental profile, and by
our smoothing, both with $\sigma_{gauss}\about8.07$\kmsno, our measured \bvalues are related to our smoothed
observed \bvalues (\bobs$ = \sqrt{2}$~\WG) by:
\begin{equation}{\label{b_equation}
 \bmsd= \sqrt{2} \sqrt{{\rm \bf W}_G^2 - 2\times (8.07\kmsno)^2} = \sqrt{ \bobs^2 - (16.14\kmsno)^2} .
}\end{equation}
As discussed in Paper~II, the measured \bvalues in pre-COSTAR data are occasionally artificially large due to spacecraft ``jitter''.
In all Tables reporting \bvaluesno, we always list the \bvalues corrected by equation~\ref{b_equation} above.
\subsubsection{Continuum Normalization: Global Continuum Method}
\label{sec:global}
Because a major source of uncertainty in all EW determinations is the
continuum placement, we also performed a global continuum normalization to give two independent
estimates (``local'' vs. ``global'') of the continuum placement.
	We used polynomial fits to model a global continuum
and normalize each spectrum.   The order of the continuum-fitting polynomial was
selected by  performing F-tests on increasing polynomial orders until adding an additional order 
 is not justified \citep{Mandel84}. 
For each spectra, the order of the normalizing polynomial is indicated in Table~\ref{objects} and 
is restricted to be less than 30.  For objects with broad intrinsic \lya emission 
(Akn~120, \ESO, Fairall~9, Mrk~279, Mrk~290, Mrk~335, Mrk~509, and Mrk~817), we include a single Gaussian
to improve the continuum fit near the emission feature.
%This procedure occasionally gives less than ideal results near the \lya emission peak of the
%targets. This is not a serious concern in our studies, since we are concerned only with the \lya
%absorbers that are not intrinsic to the targets.  
The continuum fits are made on the raw data, with $3 \times $FWHM 
(Full Width at Half Maximum) regions 
centered on  each 3$\sigma$ absorption feature replaced by continuum 
estimates plus added Gaussian noise. Including noise in the replaced spectral regions was
necessary to avoid biasing the global continuum modeling. The unnormalized spectra
of each target and the polynomial fits  are shown in the spectral plots (Appendix~A). 

\begin{figure} 
\plotone{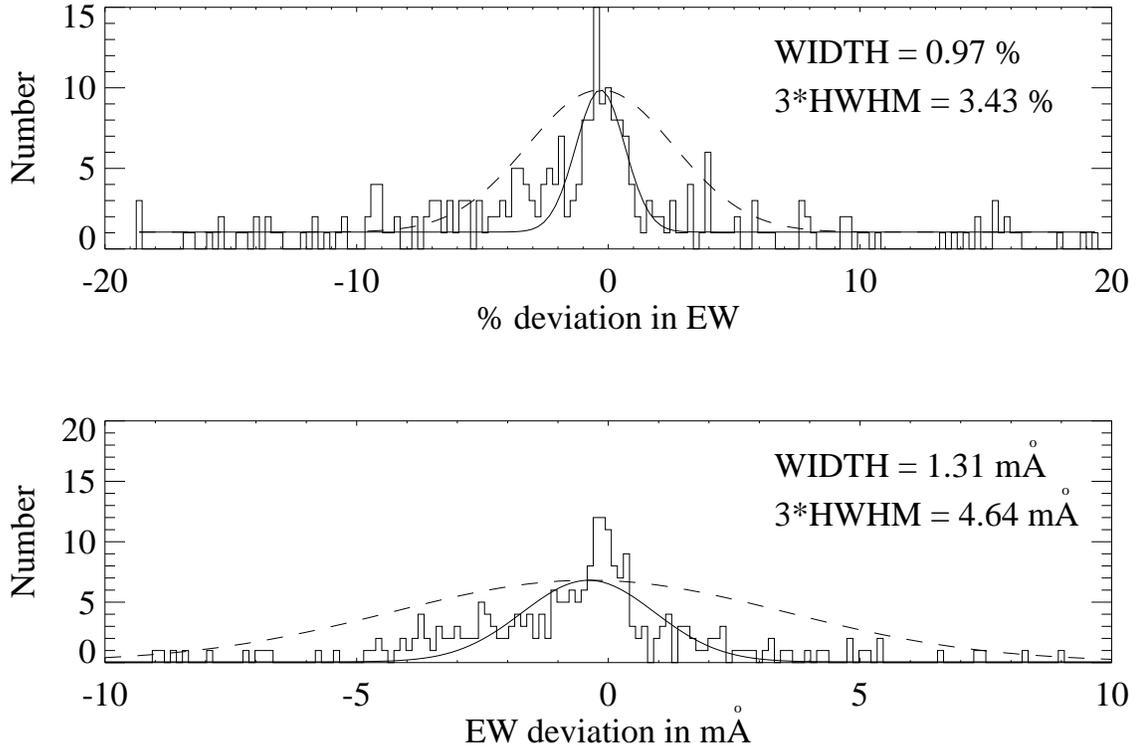}
\caption{\label{local_vs_global} The upper panel shows the
percent deviation in absorption feature EWs of the ``local'' continuum
determinations versus the ``global'' determinations. WIDTH~indicates
the width of the Gaussian that best fits the distribution (shown as a solid curve). The
dashed curve represents a Gaussian with a width of $3\times$ the HWHM (half width at half
maximum) of the best-fit distribution. This 3$\sigma$ deviation represents the additional 
systematic EW uncertainty introduced in the spectral normalization.  The bottom panel shows the same
deviations in units of\mang.}
\end{figure}

 From the difference in the measured values of EW obtained from the ``local'' and ``global'' continuum methods,
we conservatively overestimate the contribution of the continuum uncertainty to the EW and significance level (SL)
determinations. In Figure~\ref{local_vs_global}, we compare the deviations of EWs using the
local continuum fits from the global continuum results. The differences are well-fitted by a Gaussian with a
HWHM (half width at half maximum) of 1.1\%. Taking this value to be the 1$\sigma$
error estimate, we conservatively add a 3$\sigma$ (3.4\%) additional continuum
placement uncertainty in quadrature with the statistical uncertainties of the EW, and hence the significance levels, of
all features.  Note that this systematic continuum placement error is an additional uncertainty 
above and beyond the
statistical uncertainty of absorption feature plus continuum included in the error estimates. 
%This error represents the additional uncertainty introduced during global continuum normalization. 
Although it is clear from Figure~\ref{local_vs_global} that our continuum normalization correction is a considerable
overestimate for most absorption features, a few words are needed to explain the ``outliers'' 
 at $> 10$\% difference in Figure~\ref{local_vs_global}. 
These outliers occur mostly from blended lines or
clustered absorption features located on the wings of significant continuum features, such as the 
target's \lya emission. Multi-component fits of blended  absorption lines find a \Xtwo solution that shares
the EW between the absorption components. 
This increases the uncertainty in determining the EW of the individual blended components, but not the total EW.
Even slight continuum adjustments can significantly alter the EW absorption ratio between the blended features,
while leaving the total absorption almost unchanged. This produces outliers in
Figure~\ref{local_vs_global} that are not associated with the continuum normalization error, but rather 
arise from special individual
circumstances. Hence, our procedure does not underestimate the total errors for these outlier features.
Indeed, because a continuum placement error is rarely included in the error budget for absorption-line EWs, we have been
conservative, but we believe correct, in including this additional uncertainty.
\subsubsection{Wavelength Adjustment to Local Standard of Rest (LSR)} \label{sec:LSR}
Once all features have been modeled, we use the wavelength of the strongest 
components of the Galactic \ion{S}{2} triplet, 1250.584\ang, 1253.811\ang, 1259.519\Ang to determine the offset
$\Delta \vlsr = (\vlsr - \vobsno)$ that can be applied to adjust 
the observed, approximately heliocentric, wavelength  scale of the GHRS spectra to the LSR frame. These adjustments are
given in column 10 of Table~\ref{objects}.
This technique assumes that the \ion{S}{2} lines occur at the same LSR velocity as the dominant component of the
Galactic \ion{H}{1}, determined from the Leiden/Dwingeloo survey \citep{Dwing}
for all objects except Fairall 9 and \ESO. 
For these two objects, the Galactic \hone velocities relative to the LSR
were determined from the Parkes multibeam survey \citep{Haynes99}. 

The Galactic \hone velocities relative to the
LSR along each sightline are given in column 9  of Table~\ref{objects}. 
It is our opinion that this LSR-corrected wavelength
scale is more accurate than the heliocentric scale provided by the GHRS recalibration because the precise location of the target
in the LSA is unknown. Owing to target motion in the LSA, pre-COSTAR observations can have velocity offsets as large as 70\kms 
(1\arcsec\ in the LSA), while post-COSTAR observations can have velocity offsets as large as 60\kms (0.85\arcsec\ in the LSA). As
described in the HST Data Handbook\footnote{Version 3.1: March 1998, Section 37.6.5, ``Wavelength Data Uncertainty",  available
from \url{http://www.stsci.edu} .},  there are a number of other effects, such as thermal motion, that can significantly
affect the wavelength scale of an observation.  By consistently applying our LSR adjustment, we obtain a
more precise wavelength scale. 
We estimate that the accuracy  with which H~I and S~II velocities can be matched is $\pm$ 5\kmsno, limited by the
width and structure of the H~I 21~cm profile. This amount of velocity uncertainty is added in quadrature to the individual
uncertainties of the features based upon the Gaussian fits. This total velocity uncertainty of individual absorption 
features is $\le$ 10\kmsno, and it is this total value which is  listed in all tables.
Column 8 of Table~\ref{objects} lists these velocity offsets for each target.

We construct the final line list by refitting all features in the  manner described above in the normalized 
(``global'' continuum method) LSR. We use the  Gaussian centers, heights, and widths (\WG, related to the \bvalueno) 
from the ``local'' continuum method  as the
initial values for the final \Xtwo ``global continuum'' minimization to determine the final line lists. As described above,
we determine the statistical significances of all lines  using an error estimate that includes both the statistical
error in the
\Xtwo fitting technique and the systematic error associated with continuum placement.
\subsubsection{Galactic Spectral Features}\label{sec:gal}
The \wrange\ range contains only a few strong Galactic interstellar lines \citep[see Table~\ref{linelist_galactic},][]{Morton91}. These lines, although useful for correcting the wavelength scale to LSR, also
eliminate a small portion of our available pathlength for \lya detection. 
A full discussion of the pathlength availability appears in Appendix~A. Table~\ref{linelist_galactic} lists the rest wavelength
for each of the Galactic lines in our waveband, as well as $\log$($\lambda f$) and $\lambda f$. Unsaturated lines of the same
species and ionization state are expected to have EW ratios that match the $\lambda f$ ratios. For
example,
\NV{1238} and $\lambda$1242 are expected to have \Ws\ in the ratio of 2:1. These ratios allow for distinction
between Galactic absorption and intervening \lya absorption at expected Galactic line positions. For many of these
features, the appearance of multiple lines of the same species in our waveband allows us to identify Galactic features
based upon expected wavelengths and strengths. These features are denoted by a ``{\bf g:}'' in the line lists. Since
highly ionized species may have different velocities within our Galaxy, we allow an offset of up to 0.2\Ang (\about50\kmsno) for definite (\real) Galactic feature identifications. Possible (\tent) Galactic
identifications are allowed a larger offset range up to 0.5\Ang (\about120\kmsno) and are denoted by a question mark in
the line identifications. 

Often, Galactic absorption is detected at velocities other than the LSR, due to well-known high-velocity clouds (HVCs) within
our Galaxy. These detections are indicated by a ``{\bf
h:}'' in the identification field of our line lists. Our observations are unique in that they are able to place upper
limits on  metallicity of the HVCs in several species, \ion{Mg}{2}, \ion{S}{2}, and, in some cases, \ion{Si}{2}. Individual detections
of HVCs will be reported both in the line lists and in later sections.  A detailed analysis of the HVC
results toward Complex C and the Magellanic Stream will appear elsewhere \citep{GibsonHVCC,GibsonMS}.

Galactic and HVC absorption lines remove a portion of our available pathlength for \lya detection. At the
location of each definite, or possible, Galactic and HVC absorption line, we remove a portion of the spectrum equal 
to the FWHM of the absorption feature  on each side of the identified line center.
 We suspect that occasional weak intervening \lya features are superimposed upon Galactic/HVC features.
However, if we cannot conclusively determine this occurrence based upon expected line strengths, we consider the
superimposed portion of the spectrum unusable for \lya detection. In such cases, we report the total absorption as
Galactic/HVC and remove this portion of the available pathlength for intervening \lya detection.
\setlength{\tabcolsep}{2mm}
\begin{center}
\begin{deluxetable} {lclr}
%\tabletypesize{\small}
\tablecaption{Galactic Lines\label{linelist_galactic}}
\tablewidth{0pt}
\tablehead{
\colhead{Ion} &
\colhead{Wavelength (\AA)} &
\colhead{$\log(\lambda f)$} &
\colhead{$\lambda f$}
}
\startdata
%O~I & 971.738 & 1.082 & 12.08\\
%H~I & 972.537 & 1.450 & 28.18\\
%C~III & 977.020 & 2.872 & 744.69 \\
%H~I & 1025.722 & 1.909 & 81.10 \\
%O~VI & 1031.926 & 2.137 & 137.09 \\
%C~II & 1036.337 & 2.106 & 127.64\\
%NII & 1083.990 & 2.048 & 111.73 \\
%SiII & 1190.416 & 2.474 & 297.85\\
%SiII & 1193.290 & 2.775 & 595.66\\
%NI & 1199.550 & 0.163 & 1.45 \\
%NI & 1200.223 & 0.108 & 1.28 \\
%SiIII &1206.500 & 3.304 & 2013.7 \\
%HI & 1215.670 & 2.704 & 505.82 \\
N~V & 1238.821 & 2.289 &194.54 \\
Mg~II & 1239.925 & -0.106 & 0.784 \\
Mg~II & 1240.395 & -0.355 & 0.442 \\
N~V & 1242.804 & 1.988 & 97.28 \\
S~II & 1250.583 & 0.834 & 6.82\\
S~II & 1253.811 & 1.135 & 13.65 \\
S~II & 1259.519 & 1.311 & 20.46 \\
Si~II & 1260.422 & 3.104 &1270.57 \\
Fe~II & 1260.533 & 1.499 & 31.55\\
C~I & 1260.736 & 1.696 & 49.66 \\
C~I & 1277.245 & 2.091 & 123.31\\
C~I & 1280.135 & 1.493 & 31.12 \enddata
\end{deluxetable}
\end{center}
\normalsize
\subsubsection{Intrinsic, Intergalactic, and Non-\lya Absorbers} \label{sec:IIA}
Many AGN are observed to possess ``intrinsic'' absorbers at $z_{\rm abs}$\about\zem.
One measure of the appropriateness for excluding potential intrinsic absorption
systems from our line list is the ``Proximity Effect'' \citep{Bajtlik88} or ``Inverse Effect'' \citep{Murdoch86},
which refer to the observed
decrease of absorption line density, \dndzno, when approaching the target AGN's emission redshift.  This decrease has been
interpreted as due to increased photoionization of the absorption-line systems by the
target AGN's  UV radiation field \citep{Carswell82}. To avoid any possible skewing
of our statistics by this effect, we include only absorption systems in which the
ionizing radiation from the target AGN is 10\% or less of the metagalactic 
ionization radiation field. At low redshift,  the metagalactic
ionizing intensity at 912\Ang has been estimated to be $I_0 = 1.3_{-0.5}^{+0.8} \times 10^{-23} $
ergs\percmtwo\persecond\perhz sr$^{-1}$ \citep{Shull99b}. Assuming that the mean value of the specific ionizing luminosity
at 912\Ang from the IUEAGN database,  
$L_{912} = 6 \times 10^{29}$ erg\persecond\perhzno, is representative of our HST targets, we can
calculate the ``standoff'' distance ($D$) from the target QSO at which the QSO's ionizing flux decreases to $I_0/10$ :
\begin{equation}
{	L_{912} \over (4 \pi D^2) 4 \pi} = {I_0 \over 10} \Rightarrow D = {1 \over 4 \pi} \sqrt{10~L_{912} \over I_0} =
17.5~\hsfi~{\rm Mpc,}
\end{equation}
which is equivalent to $\cz \approx1,200$\kmsno, where \hseventy is the Hubble constant
in units of 70\kms\perMpcno.

Therefore,	we consider non-Galactic absorption features within 1,200\kms of the redshift of the target to
be either \lya clouds near the AGN affected by the ``proximity effect'' or intrinsic absorption near the target. 
For consistency, we will use the term ``intrinsic'' for all \lya absorptions within 1,200\kms of the target, or metal lines
associated with the target AGN.  Lines identified as \lya outside of this limit will be termed ``intergalactic''. Non-\lya
lines associated with known higher-\z\ systems, such as in H~1821+643, will be termed ``non-\lyano'' absorbers.

\begin{figure} 
\plotone{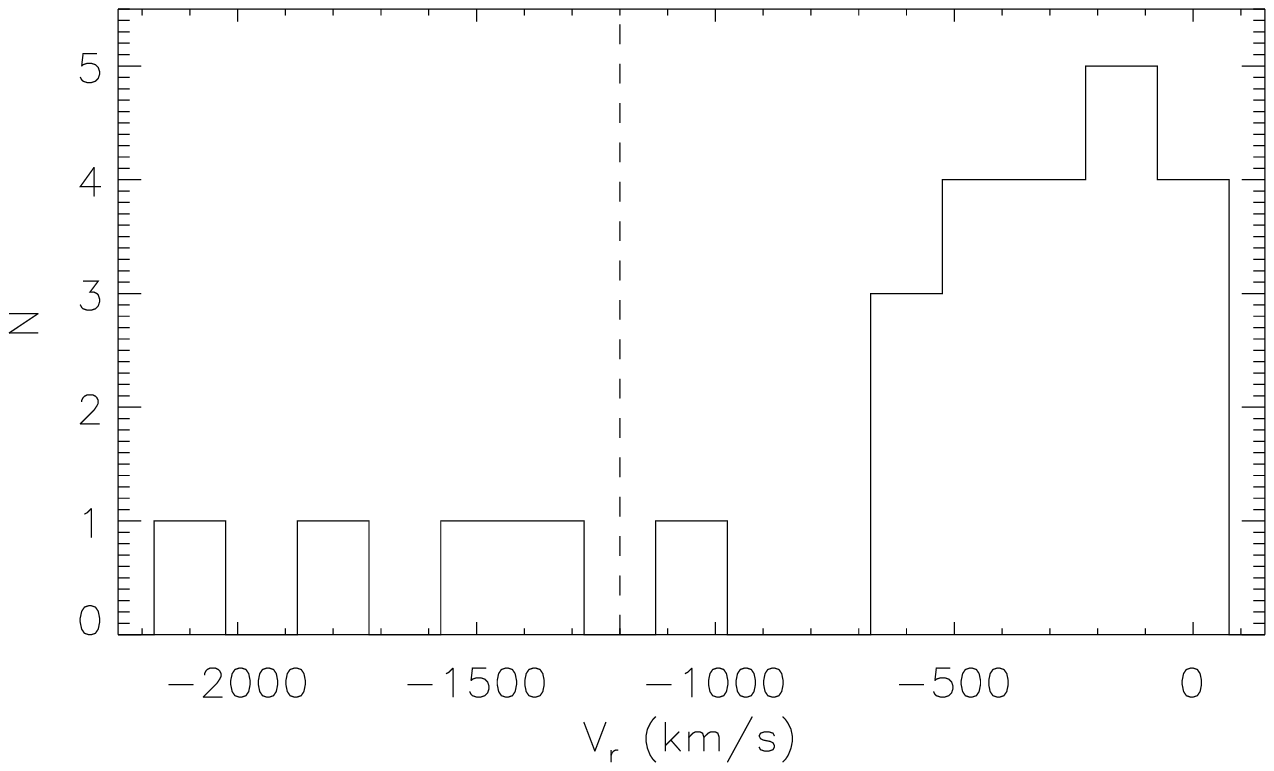}
\caption{\label{crenshaw} Radial velocity
distribution relative to the target (\Vr in\kmsno) of the HST UV
sample of Seyfert 1 galaxy ``intrinsic'' absorbers  presented by
\protect{\citet{Crenshaw99}}. The dashed vertical line at $-$1,200\kms indicates
our blueward radial velocity limit for determining which \lya features
are intrinsic or ``associated'' with the target.}
\end{figure}

\begin{figure}
 \plotone{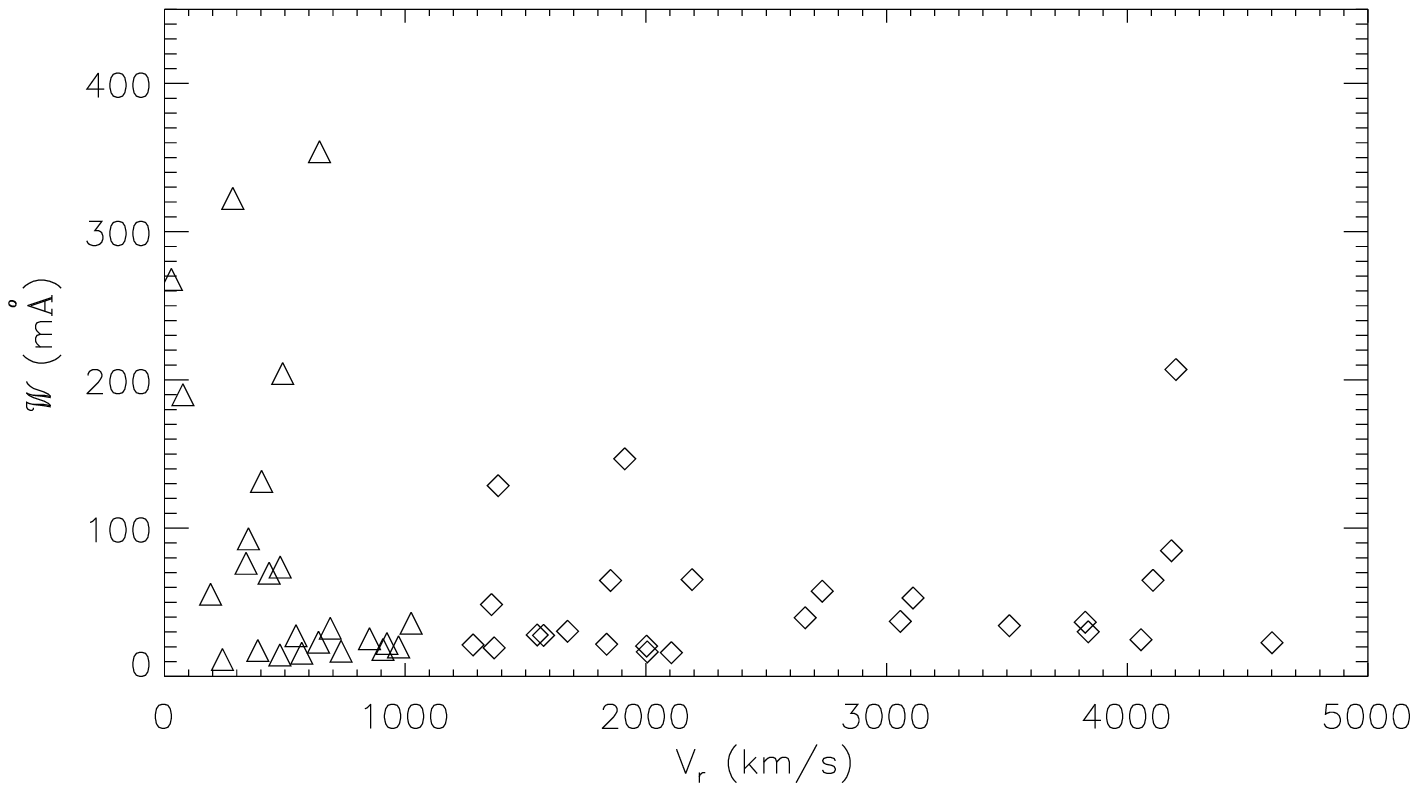}
\caption{\label{beta} Distribution of
rest-frame EW (\Wno) versus radial velocity relative to target
(\AVr\ in\kmsno)  of our sample. The triangle absorbers are
those determined by our criterion to be intrinsic (\AVr\ \lt~1,200\kmsno), while the diamonds indicate intergalactic
\lya absorbers.}
\end{figure}

Lines intrinsic to the target AGN are not included in any of our \lowzya forest statistics and are marked by an ``{\bf i:}'' prefix in the 
identification fields of the target line lists. Likewise, the
pathlength within 1,200\kms of the AGN is excluded from our total observed pathlength. Our choice of a 1,200\kms
radial velocity limit relative to the target (\Vr) limit is supported by the analysis of intrinsic absorption lines in
Seyfert 1 galaxies by \citet{Crenshaw99}.  In Figure~\ref{crenshaw}, we display this sample, obtained from ultraviolet HST/FOS 
and GHRS \lyano, \CIVdoublet, \NVdoublet, and \SIVdoublet\ detections, in terms of number (N) of intrinsic  
systems versus \Vr.
Five of the 17 targets in \citet{Crenshaw99} are also in our sample. As
shown by the vertical line at $-$1,200\kms in Figure~\ref{crenshaw}, our radial velocity limit includes the vast majority,
but not quite all, of the detected intrinsic systems.  These results support the claim
that our blueward velocity limit is reasonable, but not perfect, in protecting against any outflowing ``intrinsic''
absorbers associated with the target. In Figure~\ref{beta}, we plot 
\AVr\ versus rest-frame EW (\Wno) for all \lya absorbers in our sample with \AVr\ \lt~5,000\kmsno.
Notice that several absorbers with \AVr\ \lt~1,200\kms have \W \gt~200\mang, while
only one of the absorbers identified as intergalactic has \W \gt~200\mang. This further supports our assumption that a 1,200\kms
blueward velocity limit accounts for the majority of intrinsic or outflowing absorbers.  In terms of pure Hubble flow
distance, this blueward velocity limit is well in excess of the 230\ \hsfi kpc galactic halo size inferred by
\citet{Lanzetta95}, and probably eliminates any possible \lya absorptions associated with the
targets.

We stress that any velocity limit that purports to separate intergalactic from intrinsic systems is somewhat
uncertain.  While we have presented several pieces of evidence that 1,200\kms is a reasonable velocity limit, it is
impossible  unambiguously to classify any individual \lya absorber as either intrinsic or intergalactic. Any use of the
intrinsic or intergalactic line lists must take this uncertainty into account.

	A few of our targets, such as H~1821+643, have known intergalactic \lya absorption systems with redshifts greater than our
survey cutoff. Higher Lyman series lines or metal-line absorptions associated with these systems that fall within our
wavelength range (\wrange) are termed ``non-\lyano''. These features are indicated by a ``{\bf z:}'' in the identification
field of the line lists. As with the intrinsic features, these features are not included in any of our
\lowzya forest 	statistics. Additionally, the pathlength obscured by these features is excluded from the available
pathlength for \lya detection using the same method as the Galactic/HVC features.
\section{Results by Object}\label{sec:each}
\begin{figure}
 \plotone{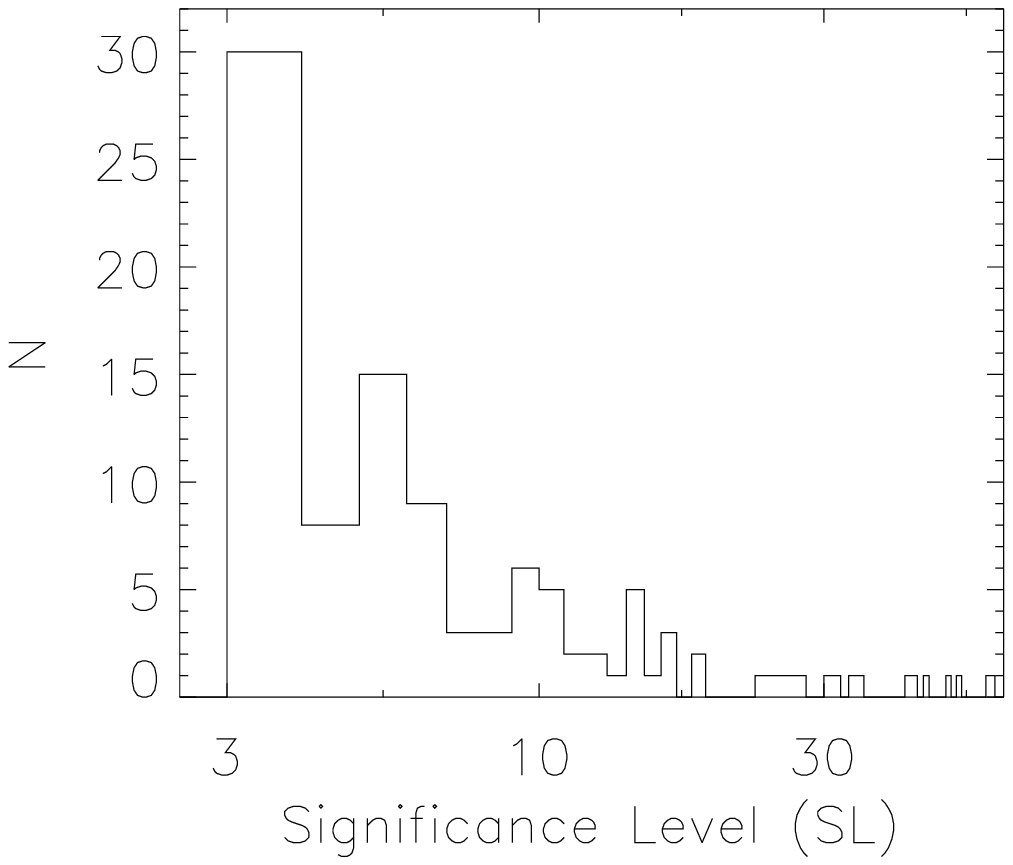}
\caption{\label{N_SL} Significance level (SL) distribution of our detected intergalactic \lya absorbers.}
\end{figure}

In this section, the observations of each target are discussed in detail. 
Composite spectra (Figures~\ref{3C273}--\ref{Q1230+0115}) and 
a spectral line identification table (Table~\ref{linelist_all}) appear in Appendix~A, for our sightlines in the same
order presented here. The wavelength scales have been corrected to the LSR by equating the Galactic
\hone LSR velocity in each sightline (Table~\ref{objects}, column 7) 
with the location of the \SIItriplet\ Galactic interstellar lines in the
observed waveband.  To provide an overall sense of the number and strength of our detections, as well as the quality of our
spectra, Figure~\ref{N_SL} shows the number of reported intergalactic \lya absorbers as a function of significance level (SL).
%
% 3C273
%
\subsection[3C~273]{3C~273}
The sightline to \objectname[]{3C~273} is perhaps the most studied of all \lowz QSO targets 
\citep{Bahcall91a,Bahcall91b,Bahcall93,Brandt93,Brandt97,Morris91,Morris93,Weymann95}.
 \citet{Morris93} provide a summary of all HST
spectra available as of 1993, including those observed with the GHRS+G160M, GHRS+G140L, and  FOS+G130.  Refitting all features with
their JASON software, \citet{Morris93} report 12 features in our wavelength range of interest ($z < 0.07$).
Our spectrum combines the two GHRS G160M spectra summarized in \citet{Morris93} with three additional
1218--1250\Ang spectra obtained in 1994 (see Table~\ref{observations}). This results in a 1218--1300\Ang spectrum shown
in Figure~\ref{3C273}, the only one in our sample that covers our entire spectral range of interest. Our analysis of these
spectra, recalibrated as discussed earlier, reveals 18
\lya absorption features at \expanded, of which 13 have \real.  Our results differ from those of
\citet{Morris93} in the following regards:
\begin{enumerate}
\item At $\lambda \approx$ 1224.2--1225.0\Ang we see a broad absorption trough at the red edge of the
Galactic damped \lya line that we model with two
\lya absorption features greater than
4$\sigma$ totalling \about64\mang. \citet{Bahcall91a} report a 240\Mang feature at 1224.52\ang,
while \citet{Bahcall93}, \citet{Brandt93}, and \citet{Bahcall91b} do not report this feature. 
\item The 1280.3\Ang feature (not given in any other line lists) was determined to be a \lya absorber, not Galactic
\CI1280, owing to the absence of the expected \CI1277 in our spectrum.
%\item \citet{Morris93} report a 68\Mang \lya feature at 1276.5\Ang (4.5\signo). We detect this feature, but only at the 2.3
%\sig (23\mang) level. Therefore, it is not included in any of our tables or statistics.
%\item We note that the 130\Mang feature at 1255.6\Ang could possibly be redshifted \ion{N}{2}\ $\lambda1084$ intrinsic to
%3C~273 at \z=0.15831. However, due to its strength we consider this to most likely be intergalactic \lya line and we include this
%line in our \lya sample.
\end{enumerate}
Also detected along this sightline are the Galactic features \MgII{1239.9}, \NVdoublet, \SIItriplet, and
\SFsixty~(which is possibly blended with \CI1260).  Based upon the absence of a \MgII{1240.4} detection,
it is possible that the 1240.01\Ang detection could be
\lya instead of the reported \MgII{1239.9}.

As discussed in \citet{Sembach96} and \citet{Burks94}, the sightline towards 3C~273 passes through the edges of Galactic radio loops I
and IV near the  North Polar Spur ($l,b$)=(290\degr, 64.4\degr) and near the
Sagittarius spiral arm.  Although no HVCs are reported in this direction, we do detect Galactic
\NV{1238}, \SIItriplet, and \SFsixty~at infalling velocities up to 50\kmsno. This is most prevalent in
the peculiarly shaped \NV{1242} absorption.

In \citet{Weymann95}, the 1220\Ang and 1222\Ang \lya features were measured to have \bvalues of $40.7 \pm 3.0$\kms and $34.3
\pm 3.3$\kmsno, respectively.  They analyzed the same GHRS observations as we did, using Voigt profile
fitting on spectra that were carefully cross-correlated at the subexposure level. We elected not to include such
cross-correlations in our standard data reduction because the vast majority of our GHRS subexposures were of insufficient S/N
to accomplish this  procedure with confidence.
Positional shifts of the target in the HST aperture can cause sub-diode spectral shifts in the subexposures. This causes us
occasionally to overestimate the \bvalues of the absorption features, but it does not affect our measurements of the centers,
EWs, or significance levels of the absorption features.  The consequences of this drift on our measured
\bvalues will be explored in Paper~II.   The LSR offset for this spectrum was 0.044\Ang or 10.5\kms at 1253\ang. 
%
% Akn120
%
\subsection[Akn~120]{Akn~120}
The spectrum of \objectname[]{Akn~120} (Arakelian~120, also known as Mrk~1095) covers 1223 to 1259 \Ang and has an average S/N per 
resolution element of 22. Our spectrum of Akn~120 contains 5 definite and 2 possible \lya absorbers,
 one intrinsic \lya absorption, and strong Galactic detections in \MGdoublet, \SIIdoublet, and \NV{1238}. 
Based upon only a 0.9\sig detection of \NV1242, it appears that
\NV{1238} (which appears extremely broad) is partially contaminated by an intergalactic \lya system. The 0.9\sig \NV1242
feature is below our limit for inclusion. \NV1242 is listed as the alternate
identification for the 1242.9\Ang \lya feature because its spectral 
proximity falls within our generous limits
(\S~\ref{sec:gal}). However, due to the strength of the 1242.9\Ang feature, we are confident of the identification as \lyano.
Of special note is the well-resolved and closely spaced ($\Delta v = 60$\kmsno) pair of strong \lya 
absorbers at 7960 and 8020\kmsno.  
This pair is flanked to the blueward by two additional \about4\sig absorbers at 7792 and 7867\kmsno.
The dominant HVC \hone velocity, relative to the LSR, in this direction \citep{Wakker91} is $-$116\kmsno. We
not detect this HVC in any of the Galactic lines in our waveband. For \SIIdoublet, we do not detect 
this HVC at the 4$\sigma$ \W limit of 20\mang. The Akn~120 sightline was corrected to the LSR using the
Galactic H~I and only two \ion{S}{2} lines (1250, 1253\ang).  The LSR offset for this spectrum was $-0.011$\Ang or $-2.7$\kms at 1253\ang. 
%
% ESO
%
\subsection[ESO141-G55]{\ESO}
Our spectrum of \objectname[]{ESO~141-G55} contains no definite absorbers and only one possible intergalactic \lya absorber.
 This spectrum has a good S/N (\about26 per RE), so that
we are sensitive to 4$\sigma$ \lya absorptions below 40\Mang for a large portion of the spectrum.
This sightline does contain strong Galactic \NVdoublet, \MGdoublet, \SIItriplet, and \SFsixty. 
We also detect an HVC at \vlsr $\cong -65$\kms in \SIItriplet, \SFsixty, and possibly \NVdoublet\ (1.7\sig and 1.4\sig
respectively).  We also detect a very  strong intrinsic \lya absorption at 1258\ang, or
 \vlsr $\cong -640$\kms relative to \ESO.  This sightline does not pass through any known radio loops or Galactic
\ion{H}{1}-HVCs. The flux level of the z3i70105t HST exposure was scaled to be consistent with the other \ESO\ exposures (see
Table~\ref{observations}). The scaling was applied linearly across the overlap region with the other exposures to compensate
for a noticeable ``Baldwin Effect'' between exposures.  
The LSR offset for this spectrum was $-0.011$\ang ~or $-2.6$\kms at 1253\ang. 
%
% F9
%
\subsection[FAIRALL9]{FAIRALL~9}
We detect 8 definite, 1 possible, and 6 intrinsic \lya absorption features in our spectrum of Fairall 9.
The spectrum of Fairall 9  includes 3 separate post-COSTAR exposures, one of which was reported in
\citet{Lu94}. The two additional exposures were obtained in 1996, two
years after our first exposure. The additional exposures allow us to confirm and refine the earlier results on both the \lya
absorptions and the Galactic/HVC absorptions.	The flux levels of the additional HST/G160M exposures,
z3e70404m and z3e70406t, were scaled to be consistent with the initial Fairall~9 exposure. The scaling was applied linearly
across the overlap region with the initial exposure. The feature identified as  \NV{1238.8} in Table~\ref{linelist_all},
is probably blended with a \lya feature based upon the marginal 1.8\sig (\W$ = 8 \pm 6$\mang) detection of \NV1242. Given the
expected 2:1 ratio of these Galactic absorption features, the 48$\pm$11\Mang absorption feature at 1238.8\Ang is anomalously
strong and asymmetric.  However, given the uncertainty in determining the relative strength and \bvalue for the possible \lya absorption, we
have chosen to exclude this feature and the associated pathlength from our \lya survey. 

The Fairall 9 sightline passes through the Magellanic Stream (MS), which is detected in our waveband
 (1220--1276\ang) in \SFsixty~and \SIIdoublet. 
In addition, the Galactic \SFsixty~absorption feature appears to be blended with a MS
\SII{1259} absorption. \hone 21~cm emission \citep{Morras83} reveals two high velocity MS components at \about160\kms
(\nh\about\Exp{2}{19} cm$^{-2}$) and 200\kms (\nh\about\Exp{6}{19}\percmtwo). 	
\citet{Lu94} report two high velocity components of  \SiII1526 at +170\kms and +200\kmsno. In our \SIItriplet\ detections of the MS,  we are unable to resolve the MS clearly into separate components.
\citet{Savage97} report detection of this HVC in \SFsixty. We confirm this detection
centered at +180\kms and covering the range +100 to +260\kmsno.  We were able to model this absorption with two features centered at
\vlsr~=~+128\kms and +190\kmsno. 	 We do not detect the MS in  \NVdoublet\  or in \MGdoublet\ at the
10\Mang level.
% A velocity ``stack" plot of the Galactic features in the \objectname[]{Fairall 9} sightline is shown
%in Figure~\ref{GAL_FAIRALL9}.  
Assuming that the MS~\SIIdoublet\ absorption features are on the linear portion of the
curve of growth, \citep{GibsonMS} find a MS metallicity of [S/H]$ = -0.52 \pm 0.04$. This
compares to the LMC and SMC values of $-0.57 \pm 0.09$ and $-0.68 \pm 0.15$, respectively \citep{Russell92}. 

While several absorbers are detected in the wavelength range 1262--1266\ang, the rapidly rising continuum on the blue
wing of \lya calls their reality into question. Specifically, due to our concerns about the reality of these features, 
we have refit the continuum for Fairall~9 by altering the order of the global fit significantly. The statistical significances
of these features remains robust despite these changes, so we list them as definite or possible \lya features as shown in 
Figure~\ref{FAIRALL9}.

The z3e70404m exposure of Fairall~9 contains a ``medium" photocathode blemish in the 1254--1256\Ang range. This blemish,
combined with the coaddition of multiple exposures with different wavelength ranges, produces a notch in the sensitivity
detection limit  near 1255\ang. The LSR offset for this spectrum was $-0.037$\Ang or $-8.8$\kms at 1253\ang.
%
% H1821+643
%
\subsection[H1821+643]{H1821+643} The GHRS/G160M spectrum of  \objectname[]{H~1821+643} (as known as QSO E1821+643) covers
1231.66--1267.75\Ang or $0.01315<\z<0.04284$. 
In this spectrum, we detect five definite and seven possible \lya lines,
along with very strong Galactic \NVdoublet\ and
\MGdoublet. \citet{Savage95} report \ion{C}{4}-HVCs along this
sightline at \vlsr$ = -$70, $-$120, and $-$213\kms with column densities of $N_{\rm CIV} = 6, 6$~and \Exp{1}{13}
cm$^2$, respectively. We detect the \vlsr$ = -$70\kms component in \NV{1238} and \SII1250, and the \vlsr
$ = -120$\kms component in 
%\NVdoublet,
S~II~\ensuremath{\lambda\lambda1253, 1259}, and \SFsixty. The individual detection of the $-$70 and $-$120\kms
components is complicated by their spectral proximity.  The LSR offset for this spectrum was 0.187\ang ~or 44.7\kms at 1253\ang. 

Many \lya lines outside of our wavelength coverage have been reported along this sightline. This complicates detection
of \lya absorption features in our spectrum due to the possibility of non-\lya lines associated with the higher \cz\ \lya
absorbers occurring in our waveband.
As reported in \citet{Bahcall93},  \citet{Savage95}, and \citet{Tripp98a,Tripp98b}, this sightline contains 9 known
intergalactic \lya absorbers with \W
\gt~200\Mang that are outside our redshift coverage.
% at \z=[0.12123,0.12157,0.14760,0.16990,0.21325,0.22484,0.22621,0.26163,0.26660].
Also outside our coverage are 19 weaker intergalactic \lya absorbers. Six of these have 100\Mang $\le$ \W \lt~200\Mang
and 13 with  50\Mang $\le$ \W \lt~100\mang.
%\z=[0.06722,0.11974,0.21176,0.21577, 0.21656, 0.22484, 0.25822], at \z=[0.05704, 0.06432,
%0.08910, 0.11152, 0.15727, 0.16350, 0.17915, 0.18049, 0.19794, 0.22782, 0.23864, 0.24132, 0.24514].
In addition to these 28 \lya systems, \citet{Savage95} report a \lyb absorber at  $z$=0.21656. We doubt the
reality of this identification owing to the absence of any \lya absorption at this redshift (1478.9\ang) in 
\citet{Tripp98a,Tripp98b}. We do detect the feature identified as \lyb in \citet{Savage95} at 1247.94\ang, but we identify
it as \lya at z=0.0265.

\citet{Bahcall93} report a 680\Mang intrinsic \lya absorption at 1577\Ang in their FOS data of H~1821+643. In our waveband, we expect 
to detect \CIII977 and \lyg associated with this absorber.  Indeed, we detect these strong features at $z=0.29673 \pm
0.00002$. Since both of these features are present, these detections allow us to shift the observations of \citet{Savage95} and 
\citet{Tripp98a,Tripp98b}  to our LSR velocity scale. Based upon these LSR corrected wavelengths, in Table~\ref{anticipated_1821} we
list all expected intervening absorptions lines from known intervening systems at higher redshift. 
Table~\ref{anticipated_1821} is ordered by strength of
the rest-frame\footnote{Table~6 of \citet{Tripp98a} incorrectly identifies rest-frame values as observed-frame.} \lya absorption (\Wlya),
 and breaks naturally into detections and non-detections.  Table~\ref{anticipated_1821} lists by column: 
(1) expected wavelength of the absorption based upon the redshift of the system and our LSR correction; 
(2) line identification; (3) system redshift; (4)  rest-frame EW of \lya in the system;
(5) reference for the \lya detection; (6) whether this feature was detected in our spectrum; and
(7) comments, including the observed rest-frame EW.

In addition to the \CIII{977.0} and \lyg absorptions at
\z=0.297, we detect 6 (possibly 7) of the 14 expected features, including 3 \lyb and 3 \OVI{1031.92} lines. 
Of these 6 lines, the 1244.6\Ang (\lybno), 1256.5\Ang (\lybno), and 1264.1\Ang (\OVI{1031.92})  
detections were reported by \citet{Tripp98a,Tripp98b}  or \citet{Savage95}. The 1264\Ang \OVI{1031.92} absorption system at
\z=0.225 is studied in detail by \citet{SavageOVI}, \citet{Barlow98}, and \citet{Tripp98a,Tripp98b}.
Our observations suggest that the \z=0.225 system is actually composed of two components separated
by \about 75\kmsno. This has recently been confirmed by \lyd and \lye FUSE observations \citep{Shull00,Oegerle00}.
We report the first detection of \OVI{1031.92} (1252.2\ang) associated with the detected \lyb (1244.6\ang) at \z=0.213.
The unambiguous identification of \OVI{1037.62} associated with this system is not possible due to HVC \SII1259.5 absorption.
However, based upon the strength of the \OVI{1031.92} detection (51\mang),
% and the \OVI1031.92:1037.62 line ratios obtained along this sightline in \citet{SavageOVI}, 
we would expect a \about34\Mang detection of  \OVI1037.62. The \SII{1259.5} HVC
detection (80$\pm$21\mang) is significantly stronger than our \SII{1250.6} and \SII{1253.8} HVC detections (30$\pm$17 and 33$\pm$13\mang) indicating
that we are probably detecting \OVI1037.62 at \z=0.213.
In addition, we report the detection of  \lyb (1257.7\ang) and \OVI{1031.92} (1257.7\ang) associated with the \z=0.226 \lya
detection of \citet{Tripp98a}.
As summarized in Table~\ref{anticipated_1821}, we detect all anticipated absorption features (\lybno, \OVIdoublet, or \CIII{977}) whose
associated \lya absorption has \W \gt~163\mang.

It is important to note that, due to the uncertainties associated with the redshift determinations of the FOS \lya detections and the
limited resolution of the GHRS/G160M, the possibility of feature misidentifications in this sightline is greater than in
the rest of our sample. As a followup study of this target with HST/STIS, we were awarded Cycle~8 observing time to
measure the \CIVdoublet\ absorption features associated with our \OVI1031.9 detections. We detect  \ion{C}{4} for the
strongest systems surveyed and will present these results elsewhere.
%Owing to the non-detection of \lyb associated with the 1265.4\ang, \OVI1031.92 absorption, it is possible that this feature
%is misidentified and is instead another intergalactic \lya in our redshift range. As indicated in
%Table~\ref{anticipated_1821}, \lya associated with this system has an \W of 343$\pm$25\Mang \citep{Tripp98a,Tripp98b}, and
%our 4$\sigma$ sensitivity limit in this portion of the spectrum, for \lyb detection, is \about40\mang. Therefore,
%based upon the uncertainty of the identification of this feature, we exclude the region around this feature from our
%intervening \lya absorber search.
At the extreme blue end of our spectrum, a feature which is possibly \CIII977 associated with the \z=0.261 system, is
marginally detected (2$\sigma$). Due to the systematically higher noise level at the edges of the GHRS spectrum, this feature is
not included by our \expanded\ criteria.
\setlength{\tabcolsep}{2mm}
\begin{center}
\begin{table}
\caption[Anticipated Absorption Features in the H1821+643 Spectrum]{Anticipated Absorption Features in the H1821+643 Spectrum}
\scriptsize
\label{anticipated_1821}
\begin{tabular}{llccccl}
\ \\
\multicolumn{7}{c}{ }\\
\multicolumn{7}{c}{\small Detections \tiny}\\
\tableline\tableline
$\lambda_{\rm exp}$& Identification &\z &\Wlya &Ref&Detected&Comments\\
(\ang)           &                &   &(\nomang)      &&&\\
\tableline
1256.53&Ly$\beta$&0.22489		 &739 $\pm$ 22&   1&Yes&\W = 509 $\pm$ 28\Mang, two components\\
1264.13&OVI~1031.92 &0.22489&739 $\pm$ 22&   1&Yes&\W = 190 $\pm$ 18\mang, \Dv\about75\kms\\
\tableline
1261.25&Ly$\gamma$&0.29673 &524 $\pm$ 8&    1&Yes&\W = 253 $\pm$ 19\Mang\\
1267.07&CIII~977.03&0.29673&524 $\pm$ 8&    1&Yes&\W = 259 $\pm$ 17\Mang\\
\tableline
1244.60&Ly$\beta$&0.21326			&483 $\pm$ 18&   1&Yes&\W = 153 $\pm$ 18\Mang\\
1252.12&OVI~1031.92 &0.21326&483 $\pm$ 18&   1&Yes&\W = 51 $\pm$ 17\mang, blended with \lya at 1252.5\ang\\
1259.03&O~VI~$\lambda$1037.62 &0.21326&483 $\pm$ 18&   1&Probable& \W \about 40\mang, blended with HVC S~II~$\lambda$1259.5 \\ 
\tableline
1257.75&Ly$\beta$&0.22621&280 $\pm$ 20&      2&Yes& \W = 20 $\pm$ 14\Mang \\
1265.36&OVI~1031.92 &0.22621&280 $\pm$ 20&   2&Yes& \W = 16 $\pm$ 12\Mang\\
\tableline
\multicolumn{7}{c}{ }\\
\multicolumn{7}{c}{\small Non-Detections \tiny }\\
\tableline\tableline
$\lambda_{\rm exp}$& Identification &\z &\Wlya &Ref&Detected&Comments\\
(\ang)           &                &   &(\nomang)      &&&\\
\tableline
1237.63&CIII~977.03&0.26660&163 $\pm$ 13&    1&No& \\
\tableline
1247.04&Ly$\beta$&0.21577			&146 $\pm$ 15&   2&No& 	$\lt$  2$\sigma$\\
1254.58&OVI~1031.92 &0.21577&146 $\pm$ 15&   2&Possible&blended with Ly$\alpha$\\
\tableline
1242.93&Ly$\beta$&0.21176&104 $\pm$ 14&      2&No&			Blended with N~V~$\lambda1242$\\
1250.45&OVI~1031.92 &0.21176&104 $\pm$ 14&   2&Indeterminate&If present, blended with Galactic S~II~$\lambda1250.6$\\
\tableline
1259.53&Ly$\beta$&0.22782			&71 $\pm$ 11&    1&Indeterminate&If present, blended with Galactic S~II~$\lambda1259.5$\\
\tableline
1236.32&OVI~1031.92 &0.19794	&58 $\pm$ 15&   1&No& \\
\tableline
\multicolumn{7}{c}{ }\\
\end{tabular}
\tablecomments{All equivalent widths are rest-frame values. These spectral features are anticipated based upon the detection of \lya absorption at a redshift that places
strong non-\lya lines (e.g., \lybno) in our waveband. The table is ordered by strength of the \lya
absorption and breaks naturally into detections and non-detections.}
 \tablerefs{1: (GHRS+G160M) \citet{Savage95} or 2: (GHRS+G140L) \citet{Tripp98a,Tripp98b}.}
\end{table} 
\end{center}  \normalsize 
%
% IZW1
%
\subsection[IZW1]{IZW1} 
The GHRS/G160M spectrum of \objectname[]{IZW1} was first reported in \citet{Stocke95}. Except for Q1230+0115,
this is the lowest S/N spectrum in our sample (\about13), with an average 4$\sigma$ detectable limit of \Wno=100\mang.
We detect no intergalactic \lya absorbers, although results previously presented \citep{Stocke95} showed 3
\lya absorbers (two~\gt~4\signo, one 3--4\signo). However these spectra relied on the default 1995 IRAF/STSDAS/CALHRS
calibration, which did not correctly handle photocathode blemishes and dead diodes. In both HST exposures, z1a60404n
and z2ia0204n, approximately 1\% of the spectrum was affected by a medium photocathode blemish, and 4\% suffered from bad
diodes. 
%that affected the following wavelength ranges for the two
%subexposures:
%\begin{enumerate}
%	\item z1a60404n 53 subexposures
%	1\% affected by medium photocathode blemish
%1235.55--1235.91\Ang and 1236.29--1237.01\ang: 16 of 53 subexposures
%1235.93 -- 1236.29 5 of 53 subexposures
%	4\% affected by dead diodes
%\begin{itemize}			
%	\item \vspace{-0.1 in} 1228.94 -- 1230.35\ang: 5 of 16 subexposures
%	\item 1231.84 -- 1233.22\Ang
%	\item 1241.16 -- 1242.57\Ang
%	\item 1246.13 -- 1247.54\Ang \item 1253.34 -- 1254.72\Ang
%\end{itemize}
%\item z2ia0204n 4% OF Spectrum affected by dead diodes, 12 of 48 subexposures
%          1% affected by medium photocathode blemish (1235.56 -- 1236.82\ang) 12 of 48subexposures
%	\begin{enumerate}
%	\item 1228.72 -- 1230.12\Ang
%	\item 1230.58 -- 1233.03\Ang
% \item 1240.95 -- 1242.35\Ang
% \item 1245.94 -- 1247.32\Ang
% \item 1253.15 -- 1254.53\Ang
% \end{enumerate}
%\end{enumerate}
%GHRS spectra contaminated by the photocathode blemishes or  and imperfect diodes, which can appear
% as artifical features in the standard co-additions. 
The previously reported 1222.21\Ang absorption feature, now present at only SL=1.6\signo, is close to the edge of our
spectrum in a region of poor S/N. The previously reported 1236.5\Ang feature is marginally present at  SL=1.8\signo.
This is unfortunate, since two studies \citep{Grogin98,Stocke95} used these results
for detailed analysis of the galaxy density distribution along this sightline.  Our recalibrations show the previously
reported
\lya features to be non-existent or below the 3$\sigma$ level.  The only features with
\expanded\  in the GHRS/G160M spectra of IZW1 are the Galactic \ion{S}{2} features.  As previously reported in \citet{Stocke95},
and observed with IUE, the continuum hump at \about1245\Ang is rest-frame \CIII1175 emission at the redshift of IZW1. 
 The LSR offset for this spectrum was 0.139\Ang or 33.4\kms at 1253\ang. 
%
% MARKARIAN 279
%
\subsection[Markarian~279]{Markarian~279}
This spectrum is dominated by strong \lya emission from \objectname[]{Markarian~279} centered at 1252\Ang (\z=0.0294).
The sightline passes through the HVC complex C, which we detect in \SII1250 at $-$160\kms (4.0\signo)  and
 in \NV{1238} at $-$136\kms (4.4\signo). 
Detection of this HVC is not possible in \SII1253 owing to strong intrinsic \lya absorption at
1253.0--1253.7\ang.  
While Galactic \NV{1238} is strong in this sightline (11\signo, 51\mang), Galactic \NV1242 falls below
our 3$\sigma$ threshold for being included in our line list for this object. This alerts us to the
possibility that the Galactic \NV{1238} absorption may be superimposed on an intergalactic \lyano.  
Although we have only \about23\Ang of coverage after applying our \prox blueward limit,  
we detect six \real\ and two \tent\ \lya absorbers. 
The 1241.8\Ang \lya absorption feature could possibly be  \Sithree\ intrinsic to
Mrk~279 at  \vlsr $\cong -40$\kmsno. However, since we do not detect any intrinsic \lya absorption at this velocity,
 this is likely to be an intergalactic \lya absorption and we identify it as such. 
 The LSR offset for this spectrum was 0.116\Ang or 27.7\kms at 1253\ang. 
%Also, the 1238.3\Ang feature marked as \lya could possibly be anomolously strong HVC \NV{1238}. We find this unlikely
%
% MARKARIAN 290
%
\subsection[Markarian~290]{Markarian~290} The \objectname[]{Markarian~290} spectrum was obtained by Wakker in 1997 for the
purpose of studying the Galactic HVC complex C.  This spectrum contains 3 definite and 3 possible
(\tent) \lya absorption features. Our spectrum covers a spectral range of 1232--1263\Ang and is dominated by
\lya emission from Mrk~290 at \about1253\ang. Along this sightline we detect strong Galactic \SIItriplet, and \SFsixty.
%\NVdoublet, and \MGdoublet.
 As determined from \citet{Wakker91}, the dominant \ion{H}{1}-HVC velocity in this direction is $-$120\kmsno. 
\citet{Wakker96} report detections of high velocity gas (\hone~21~cm) at  \vlsr $\cong -138$\kms and $-115$\kmsno. 
We do not
have the spectral resolution to resolve these two HVCs.
% so we modeled this as one absorption feature at  \vlsr$ = -120$\kmsno. 
We detect this combined HVC in \SFsixty, \SIItriplet, and  \NVdoublet.
The S~II and Si~II+Fe~II detections are all heavily
blended with intrinsic \lya absorption, and their \Ws\ are not well determined.
The LSR and HVC \NVdoublet\  detections all fall below the
3$\sigma$ level and are not included in the line list. 
The HVC detection of \NV1242.8 appears to be partially blended with
\SiIII1206.5 intrinsic to Mrk~290 (which is detected at SL=3.7\signo).
The LSR offset for this spectrum was 0.116\Ang or 27.9\kms at 1253\ang. 

While several absorbers are reported in the wavelength range 1245--1248\ang, the rapidly rising continuum on the blue
wing of the target's \lya emission calls their reality into question. Due to our concerns about the reality of these features, 
we have refit the continuum for Mrk~290 by altering the order of the global fit significantly. The statistical significances
of these features remains robust despite these changes, so we list them as definite or possible \lya features as shown in 
Figure~\ref{MARK290}.

% We also have a very marginal 1$\sigma$ detection of the combined HVC in \MgII{1239.9}.
%
%
% MARKARIAN 335
%
\subsection{Markarian~335} 
This sightline was previously reported in \citet{Stocke95}. It contains 4 definite \lya absorption features along with strong
\lya emission from \objectname[]{Markarian~335} and Galactic features.  Due to calibration improvements, the wavelengths and
feature significances have changed slightly from our previous results. 
Although not listed in Table~\ref{linelist_all}, we also report possible weak detections of Galactic \MGdoublet\ (2.5\sig and 2.7\signo). 
The Galactic \MGdoublet\ absorption occurs at a wavelength at which \lya would occur if it were
associated with the galaxy identified as 00036+1928 in the CfA redshift survey and in Table~\ref{nearest_master}.
 This galaxy  is only \about700\hsfi kpc from the sightline and appears as the small `v' at \about 6,000\kms in
the Mrk~335 ``pie diagrams'' of Figure~\ref{PIE_MARK335}.
%We are now  able to report detections of the Galactic \MGdoublet. 
All \lya absorption features are
well-modeled by a single Gaussian component, except for the 1232.9\Ang absorption feature, which at higher resolution may
actually be two features separated by \about45\kmsno. But, at the resolution and S/N of our spectrum, it must be considered as a single
absorber. 
%The EWs of the 1241.1\Ang \lya  and 1239.98\Ang \ion{Mg}{2} are somewhat uncertain due to
%a poor global continuum fit in this region. The reality of these features is confirmed by the ``local continuum method'' fit
%in this region, which actually gives a slightly higher \W for the 1241.1\Ang \lya feature, 167$\pm$20\Mang versus 144$\pm$16\Mang for the 
%global fit. 
Although this sightline passes through Galactic Radio Loop~II, no high-velocity Galactic absorptions are detected.
The LSR offset for this spectrum was 0.119\Ang or 28.6\kms at 1253\ang. 
%
% MARKARIAN 421
%
\subsection[Markarian~421]{Markarian~421} This sightline was previously reported
in \citet{Shull96}. The \objectname[]{Markarian~421} observation (z2ia0104t) suffered from HST target acquisition
anomalies, which resulted in 11 of 42 spectral subexposures being removed from the coaddition. These
anomalies were not detected by the default calibration used in our previous report, and as a result
there are differences in our detections. Most noticeable is the reduced SL of the ``mystery absorber''
noted in \citet{Shull96} at 1257.1\ang, whose velocity was 1150\kms higher than that reported for Mrk~421
itself, if it were \lyano.
This feature, and its companion at 1256.9\ang, are now identified as intrinsic \lyano, but with SL\lt4\signo.
Table~\ref{linelist_all} indicates our new results,
which indicate one definite \lya absorption feature at 1228.0\ang. This ``void'' absorber was first reported in \citet{Shull96}. 

Although this sightline does not pass through any known Galactic radio loops, it passes between \ion{H}{1}-HVC
components of the M complex (MI and MII) as discussed in \citet{Tufte98}.  \citet{Lockman95} report an
\ion{H}{1}-HVC at
\vlsr$ = -58$\kms with \nh~=~\Exp{5.9}{19}~cm$^{-2}$ towards Mrk~421. We detect a HVC in both
\SII1250 (\vlsr$ = -52$\kmsno) and \SII1253  (\vlsr$ = -$53\kmsno). 
The rest EWs for these \ion{S}{2}-HVC detections are 28\Mang and 33\mang, respectively. 
If \bb=25\kmsno, then $N_{\rm SII}$~\about~\Exp{3.4}{14}\percmtwo and 
$N_{\rm SII}$/\nh  = \Exp{6}{-6}, or \about30\% of the solar abundance of S/H.
Although it is possible that the \ion{S}{2} HVC detections could be intrinsic \lya absorbers, we consider this unlikely due
to the presence of absorptions in both components of the \ion{S}{2} lines detectable in the observed band.

 Also of note is the possibility that the 3$\sigma$ \NV{1238} detection  has a velocity that is more consistent with the detected HVC
component.  There is an unreported 1.9$\sigma$ detection at the expected LSR position of \NV1242.   
The LSR offset for this spectrum was 0.160\Ang or 38.2\kms at 1253\ang. 
%
% MARKARIAN 501
%
\subsection{Markarian~501}
The \objectname[]{Markarian~501} sightline was previously reported in \citet[][S95]{Stocke95}
and passes through the Galactic HVC complex C.  Along this sightline, the dominant \ion{H}{1}-HVC velocity  is at  \vlsr $\cong
-115$\kms \citep{Wakker91}. However, this is one of our lower S/N spectra and we do not detect any HVC absorption.  We do detect
strong Galactic
\SIIdoublet\   absorption. 
Due to improvements in calibration, as previously described, our results  have changed slightly since S95.
 Similar to S95, we report 4 definite \lya
absorptions and no possible (\tent) detections.  Absorber ``A'' of S95 is essentially unchanged, but absorbers ``B'' and
``C'' of S95 have changed somewhat. Previously, the ``B'' absorber (1239.97\ang) was quoted as having a
significance of 3.4\signo, while the ``C'' absorber (1246.18\ang) was quoted as 4.6\signo. Our new results place the
S95~``B'' absorber's significance at 4.2\sig and the ``C'' absorber's at 4.0\signo.  
Therefore, we now believe the ``B'' (1239.97\ang) feature to be definite, however it may be  partially contaminated by Galactic \MgII{1239.9} absorption.  
No Galactic \MgII{1240.4} absorption is detected at the 80\Mang (4\signo) level. Hence, we cannot estimate the amount of \MgII{1239.9}
contamination of the 1239.97\Ang feature, if any. We note that, while absorber ``C'' was identified in
S95 as the best example of a \lya cloud in a void, absorber ``B'' is located much farther into the
void in this region. Therefore, the important conclusions of S95 are unaffected. 
Calibration improvements have also allowed us to resolve the previously unreported 1251.15\Ang \lya feature from
Galactic \SII1250.6 absorption. We have identified this absorption as \lya because there is no H~I emission seen at 
positive velocities in this direction. However, Mrk~501 lies behind complex~C which has blueshifted H~I only 
\citep{Wakker91} which may account for the blueshifted component of \SII1253.8.
 The LSR offset for this spectrum was 0.232\Ang or 55.4\kms at 1253\ang. 
%
% MARKARIAN 509
%
\subsection[Markarian~509]{Markarian~509}
The \objectname[]{Markarian~509} (1231.7\Ang -- 1268.8\ang) sightline has been previously reported in \citet{Savage97}
and \citet[][who focused on the Galactic HVCs]{Sembach95b,Sembach99}. 
 These reports were based solely upon the pre-COSTAR
HST/G160M observation z1790208m. We detect one definite intergalactic \lya along this sightline.
 Our analysis combines this observation with a post-COSTAR observation at a slightly
lower wavelength range (1219.5\Ang -- 1255.6\ang, see Table~\ref{observations}). 
The flux level of our HST exposure, z3e70704t, was scaled to be consistent with the initial Mrk~509 exposure. 
The scaling was applied linearly across the overlap region
between exposures. The large gradient in the \W detection limit shortward of 1233\Ang is because
this portion of the spectrum was observed only with the post-COSTAR exposure. Along this sightline, \citet{Sembach95b,Sembach99} 
report  \ion{C}{4}-HVCs at
$-$228\kms and $-$283\kmsno,  while Anglo-Australian telescope \citep{York82} observations of \ion{Na}{1} D lines and \ion{Ca}{2} H and K lines
indicate Galactic velocity components near \vlsr$\approx+5$\kms and +60\kmsno. With our GHRS/G160M data we detect HVC gas in \SIIdoublet, \NV1238, \MgII1239, and possibly in
\NV1242 and \MgII1240. The N~V and Mg~II Galactic lines show velocity structure over the \vlsr range of $-65$ to $+60$\kmsno.
 We do not have the spectral resolution to confirm the +5\kms component. We do not detect
\SIIdoublet\  absorption at \vlsr = $-$228 or $-$283\kmsno.
Intrinsic \lya associated with Mrk~509
prevents us from detecting HVC absorption due to \SII{1259} at these velocities (\vlsr = $-$228 and $-$283\kmsno). Similarly,
HVC detection of \SFsixty\ at these velocities is not possible due to Galactic \SII{1259}, and 
\MGdoublet\  detections are complicated by the presence of Galactic and HVC \NVdoublet.
We also detect Galactic absorption in \NVdoublet, \MgII{1239.9}, \SIItriplet, and \SFsixty, and possibly \CI{1260}\ 
(heavily blended with \SFsixty, and \SFsixty\ HVC, and a possible intrinsic \lya absorption in Mrk~509). 
%As shown in the velocity ``stack" plot in Figure~\ref{GAL_MARK509}, 
We detect high-velocity Galactic gas at +60 and $-$65\kms in \NV{1238}, \SIItriplet\  and possibly \SFsixty.  
The \SFsixty\ absorption is apparently heavily blended with  \hone absorption intrinsic to Mrk~509, and possible
Galactic \CI1260. Along with these HVCs, the spectrum of Mrk~509 contains a single definite (\real) \lya 
absorption feature at 2560\kmsno.  The LSR offset for this spectrum was 0.060\Ang or 14.4\kms at 1253\ang. 

Numerous strong intrinsic \lya absorption lines are observed in our Mrk~509 spectrum. This absorption shows
 significant structure over the velocity range of -450 to +200\kmsno. Our modelling suggests 7-9 separate
velocity components. Recent FUSE observations \citep{Kriss00} detect these absorbers in \lybno, \CIII977.0, and \OVI1031.9,
 which they model with 7 separate components. Efforts are underway to match our \lya detections with the FUSE
detections to determine the metallicity and \bvalues of these intrinsic absorbers. These results will be reported 
elsewhere \citep{Shull00}.
% \begin{figure} 
%\plotone{figure5.eps}
%\caption{\label{GAL_MARK509} Markarian~509 LSR velocity stack
%of Galactic features: 
% ``W'' indicates the mean HVC \hone absorption, 
% ``1'' indicates the location of the +60\kms  HVC absorption \protect{\citep{York82}},
% ``2'' and ``3'' indicate the positions of detected \ion{C}{4}-HVCs \protect{\citep{Sembach95b}}, and
% ``4'' indicate the location of
%$-$65\kms (an HVC in \SIIdoublet, and \NV{1238}). 
%The absorption line at $\approx-200$\kms in the \ion{Si}{2}/\ion{Fe}{2} 1260\Ang plot is Galactic \SII{1259}\ang.}
%\end{figure}
%
% MARKARIAN 817
%
\subsection[Markarian~817]{Markarian~817}
The sightline towards \objectname[]{Markarian~817} passes through HVC complex C.   The dominant
\ion{H}{1}-HVC velocity in this direction is $-115$\kms \citep{Wakker91}. 
 We detect strong Galactic \SIIdoublet, two HVCs (\vlsr$\approx-45$\kms and $-110$\kmsno) in
\SIIdoublet\  as well as \NVdoublet\  (Galactic+HVCs). 
The \NV{1238}.8 and \NV1242.8 regions are both confused with significant structure.
The \NV{1238}.8 result is the combination of LSR+HVC absorptions, while the 1242.5\Ang detection
appears to be associated with the HVC at\ \vlsr=$-$45\kmsno. The\ \vlsr=$-110$\kms HVC is also detected in \NV1242.8 at the
2.8\sig level.  Galactic \NV1242.8 at LSR is also detected, but at the anomalously low 3.1\sig level.  We do not detect any
Galactic \MGdoublet\ absorption, although it is possible that the reported 1239.5\Ang feature is contaminated with weak \MgII{1239.9} HVC
absorption. 
Note that, along this sightline the HVC N~V and Mg~II absorptions are stronger than the LSR Galactic components.
In addition to the Galactic detections, this rich sightline contains 9 definite \lya absorption features
%2 possible \lya absorption features, 
and several \lya absorptions identified as intrinsic to Mrk~817. The LSR offset for this spectrum
was 0.083\Ang or 19.8\kms at 1253\ang. 
%
% PKS 2155 - 304
%
\subsection[\PKS] {\PKS} The pre-COSTAR spectrum of
\objectname[]{PKS~2155-304}   contains 8 definite (\real) and 4 possible (\tent) \lya absorption
features,  as well as Galactic  \MGdoublet\ and \SIIdoublet\ absorption. 
An analysis of this sightline has been previously
reported in \citet{pks}. Improvements in the calibration process since the publication of that paper produce slightly different
results. Most notable is the inclusion of the 1235.8\Ang \lya feature, which is blended with the 1236.0\Ang and
1236.4\Ang features.  Our restriction of \bb~\lt~100\kmsno, which was not present in our previous
reduction, forced us to add a third component to this blend. As a result, EWs for the 1235.6\Ang
and 1236.4\Ang features are slightly different than previously reported. The 1238.7\Ang feature is 
partially blended with Galactic
\NV{1238}.  In our previous paper \citep{pks} we estimated the N~V contribution, while for this paper we have fitted
blended components to both \NV{1238} and \lyano. 
The feature marked as possible \lya at 
1256.7\Ang could also be a Galactic \SII1253 HVC. Galactic \NVdoublet, \MGdoublet, and \SIIdoublet\ are all present and strong.  
The LSR offset for the pre-COSTAR spectrum was 0.013\Ang or 3.0\kms at 1253\ang. 

The post-COSTAR spectrum of \PKS\ was also reported in \citet{pks}
and contains 8 definite and 1 possible \lya absorption features. As with the
pre-COSTAR spectrum, minor improvements to the calibration process produce slightly improved feature modelings. A detailed
discussion of the clump of 8 \lya absorbers near \cz = 17,000\kms is given in \citet{pks}. 
The third (most blueward) component of the 1284--1285\Ang \lya complex is only marginally required by our fitting routines.
Relaxing our \bvalue $\le$ 100\kms criteria provides an adequate 2 component spectral fit as originally reported in \citet{pks}.
Galactic \SII{1259} is present
along with strong
\SFsixty, and a \SFsixty\ HVC detection \citep{Sembach99}. The post-COSTAR \PKS\ spectrum was corrected to the LSR using
the Galactic H~I and only the \ion{S}{2} line (1259\ang).  
The LSR offset for the post-COSTAR spectrum was 0.073\Ang or 17.4\kms at 1253\ang. 
% The absorber at 1238.7\Ang is determined to be not entirely \NV{1238} or a high velocity component ($-$200\kmsno)
% of\MgII{1239.9}
% based upon the large \W (84\mang) compared to that of the marginally detected \NV1242 ($\le$15\mang).  On the basis of an
% expected line strength of $\le$ 30\Mang for \NV{1238}, we have adjusted the reported \W of the intergalactic \lya portion of
%this feature to be 54\Mang in our complete \lya results of Table~\ref{linelist_master}.
The \PKS\ sightline does not pass through any known \ion{H}{1}-HVCs or Galactic radio loops, although  \ion{C}{4}-HVCs were detected by
\citet{Sembach95b} at \vlsr = $-$260 and $-$145\kmsno.  We detect the \vlsr = $-$145\kms component in
\SFsixty\ (at \vlsr =$-$154\kmsno) and report a possible (2.4$\sigma$) detection of the \vlsr = $-$260\kms component
in \SII1253.
%
% Q1230+0115
%
\subsection[Q1230+0115]{Q1230+0115}
The spectrum of the QSO \objectname[]{Q1230+0115} contains 6 definite and 4 possible (\tent) \lya absorption features. This sightline does not
pass through any known radio loops or HVCs and is only 0.91\degr\ degrees from the 3C~273 sightline. 
	This sightline shows a cluster of four \lya absorbers within 1\Ang (\about200\kmsno) of 1222.5\Ang (1,700\kmsno) and an
additional \lya absorber at 1225\Ang (2,303\kmsno).  
A possible (3.1\signo) \lya absorber is also present at 1223.9\Ang (2037\kmsno).
Another possible absorber (2.0\signo) at 1221.05\Ang is not statistically significant due a minor diode problem close to its location
(see the EW detection limit panel in Figure~\ref{Q1230+0115}).
The 3C~273 sightline also contains an absorber at 1,586\kms and 2
possible absorbers at 2,200--2,300\kms
which appear to closely match detections in this sightline. The transverse separations between these sightlines at 1,500\kms
and 2,300\kms are 340 and 520 \hsfi kpc, respectively. This sightline pair is investigated further in Paper~III. Galactic
\NV{1238} is detected at the 2.7\sig level, while \NV1242 is marginally detected at the 1.7\sig level. The
\NV1242 absorption is partially blended with the possible absorber at 1242.9\ang. The \MGdoublet\ lines are not detected in
this sightline. This spectrum has the lowest S/N  (\about9) of our sample and is being reobserved in cycle 7 with HST/STIS.
 The LSR offset for this spectrum was -0.003\Ang or -0.7\kms at 1253\ang. 
\section{Master Lists of Detected Absorption Features}\label{sec:master}
We consider absorption features with significance levels \real\ to be definite, while 
absorption lines with \tent\ are classified as  possible. 
Table~\ref{linelist_master} presents all definite and Table~\ref{linelist_possible} all possible intergalactic \lya absorptions.
  Table~\ref{linelist_hvc} presents a
summary of all definite Galactic HVC clouds. Due to the overlap between the regions of the sky covered by the HVC
complex C and the initial CfA slices, we detect Galactic HVCs in 8 of our 15 sightlines. 
This fraction is considerably larger than the value of 37\% found for HVCs seen in 21~cm emission toward AGN sightlines
down to \Nh = $7\times10^{17}$\percmtwono.
\citep{Murphy95}.
Hence, we expect that the HVC detection rate will decrease with future observations due to this
unintentional overlap. Table~\ref{linelist_intrinsic} shows a list of all definite lines determined to be intrinsic to the
observed target.

The first column of each of the master tables indicates the name of the target in which the feature was detected.
An asterisk before the target name denotes a detection in pre-COSTAR data.
The second column indicates the LSR-adjusted wavelength and wavelength uncertainty for each feature.
The third column lists the recession velocity (for \lya features, Tables~\ref{linelist_master} and \ref{linelist_possible}), 
velocity relative to LSR (HVC features, Table~\ref{linelist_hvc}), or velocity relative to the narrow emission line rest frame of the target 
(intrinsic features, Table~\ref{linelist_intrinsic}). All velocities are quoted as \cz, in units of\kmsno. 
Velocity uncertainties based upon the total wavelength uncertainties (see \S~\ref{sec:LSR}) are also provided.
The fourth column provides the single-component rest-frame Doppler width (\bmsd\ in\kmsno) and its 1\sig uncertainty for 
each feature as estimated from the Gaussian width (\WG~=~\bobs/$\sqrt{2}$) of the fitted feature, 
 restricted to the range of
$12 < \bobs < 100$\kmsno. The Doppler widths have been corrected for the spectral resolution of the 
GHRS (FWHM\about19\kmsno) and our pre-fit smoothing (see \S~\ref{sec:local} and eq.~\ref{b_equation}). 
 As discussed in Paper~II, the measured \bvalues in pre-COSTAR data are occasionally artificially
large. The fifth column lists the rest-frame EW (\W in\mang) and 1$\sigma$
uncertainty  of each absorption feature. The sixth column indicates the significance level (SL in
\signo) of each feature. In Tables~\ref{linelist_hvc}, 
the seventh column indicates the atomic identification of the detected
absorption. In Table~\ref{linelist_intrinsic}, the final column gives the redshift
of the intrinsic system. 
\setlength{\tabcolsep}{1.8mm}
\begin{deluxetable}{lcrccc}
%\tabletypesize{\small}
\tablecolumns{6}
\tablecaption{Definite (\real) Intergalactic \lya Features\label{linelist_master}}
\tablewidth{0pt}
\tablehead{
\colhead{Target} &
\colhead{Wavelength} &
\colhead{Velocity	} &
\colhead{\bb} &
\colhead{\Wno} &
\colhead{SL} \\
\colhead{ } &
\colhead{(\AA)} &
\colhead{(\nokmsno)} &
\colhead{(\nokmsno)} &
\colhead{(m\AA)} &
\colhead{($\sigma$)} 
}
\startdata
$*$3C273&1219.786 $\pm$  0.024&  1015 $\pm$     6&  69 $\pm$   5& 369 $\pm$  36& 34.8 \\
$*$3C273&1222.100 $\pm$  0.023&  1586 $\pm$     6&  72 $\pm$   4& 373 $\pm$  30& 42.0 \\
$*$3C273&1224.954 $\pm$  0.029&  2290 $\pm$     7&  54 $\pm$  33&  35 $\pm$  30&  4.2 \\
$*$3C273&1247.593 $\pm$  0.046&  7872 $\pm$    11&  34 $\pm$  17&  33 $\pm$  18&  5.8 \\
$*$3C273&1251.485 $\pm$  0.032&  8832 $\pm$     8&  61 $\pm$  10& 114 $\pm$  25& 14.8 \\
$*$3C273&1255.542 $\pm$  0.069&  9833 $\pm$    17&  64 $\pm$  24&  46 $\pm$  22&  5.9 \\
$*$3C273&1275.243 $\pm$  0.031& 14691 $\pm$     7&  61 $\pm$   8& 140 $\pm$  25& 16.3 \\
$*$3C273&1276.442 $\pm$  0.059& 14987 $\pm$    14&  52 $\pm$  20&  46 $\pm$  22&  5.4 \\
$*$3C273&1277.474 $\pm$  0.136& 15241 $\pm$    33&  88 $\pm$  52&  52 $\pm$  40&  6.3 \\
$*$3C273&1280.267 $\pm$  0.077& 15930 $\pm$    19&  71 $\pm$  28&  64 $\pm$  33&  6.4 \\
$*$3C273&1289.767 $\pm$  0.098& 18273 $\pm$    24&  82 $\pm$  36&  47 $\pm$  28&  6.1 \\
$*$3C273&1292.851 $\pm$  0.051& 19033 $\pm$    12&  45 $\pm$  17&  47 $\pm$  22&  5.5 \\
$*$3C273&1296.591 $\pm$  0.025& 19956 $\pm$     6&  62 $\pm$   4& 297 $\pm$  25& 33.0 \\
AKN120&1232.052 $\pm$  0.034&  4040 $\pm$     8&  32 $\pm$  11&  48 $\pm$  18&  5.5 \\
AKN120&1242.972 $\pm$  0.028&  6733 $\pm$     7&  33 $\pm$   8&  53 $\pm$  13&  9.2 \\
AKN120&1247.570 $\pm$  0.087&  7867 $\pm$    21&  34 $\pm$  35&  20 $\pm$  25&  4.3 \\
AKN120&1247.948 $\pm$  0.023&  7960 $\pm$     5&  27 $\pm$   4& 147 $\pm$  22& 31.4 \\
AKN120&1248.192 $\pm$  0.027&  8020 $\pm$     7&  23 $\pm$   6&  65 $\pm$  17& 14.1 \\
FAIRALL9&1240.988 $\pm$  0.038&  6244 $\pm$     9&  35 $\pm$  13&  22 $\pm$   9&  5.3 \\
FAIRALL9&1244.462 $\pm$  0.034&  7100 $\pm$     8&  39 $\pm$  11&  32 $\pm$  10&  8.2 \\
FAIRALL9&1254.139 $\pm$  0.024&  9487 $\pm$     6&  43 $\pm$   6&  84 $\pm$  13& 26.5 \\
FAIRALL9&1262.864 $\pm$  0.029& 11638 $\pm$     7&  28 $\pm$  13&  16 $\pm$   8&  5.2 \\
FAIRALL9&1263.998 $\pm$  0.041& 11918 $\pm$    10&  42 $\pm$  15&  22 $\pm$   9&  7.0 \\
FAIRALL9&1264.684 $\pm$  0.073& 12087 $\pm$    18&  44 $\pm$  33&  30 $\pm$  28& 10.2 \\
FAIRALL9&1265.104 $\pm$  0.026& 12191 $\pm$     6&  23 $\pm$   7&  28 $\pm$   7&  9.5 \\
FAIRALL9&1265.970 $\pm$  0.117& 12404 $\pm$    29&  32 $\pm$  23&  19 $\pm$  23&  7.0 \\
H1821+643&1245.440 $\pm$  0.023&  7342 $\pm$     5&  47 $\pm$   3& 298 $\pm$  20& 42.4 \\
H1821+643&1246.301 $\pm$  0.036&  7554 $\pm$     9&  41 $\pm$  14&  50 $\pm$  24&  7.1 \\
H1821+643&1247.583 $\pm$  0.029&  7870 $\pm$     7&  19 $\pm$   9&  40 $\pm$  17&  5.4 \\
H1821+643&1247.937 $\pm$  0.033&  7957 $\pm$     8&  33 $\pm$  10&  68 $\pm$  38&  9.4 \\
H1821+643&1265.683 $\pm$  0.025& 12334 $\pm$     6&  28 $\pm$   6&  64 $\pm$  15& 11.8 \\
MARK279&1236.942 $\pm$  0.030&  5246 $\pm$     7&  27 $\pm$   9&  30 $\pm$  10&  5.6 \\
MARK279&1241.509 $\pm$  0.029&  6372 $\pm$     7&  18 $\pm$   4&  58 $\pm$   7& 14.2 \\
MARK279&1241.805 $\pm$  0.023&  6445 $\pm$     6&  21 $\pm$   4&  40 $\pm$   7& 10.4 \\
MARK279&1243.753 $\pm$  0.023&  6925 $\pm$     5&  26 $\pm$   3&  65 $\pm$   8& 16.8 \\
MARK279&1247.216 $\pm$  0.024&  7779 $\pm$     6&  28 $\pm$   5&  48 $\pm$   9& 15.1 \\
MARK279&1247.533 $\pm$  0.042&  7858 $\pm$    10&  34 $\pm$  14&  21 $\pm$  10&  6.7 \\
MARK290&1234.597 $\pm$  0.027&  4667 $\pm$     7&  26 $\pm$   8&  60 $\pm$  18&  7.2 \\
MARK290&1244.408 $\pm$  0.032&  7087 $\pm$     8&  23 $\pm$  11&  23 $\pm$  10&  4.3 \\
MARK290&1245.536 $\pm$  0.025&  7365 $\pm$     6&  11 $\pm$   9&  21 $\pm$   7&  4.2 \\
$*$MARK335&1223.637 $\pm$  0.026&  1965 $\pm$     6&  75 $\pm$   7& 229 $\pm$  30& 28.1 \\
$*$MARK335&1224.974 $\pm$  0.049&  2295 $\pm$    12&  73 $\pm$  17&  81 $\pm$  26& 10.2 \\
$*$MARK335&1232.979 $\pm$  0.057&  4268 $\pm$    14&  51 $\pm$  20&  33 $\pm$  16&  5.1 \\
$*$MARK335&1241.093 $\pm$  0.026&  6269 $\pm$     6&  75 $\pm$   6& 130 $\pm$  14& 30.3 \\
MARK421&1227.977 $\pm$  0.025&  3035 $\pm$     6&  35 $\pm$   5&  86 $\pm$  15& 12.4 \\
$*$MARK501&1234.572 $\pm$  0.039&  4661 $\pm$    10&  60 $\pm$  13& 161 $\pm$  43& 11.1 \\
$*$MARK501&1239.968 $\pm$  0.029&  5992 $\pm$     7&  59 $\pm$  39&  55 $\pm$  46&  4.2 \\
$*$MARK501&1246.177 $\pm$  0.069&  7523 $\pm$    17&  48 $\pm$  26&  53 $\pm$  36&  4.0 \\
$*$MARK501&1251.152 $\pm$  0.029&  8750 $\pm$     7&  77 $\pm$  49&  66 $\pm$  57&  4.9 \\
MARK509&1226.050 $\pm$  0.025&  2560 $\pm$     6&  40 $\pm$   5& 209 $\pm$  32& 14.4 \\
MARK817&1223.507 $\pm$  0.037&  1933 $\pm$     9&  34 $\pm$  13&  29 $\pm$  13&  5.3 \\
MARK817&1224.172 $\pm$  0.023&  2097 $\pm$     5&  40 $\pm$   4& 135 $\pm$  15& 25.3 \\
MARK817&1234.657 $\pm$  0.041&  4682 $\pm$    10&  40 $\pm$  15&  23 $\pm$  11&  5.6 \\
MARK817&1236.303 $\pm$  0.023&  5088 $\pm$     6&  84 $\pm$   4& 207 $\pm$  14& 56.3 \\
MARK817&1236.902 $\pm$  0.027&  5236 $\pm$     7&  24 $\pm$   7&  25 $\pm$   7&  6.6 \\
MARK817&1239.159 $\pm$  0.029&  5793 $\pm$     7&  39 $\pm$  12&  34 $\pm$  13&  8.7 \\
MARK817&1241.034 $\pm$  0.024&  6255 $\pm$     6&  29 $\pm$   5&  37 $\pm$   8& 11.6 \\
MARK817&1245.395 $\pm$  0.051&  7330 $\pm$    13&  51 $\pm$  18&  17 $\pm$   7&  5.9 \\
MARK817&1247.294 $\pm$  0.044&  7799 $\pm$    11&  56 $\pm$  16&  28 $\pm$   9& 10.1 \\
$*$PKS2155-304&1226.345 $\pm$  0.060&  2632 $\pm$    15&  61 $\pm$  33&  42 $\pm$  40&  9.2 \\
$*$PKS2155-304&1226.964 $\pm$  0.065&  2785 $\pm$    16&  64 $\pm$  26&  36 $\pm$  22&  7.9 \\
$*$PKS2155-304&1232.016 $\pm$  0.049&  4031 $\pm$    12&  39 $\pm$  17&  21 $\pm$  11&  4.2 \\
$*$PKS2155-304&1235.748 $\pm$  0.029&  4951 $\pm$     7&  68 $\pm$  15&  64 $\pm$  23& 14.3 \\
$*$PKS2155-304&1235.998 $\pm$  0.029&  5013 $\pm$     7&  58 $\pm$  11&  82 $\pm$  22& 18.1 \\
$*$PKS2155-304&1236.426 $\pm$  0.029&  5119 $\pm$     7&  80 $\pm$   5& 218 $\pm$  20& 48.1 \\
$*$PKS2155-304&1238.451 $\pm$  0.029&  5618 $\pm$     7&  33 $\pm$  14&  29 $\pm$  15&  6.7 \\
$*$PKS2155-304&1238.673 $\pm$  0.031&  5673 $\pm$     8&  30 $\pm$  12&  39 $\pm$  16&  9.1 \\
PKS2155-304&1270.784 $\pm$  0.027& 13591 $\pm$     6&  39 $\pm$   6& 101 $\pm$  18& 12.5 \\
PKS2155-304&1281.375 $\pm$  0.024& 16203 $\pm$     5&  58 $\pm$   3& 346 $\pm$  23& 44.8 \\
PKS2155-304&1281.867 $\pm$  0.061& 16325 $\pm$    15&  49 $\pm$  22&  62 $\pm$  34&  8.2 \\
PKS2155-304&1284.301 $\pm$  0.030& 16925 $\pm$     7&  19 $\pm$  13&  43 $\pm$  37&  5.7 \\
PKS2155-304&1284.497 $\pm$  0.039& 16973 $\pm$     9&  63 $\pm$   6& 389 $\pm$  68& 50.9 \\
PKS2155-304&1285.086 $\pm$  0.038& 17119 $\pm$     9&  87 $\pm$  11& 448 $\pm$  79& 57.9 \\
PKS2155-304&1287.497 $\pm$  0.024& 17713 $\pm$     6&  35 $\pm$   5& 139 $\pm$  21& 18.4 \\
PKS2155-304&1288.958 $\pm$  0.029& 18073 $\pm$     7&  47 $\pm$   8&  99 $\pm$  20& 13.2 \\
Q1230+0115&1221.711 $\pm$  0.026&  1490 $\pm$     6&  21 $\pm$   8& 138 $\pm$  42&  6.3 \\
Q1230+0115&1222.425 $\pm$  0.035&  1666 $\pm$     9&  54 $\pm$  10& 385 $\pm$  94& 16.9 \\
Q1230+0115&1222.747 $\pm$  0.029&  1745 $\pm$     7&  40 $\pm$  12& 241 $\pm$  99& 10.9 \\
Q1230+0115&1223.211 $\pm$  0.051&  1860 $\pm$    13&  48 $\pm$  21& 142 $\pm$  81&  6.6 \\
Q1230+0115&1225.000 $\pm$  0.024&  2301 $\pm$     6&  55 $\pm$   6& 439 $\pm$  57& 23.9 \\
Q1230+0115&1253.145 $\pm$  0.031&  9242 $\pm$     8&  72 $\pm$   8& 301 $\pm$  49& 24.8 \enddata
\end{deluxetable}
\normalsize
\setlength{\tabcolsep}{2mm}
\begin{deluxetable}{lcrccc}
%\tabletypesize{\small}
\tablecolumns{6}
\tablecaption{Possible (\tent) Intergalactic \lya features\label{linelist_possible}}
\tablewidth{0pt}
\tablehead{
\colhead{Target} &
\colhead{Wavelength} &
\colhead{Velocity	} &
\colhead{\bb} &
\colhead{\Wno} &
\colhead{SL} \\

\colhead{ } &
\colhead{(\AA)} &
\colhead{(\nokmsno)} &
\colhead{(\nokmsno)} &
\colhead{(m\AA)} &
\colhead{($\sigma$)} 
}
\startdata

$*$3C273&1224.587 $\pm$  0.150&  2199 $\pm$    37&  57 $\pm$  55&  29 $\pm$  35&  3.5 \\
$*$3C273&1234.704 $\pm$  0.029&  4694 $\pm$     7&  69 $\pm$  61&  25 $\pm$  29&  3.5 \\
$*$3C273&1265.701 $\pm$  0.064& 12338 $\pm$    16&  33 $\pm$  26&  21 $\pm$  18&  3.3 \\
$*$3C273&1266.724 $\pm$  0.084& 12590 $\pm$    21&  52 $\pm$  30&  24 $\pm$  18&  3.6 \\
$*$3C273&1268.969 $\pm$  0.076& 13144 $\pm$    19&  43 $\pm$  28&  18 $\pm$  15&  3.1 \\
AKN120&1223.088 $\pm$  0.039&  1829 $\pm$    10&  25 $\pm$  17&  64 $\pm$  48&  3.7 \\
AKN120&1247.267 $\pm$  0.104&  7792 $\pm$    26&  34 $\pm$  32&  19 $\pm$  22&  3.9 \\
ESO141-G55&1249.932 $\pm$  0.036&  8449 $\pm$     9&  17 $\pm$  15&  12 $\pm$   8&  3.0 \\
ESO141-G55&1252.483 $\pm$  0.041&  9078 $\pm$    10&  23 $\pm$  16&  12 $\pm$   7&  3.1 \\
FAIRALL9&1265.407 $\pm$  0.029& 12265 $\pm$     7&  19 $\pm$  18&  11 $\pm$   8&  3.8 \\
H1821+643&1238.014 $\pm$  0.036&  5510 $\pm$     9&  21 $\pm$  14&  23 $\pm$  13&  3.3 \\
H1821+643&1240.569 $\pm$  0.036&  6140 $\pm$     9&  14 $\pm$  16&  24 $\pm$  16&  3.4 \\
H1821+643&1244.966 $\pm$  0.031&  7225 $\pm$     8&  15 $\pm$  12&  25 $\pm$  13&  3.5 \\
H1821+643&1247.362 $\pm$  0.029&  7815 $\pm$     7&  36 $\pm$  34&  27 $\pm$  29&  3.7 \\
H1821+643&1252.477 $\pm$  0.042&  9077 $\pm$    10&  21 $\pm$  15&  23 $\pm$  15&  3.6 \\
H1821+643&1254.874 $\pm$  0.099&  9668 $\pm$    24&  24 $\pm$  31&  21 $\pm$  25&  3.6 \\
MARK279&1237.915 $\pm$  0.029&  5486 $\pm$     7&  30 $\pm$  29&  17 $\pm$  18&  3.5 \\
MARK279&1238.502 $\pm$  0.047&  5631 $\pm$    12&  21 $\pm$  23&  18 $\pm$  18&  3.7 \\
MARK290&1232.797 $\pm$  0.064&  4224 $\pm$    16&  21 $\pm$  29&  41 $\pm$  49&  3.6 \\
MARK290&1235.764 $\pm$  0.044&  4955 $\pm$    11&  26 $\pm$  18&  28 $\pm$  19&  3.4 \\
MARK290&1245.869 $\pm$  0.026&  7447 $\pm$     6&   8 $\pm$  12&  18 $\pm$   7&  3.6 \\
$*$PKS2155-304&1234.767 $\pm$  0.051&  4709 $\pm$    12&  32 $\pm$  24&  15 $\pm$  14&  3.3 \\
$*$PKS2155-304&1246.990 $\pm$  0.029&  7724 $\pm$     7&  31 $\pm$  35&  13 $\pm$  16&  3.0 \\
$*$PKS2155-304&1247.510 $\pm$  0.029&  7852 $\pm$     7&  30 $\pm$  31&  13 $\pm$  16&  3.1 \\
$*$PKS2155-304&1255.084 $\pm$  0.041&  9720 $\pm$    10&  24 $\pm$  16&  13 $\pm$   9&  3.0 \\
$*$PKS2155-304&1256.636 $\pm$  0.042& 10102 $\pm$    10&  25 $\pm$  15&  14 $\pm$   8&  3.3 \\
PKS2155-304&1264.806 $\pm$  0.058& 12117 $\pm$    14&  39 $\pm$  20&  31 $\pm$  19&  3.6 \\
Q1230+0115&1236.045 $\pm$  0.041&  5025 $\pm$    10&  24 $\pm$  15&  53 $\pm$  31&  3.1 \\
Q1230+0115&1242.897 $\pm$  0.044&  6714 $\pm$    11&  19 $\pm$  19&  45 $\pm$  35&  3.2 \\
Q1230+0115&1246.254 $\pm$  0.049&  7542 $\pm$    12&  31 $\pm$  18&  44 $\pm$  27&  3.3 \enddata
\end{deluxetable}
\normalsize

\begin{deluxetable}{lcrcccl}
%\tabletypesize{\small}
\tablecolumns{7}
\tablecaption{Galactic HVC features\label{linelist_hvc} with \real.}
\tablewidth{0pt}
\tablehead{
\colhead{Target} &
\colhead{Wavelength} &
\colhead{Velocity	} &
\colhead{\bb} &
\colhead{\Wno} &
\colhead{SL} &
\colhead{ID}\\
\colhead{ } &
\colhead{(\AA)} &
\colhead{(\nokmsno)} &
\colhead{(\nokmsno)} &
\colhead{(m\AA)} &
\colhead{($\sigma$)} & 
\colhead{}
}
\startdata
ESO141-G55&1253.444 $\pm$  0.038&   -85 $\pm$     9&  27 $\pm$  13&  18 $\pm$   9&  4.7&SII1253.8 \\
ESO141-G55&1259.304 $\pm$  0.029&   -47 $\pm$     7&  48 $\pm$  23&  23 $\pm$  14&  5.8&SII1259.5 \\
ESO141-G55&1260.204 $\pm$  0.029&   -70 $\pm$     7&  44 $\pm$   6&  83 $\pm$  73& 21.1&SiII+FeII1260.5 \\
FAIRALL9&1251.367 $\pm$  0.025&   188 $\pm$     6&  32 $\pm$   5&  41 $\pm$   7& 11.3&SII1250.6 \\
FAIRALL9&1254.572 $\pm$  0.023&   182 $\pm$     5&  27 $\pm$   4&  56 $\pm$   7& 17.6&SII1253.8 \\
FAIRALL9&1260.338 $\pm$  0.029&   199 $\pm$     7&  15 $\pm$   5&  34 $\pm$   9&  8.9&SII1259.5 \\
FAIRALL9&1261.039 $\pm$  0.035&   128 $\pm$     8&  54 $\pm$   6& 379 $\pm$  75&105.2&SiII+FeII1260.5 \\
FAIRALL9&1261.300 $\pm$  0.028&   190 $\pm$     7&  36 $\pm$   4& 180 $\pm$  62& 51.3&SiII+FeII1260.5 \\
H1821+643&1238.608 $\pm$  0.029&   -47 $\pm$     7&  38 $\pm$  22&  42 $\pm$  27&  5.9&NV1238.8 \\
H1821+643&1242.616 $\pm$  0.029&   -44 $\pm$     7&  42 $\pm$  20&  56 $\pm$  30&  8.2&NV1242.8 \\
H1821+643&1250.314 $\pm$  0.029&   -65 $\pm$     7&  26 $\pm$  13&  30 $\pm$  17&  4.4&SII1250.6 \\
H1821+643&1253.314 $\pm$  0.029&  -119 $\pm$     7&  17 $\pm$  10&  31 $\pm$  13&  5.0&SII1253.8 \\
H1821+643&1259.157 $\pm$  0.029&   -86 $\pm$     7&  59 $\pm$  10&  99 $\pm$  24& 16.2&SII1259.5 \\
H1821+643&1260.052 $\pm$  0.029&  -107 $\pm$     7&  59 $\pm$   4& 440 $\pm$  35& 74.4&SiII+FeII1260.5 \\
MARK279&1238.260 $\pm$  0.037&  -136 $\pm$     9&  20 $\pm$  18&  22 $\pm$  16&  4.3&NV1238.8 \\
MARK279&1249.933 $\pm$  0.034&  -160 $\pm$     8&  23 $\pm$  14&  12 $\pm$   8&  4.0&SII1250.6 \\
MARK290&1250.019 $\pm$  0.029&  -139 $\pm$     7&  31 $\pm$  22&  19 $\pm$  23&  5.6&SII1250.6 \\
MARK290&1253.327 $\pm$  0.029&  -113 $\pm$     7&  28 $\pm$   5&  34 $\pm$   6& 13.7&SII1253.8 \\
MARK290&1259.058 $\pm$  0.029&  -110 $\pm$     7&  39 $\pm$   7&  72 $\pm$  16& 15.2&SII1259.5 \\
MARK290&1260.008 $\pm$  0.029&  -117 $\pm$     7&  63 $\pm$   3& 479 $\pm$  28& 97.3&SiII+FeII1260.5 \\
MARK421&1250.385 $\pm$  0.031&   -52 $\pm$     7&  11 $\pm$  16&  28 $\pm$  31&  4.5&SII1250.6 \\
MARK421&1253.578 $\pm$  0.031&   -53 $\pm$     8&  11 $\pm$  13&  33 $\pm$  29&  4.8&SII1253.8 \\
$*$MARK501&1253.069 $\pm$  0.264&  -175 $\pm$    63&  99 $\pm$   3&  64 $\pm$  33&  4.7&SII1253.8 \\
MARK509&1239.182 $\pm$  0.029&  -180 $\pm$     7&  21 $\pm$  18&  24 $\pm$  29&  5.1&MgII1239.9 \\
MARK509&1243.060 $\pm$  0.734&    62 $\pm$   177&  35 $\pm$  38&  17 $\pm$  21&  4.3&NV1242.8 \\
MARK509&1250.314 $\pm$  0.029&   -69 $\pm$     7&  24 $\pm$   9&  20 $\pm$   7&  7.2&SII1250.6 \\
MARK509&1250.819 $\pm$  0.029&    56 $\pm$     7&  42 $\pm$   5&  58 $\pm$   8& 21.3&SII1250.6 \\
MARK509&1253.561 $\pm$  0.029&   -57 $\pm$     7&  41 $\pm$   9&  27 $\pm$   7& 11.7&SII1253.8 \\
MARK509&1254.079 $\pm$  0.032&    64 $\pm$     8&  40 $\pm$   6&  85 $\pm$  19& 37.5&SII1253.8 \\
$*$MARK509&1258.365 $\pm$  0.029&  -270 $\pm$     7&  53 $\pm$  13& 237 $\pm$ 151&122.7&SiII1259.5 \\
$*$MARK509&1258.565 $\pm$  0.029&  -222 $\pm$     7&  33 $\pm$   8&  43 $\pm$  46& 22.6&SiII1259.5 \\
$*$MARK509&1259.858 $\pm$  0.029&    85 $\pm$     7&  34 $\pm$  19&  30 $\pm$  37& 16.1&SII1259.5 \\
$*$MARK509&1260.257 $\pm$  0.029&   -48 $\pm$     7&  68 $\pm$  64&  43 $\pm$  52& 22.2&SiII+FeII1260.5 \\
$*$MARK509&1260.710 $\pm$  0.029&    50 $\pm$     7&  72 $\pm$   4& 372 $\pm$  73&186.9&SiII+FeII1260.5 \\
MARK817&1238.529 $\pm$  0.029&   -65 $\pm$     7&  75 $\pm$  28&  54 $\pm$  41& 13.5&NV1238.8 \\
MARK817&1239.509 $\pm$  0.029&  -101 $\pm$     7&  46 $\pm$  18&  25 $\pm$  12&  6.3&MgII1239.9 \\
MARK817&1250.071 $\pm$  0.034&  -127 $\pm$     8&  24 $\pm$  13&  13 $\pm$   6&  5.1&SII1250.6 \\
MARK817&1250.415 $\pm$  0.029&   -44 $\pm$     7&  30 $\pm$  11&  26 $\pm$   9&  9.8&SII1250.6 \\
MARK817&1253.355 $\pm$  0.027&  -106 $\pm$     6&  27 $\pm$  12&  32 $\pm$  13& 12.9&SII1253.8 \\
MARK817&1253.653 $\pm$  0.029&   -35 $\pm$     7&  27 $\pm$   8&  53 $\pm$  16& 21.7&SII1253.8 \\
PKS2155-304&1259.848 $\pm$  0.028&  -155 $\pm$     7&  31 $\pm$   9&  76 $\pm$  23&  8.5&SiII+FeII1260.5 \enddata
\end{deluxetable}
\normalsize

\begin{deluxetable}{lcrcccc}
\tablecolumns{8}
\tablecaption{Intrinsic \lya features\label{linelist_intrinsic} with \real.}
\tablewidth{0pt}
\tablehead{
\colhead{Target} &
\colhead{Wavelength} &
\colhead{Velocity} &
\colhead{\bb} &
\colhead{\Wno} & 
\colhead{SL} &
\colhead{z} \\
\colhead{ } &
\colhead{(\AA)} &
\colhead{(\nokmsno)} &
\colhead{(\nokmsno)} &
\colhead{(m\AA)} & 
\colhead{($\sigma$)} &
\colhead{} 
}
\startdata
AKN120&1252.135 $\pm$  0.034&  -907 $\pm$     8&  28 $\pm$  11&  18 $\pm$   7&  5.1& 0.03000 $\pm$  0.00003 \\
ESO141-G55&1257.716 $\pm$  0.029&  -733 $\pm$     7&  53 $\pm$  38&  18 $\pm$  20&  5.7& 0.03459 $\pm$  0.00002 \\
ESO141-G55&1258.093 $\pm$  0.023&  -643 $\pm$     5&  46 $\pm$   3& 353 $\pm$  11&116.3 & 0.03490 $\pm$  0.00002\\
FAIRALL9&1268.850 $\pm$  0.029&  -688 $\pm$     7&  81 $\pm$  22&  32 $\pm$  12& 11.2 & 0.04375 $\pm$  0.00002\\
FAIRALL9&1269.351 $\pm$  0.029&  -570 $\pm$     7&  46 $\pm$  19&  16 $\pm$   9&  5.4 & 0.04416 $\pm$  0.00002\\
FAIRALL9&1269.731 $\pm$  0.029&  -480 $\pm$     7&  99 $\pm$   3&  74 $\pm$   7& 26.7 & 0.04447 $\pm$  0.00002\\
FAIRALL9&1270.288 $\pm$  0.032&  -349 $\pm$     8&  89 $\pm$  12&  93 $\pm$  16& 32.2 & 0.04493 $\pm$  0.00003\\
FAIRALL9&1270.746 $\pm$  0.027&  -241 $\pm$     6&  11 $\pm$  11&  11 $\pm$   6&  4.2 & 0.04531 $\pm$  0.00002\\
FAIRALL9&1270.955 $\pm$  0.029&  -191 $\pm$     7&  42 $\pm$   5&  55 $\pm$   9& 20.3 & 0.04548 $\pm$  0.00002\\
MARK279&1249.023 $\pm$  0.036&  -923 $\pm$     9&  43 $\pm$  12&  22 $\pm$   8&  7.5& 0.02744 $\pm$  0.00003 \\
MARK279&1251.056 $\pm$  0.025&  -436 $\pm$     6&  99 $\pm$   3& 646 $\pm$  18&240.5& 0.02911 $\pm$  0.00002 \\
MARK279&1251.261 $\pm$  0.029&  -387 $\pm$     7&  19 $\pm$  10&  17 $\pm$  10&  6.6 & 0.02928 $\pm$  0.00002\\
MARK279&1251.694 $\pm$  0.023&  -284 $\pm$     5&  71 $\pm$   3& 322 $\pm$  14&123.9& 0.02963 $\pm$  0.00002 \\
MARK279&1253.051 $\pm$  0.023&    41 $\pm$     6&  25 $\pm$   5& 115 $\pm$  65& 45.3& 0.03075 $\pm$  0.00002\\
MARK279&1253.284 $\pm$  0.031&    97 $\pm$     7&  51 $\pm$  12& 375 $\pm$ 117&146.8& 0.03094 $\pm$  0.00003 \\
MARK279&1253.861 $\pm$  0.040&   235 $\pm$    10&  26 $\pm$   6&  94 $\pm$  53& 35.4& 0.03142 $\pm$  0.00003\\
MARK290&1248.079 $\pm$  0.028&  -850 $\pm$     7&  20 $\pm$   9&  22 $\pm$   9&  4.9& 0.02666 $\pm$  0.00002 \\
MARK290&1249.349 $\pm$  0.029&  -545 $\pm$     7&  50 $\pm$  26&  25 $\pm$  23&  6.4& 0.02770 $\pm$  0.00002 \\
MARK290&1249.579 $\pm$  0.029&  -490 $\pm$     7&  97 $\pm$  12& 225 $\pm$  48& 60.4& 0.02789 $\pm$  0.00002 \\
MARK290&1251.115 $\pm$  0.039&  -122 $\pm$     9&  99 $\pm$   3& 475 $\pm$  37&185.6& 0.02916 $\pm$  0.00003 \\
MARK290&1251.664 $\pm$  0.025&     9 $\pm$     6&  50 $\pm$   4& 105 $\pm$  17& 45.2& 0.02961 $\pm$  0.00002 \\
MARK290&1252.109 $\pm$  0.029&   116 $\pm$     7&  61 $\pm$  12&  71 $\pm$  17& 31.3& 0.02997 $\pm$  0.00002 \\
MARK290&1252.479 $\pm$  0.029&   204 $\pm$     7&  35 $\pm$   6&  33 $\pm$  10& 14.5& 0.03028 $\pm$  0.00002\\
MARK290&1254.353 $\pm$  0.029&   653 $\pm$     7&  81 $\pm$  29&  50 $\pm$  24& 17.3& 0.03182 $\pm$  0.00002 \\
MARK290&1254.653 $\pm$  0.029&   725 $\pm$     7&  35 $\pm$  14&  24 $\pm$  12&  7.8& 0.03207 $\pm$  0.00002 \\
MARK290&1254.953 $\pm$  0.029&   797 $\pm$     7&  23 $\pm$  13&  16 $\pm$  10&  4.8& 0.03231 $\pm$  0.00002 \\
$*$MARK501&1254.773 $\pm$  0.029&  -434 $\pm$     7&  99 $\pm$   3&  70 $\pm$  28&  5.4& 0.03217 $\pm$  0.00002 \\
$*$MARK509&1255.795 $\pm$  0.029&  -403 $\pm$     7&  46 $\pm$   5& 132 $\pm$  67& 64.4& 0.03301 $\pm$  0.00002 \\
$*$MARK509&1256.065 $\pm$  0.029&  -339 $\pm$     7&  44 $\pm$  14&  76 $\pm$  92& 37.9& 0.03323 $\pm$  0.00002 \\
$*$MARK509&1256.285 $\pm$  0.029&  -286 $\pm$     7&  93 $\pm$  31& 505 $\pm$ 425&147.7& 0.03341 $\pm$  0.00002 \\
$*$MARK509&1257.165 $\pm$  0.029&   -76 $\pm$     7&  69 $\pm$  21& 191 $\pm$ 229& 61.9& 0.03413 $\pm$  0.00002 \\
$*$MARK509&1257.365 $\pm$  0.029&   -29 $\pm$     7&  65 $\pm$  62& 265 $\pm$ 318&121.5& 0.03430 $\pm$  0.00002\\
$*$MARK509&1257.605 $\pm$  0.029&    29 $\pm$     7&  44 $\pm$  16& 140 $\pm$ 168& 73.7& 0.03450 $\pm$  0.00002 \\
$*$MARK509&1257.785 $\pm$  0.029&    71 $\pm$     7&  28 $\pm$  17&  26 $\pm$  32& 13.9& 0.03464 $\pm$  0.00002\\
$*$MARK509&1257.985 $\pm$  0.029&   119 $\pm$     7&  72 $\pm$  25& 449 $\pm$ 240&238.6& 0.03481 $\pm$  0.00002 \\
MARK817&1249.477 $\pm$  0.041& -1023 $\pm$    10&  26 $\pm$   9&  36 $\pm$  17& 13.8& 0.02781 $\pm$  0.00003\\
MARK817&1249.694 $\pm$  0.077&  -971 $\pm$    18&  28 $\pm$  21&  20 $\pm$  19&  7.7& 0.02799 $\pm$  0.00006 \\
MARK817&1251.752 $\pm$  0.025&  -479 $\pm$     6&  18 $\pm$   7&  14 $\pm$   4&  6.0& 0.02968 $\pm$  0.00002 \enddata
\end{deluxetable}
\normalsize
\clearpage
	\section{CfA Galaxy pie diagrams} \label{sec:pie}
The very local redshift range of this absorption line survey  ($\cz \leq 20,000$\kmsno), allows
us to present an immediate ``first look''  at the relationship between local
\lya absorbers and galaxies using bright galaxy redshift information currently available.
Several surveys  are complete to $m_{B} \leq 15.5$ over large sky areas
\citep[e.g.,][]{CFA,Marzke96,Grogin98,Huchra99,DaCosta98}. Seven of our targets
lie in the sky areas of these surveys (3C~273, IZW1, Mrk~335, Mrk~421,
Mrk~501, Q~1230+011 and PKS~2155-304). All of these redshifts surveys are
included in the latest (Feb 8, 2000) on-line version of the CfA Redshift Catalogue
\citep{CFA} so that the environments of the \lya absorbers found along 
these sightlines can be investigated using available data. Also,
pencil-beam surveys have been conducted along several sightlines to
address the relationship between galaxies and \lya clouds using smaller
numbers of absorbers than presented herein \citep[e.g.,][]{Morris93,Tripp98a}. 
In Paper~III, we will use our \lya survey and all available
galaxy redshift survey data to investigate the relationship between \lya clouds
and galaxies. One advantage of the current work is that the
absorber redshifts are known to much better accuracy (i.e., $\pm  10$\kms
 compared to $\pm 200$\kmsno) than the absorbers detected in
the QSO Absorption Line Key Project, which used the FOS at 1\Ang resolution. 

In Figures\ \ref{PIE_3C273}-\ref{PIE_Q1230+0115}, we present 
heliocentric ``pie diagrams'' (right ascension [RA, $\alpha$] or 
declination [DEC, $\delta$] versus heliocentric radial velocity) for all 
15 HST targets in our GHRS survey. These pie diagrams indicate the
spatial positions of the CfA Redshift Catalogue galaxies relative to 
our target sightlines and the detected \lya absorbers, assuming a pure Hubble flow. 
The  2000 February 8 version of  this catalog \citep{CFA} contains \about120,000 galaxies with 
velocities less than 100,000\kmsno. 
We present pie diagram in $\pm 7$\degr\  extent as a 2D representation of the 3D 
distribution of the known low-\z\ galaxy distribution.  
The sightline towards each target is indicated by the 
dashed line that starts at the apex of each pie diagram (the Sun) and terminates with either a large 
circle, indicating the position of the HST target, or  with a dashed arrow
indicating that  target lies off the pie diagram.  Each CfA galaxy is indicated by a ``{\bf c}''.
CfA galaxies that have a sightline impact parameter less than 1 \hsfi Mpc 
are indicated by ``{\bf v}''s (the {\bf v} indicates that the galaxy is 
{\bf very} close to the sightline). The orientation of the `{\bf v}'s and  `{\bf c}'s are preserved between the RA and DEC
versions of the pie diagrams. 

In the previous sections, all \lya absorber velocities were reported in the LSR velocity scale, established by
aligning the velocity centroids of the Galactic \SIItriplet~absorption lines with the dominant Galactic \hone emission. 
We selected this velocity scale instead of the HST+GHRS wavelength solution due to the possibility of an improper 
wavelength scale caused by poor target centering in the large science aperture (LSA). 
To convert our LSR \lya absorber velocities to heliocentric
velocities, we assumed that the solar velocity with respect to LSR is +20.0\kms towards  
($\alpha$=18:03:50.3; $\delta$=+30:00:17, J2000).
The heliocentric velocities of our definite \lya absorbers (\real), 
if present, are indicated on the pie diagrams  by medium-sized dashed circles. 
Possible \lya absorbers (\tent), if  present, are indicated by small dashed circles. 
If the target is present within the pie diagram, its location is given by a large circle;
otherwise a dashed arrow indicates that the target is beyond the limit of the pie diagram.

Also in Appendix~B, Table \ref{nearest_master} lists the 
nearest 3 galaxies for all  \lya absorbers in our \expanded\ sample. For each absorber, 
we indicate absorber type (D=definite, or P=possible), heliocentric absorber velocity 
in\kmsno, name of the nearest galaxy, distance (\Dperp) from the 
galaxy to the line of sight (LOS)  in \hsfi Mpc, heliocentric recession velocity (\cz) of the galaxy, 
distance (D$_{cz}$) along the LOS from the absorber to the 
galaxy allowing for peculiar velocities of $\pm 300$\kmsno,
the Euclidean 3D absorber-galaxy distance (D$_{\rm tot}$),
 and the Zwicky blue (B) magnitude of the galaxy (if available).
All information on galaxy  location, velocity,  or magnitude is taken from the CfA redshift
survey unless otherwise indicated. The use of a ``retarded'' Hubble 
flow accounts for peculiar galaxy motion and galaxy rotation and always decreases the reported LOS and total 
absorber-galaxy distances. With our ``retarded Hubble flow" peculiar velocity allowance,
we consider any Galaxy within  $\pm 300$\kms of  the recession 
velocity of an absorber to be at the same distance  as the 
absorber (see S95 for justification of this somewhat arbitrary $\pm 300$\kms allowance). 
In Paper~III, we will use the data presented here to quantify the environment of local \lya clouds.
\section{Conclusion}\label{sec:conclusion}
This paper is the first in a series devoted to a detailed study of the
``local \lya forest'' using observations obtained with the Hubble Space
Telescope (HST) and Goddard High Resolution Spectrometer (GHRS) in its medium
resolution mode (G160M grating). The current paper has 
presented a detailed description of the target selection, 
observational strategy and data reduction (\S~\ref{sec:setup}),
along with a presentation of the reduced spectra and lists
of the all absorption lines detected (\S~\ref{sec:master} and Appendix~\ref{sec:ApA}).
 We discovered \Nabs\ definite ($\geq 4\sigma$) and \Npos\ possible ($3-4\sigma$) intervening \lya absorption lines
 along with numerous Galactic metal lines,
including metal absorption from several ``high-velocity clouds'', and a
few absorption systems judged to be ``intrinsic'' to the target AGN. 
 In Paper~II of this series  we will present
a detailed analysis of the absorption line data presented herein,
including the ``\bvalueno'' distribution, the column density distribution,
and the two-point correlation function for local \lya absorbers.
In Paper~III we will present an analysis of the
relationship between these \lya absorbing clouds and galaxies based upon
the bright galaxy survey data also presented herein. The 
\lowz galaxy redshift survey data presented in the form of ``pie diagrams''
 in \S~\ref{sec:pie} and Appendix~\ref{sec:ApB}  provide a valuable ``first look'' at the
relationship between the local \lya clouds and bright galaxies.
In Paper~IV we will present the two-point correlation function of the \lowzya forest based
upon the HST/GHRS data presented here and 13 additional HST/STIS sightlines obtained in Cycle 7.

The GHRS+G160M observing combination has two primary
advantages for doing this study: (1) it is sensitive to the weakest
\lya absorbers detected at \lowz thus far (\logNh\ $\geq 12.5$;
comparable to the weakest lines found in the high-z \lya forest with the
Keck Telescope); and (2) the \about19\kms spectral resolution resolves
almost all of the \lya lines detected.  This combination also yields extremely accurate 
velocities for detected absorbers ($\le 10$\kmsno). The disadvantages of this approach are
that only the UV-brightest AGN in the sky are possible targets for this 
work (typically $V \leq 14.5$) and that only a small
pathlength can be observed with this spectrograph configuration. The
fifteen GHRS target sightlines sample only 116,000\kms of total 
pathlength for \lya absorber detection (details of the
pathlength surveyed as a function of \hone column density are 
presented in Paper~II). However, the discovery of a large number of \lya
absorbers by this GHRS survey makes it clear that the observational
approach adopted here of obtaining the highest resolution spectra
possible to investigate the local \lya forest has been successful.
Furthermore, even though the total observed pathlength is small
compared to other \lya surveys, the pathlength observed by this work is
the most important piece of any individual sightline,  the nearest \about10,000\kmsno. 
This has enabled the discovery of the nearest \lya absorbers and given us the opportunity to relate these 
clouds to the nearest known galaxies.  

Owing to the success of this GHRS survey, our group has obtained HST +
Space Telescope Imaging Spectrograph (STIS) spectra of an additional 13 
UV bright targets to complete our survey of the \lowzno, low-\nh local \lya forest. The STIS observations will be presented 
in a later paper. 
\acknowledgements
For their assistance in obtaining the HST/GHRS data over several
cycles, we are grateful to the staff at the Space Telescope Science
Institute, particularly Ray Lucas.   We would like to thank Brad Gibson for his assistance in 
deciphering the HVCs and his helpful discussions on \bvaluesno. 
In addition, we would like to thank Mark Giroux and Greg Bothun for their helpful comments on this manuscript.
This work was supported by HST guest observer 
grant GO-06593.01-95A, the HST COS project (NAS5-98043), and by the Astrophysical Theory Program
(NASA grant NAGW-766 and NSF grant AST96-17073). 
\clearpage
\appendix
\section{GHRS/G160M Spectra and Absorption Line Lists}\label{sec:ApA}
In this Appendix we present composite HST/GHRS/G160M spectra, and line lists (Table~\ref{linelist_all}) describing the detected
absorption features. Descriptions of features in these data unique to individual objects are presented in \S~\ref{sec:each} of the
text.

\begin{figure} \epsscale{0.9} \plotone{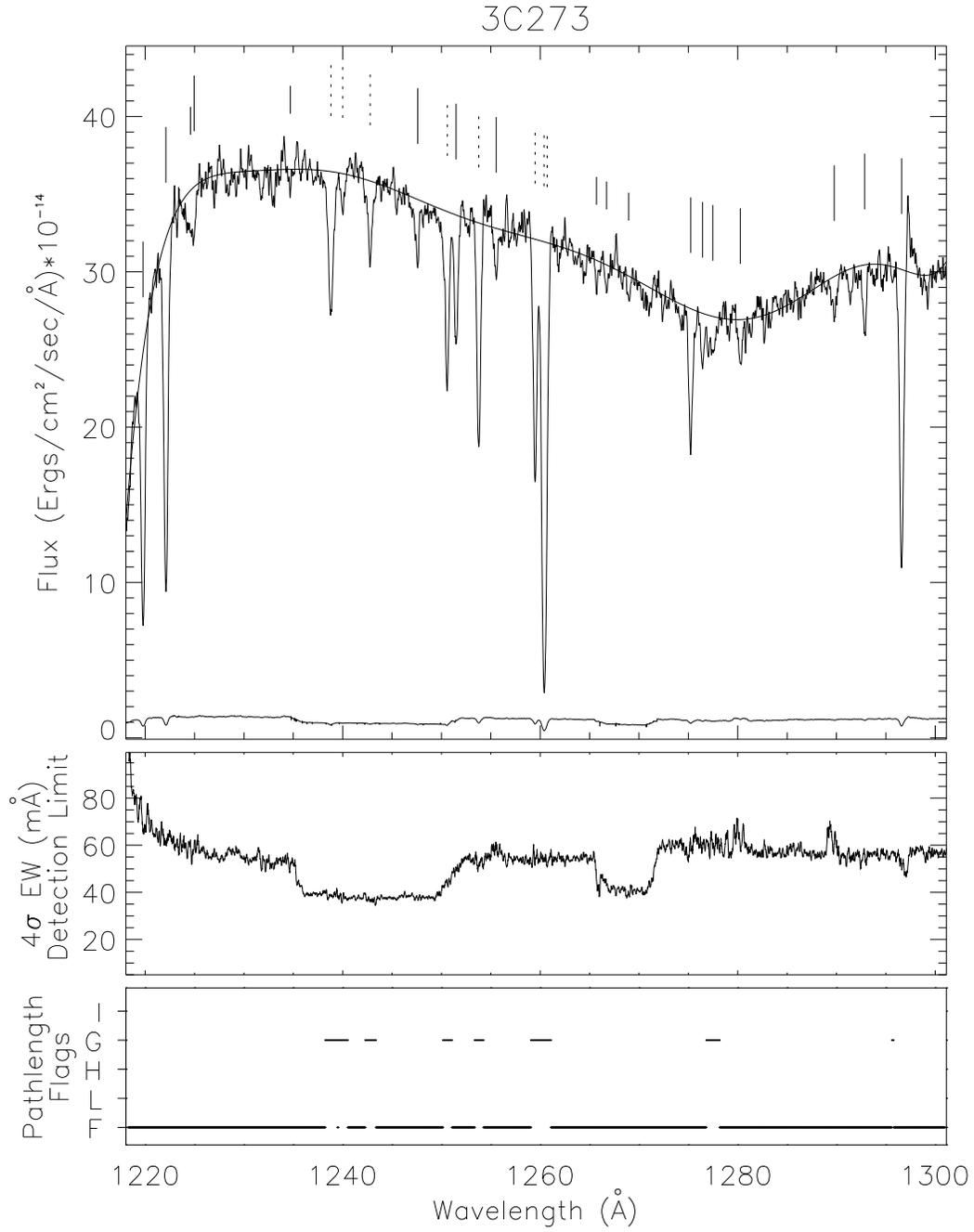}
\caption{\label{3C273}GHRS/G160M 3C~273 spectrum, sensitivity limits, and pathlength flags.}
\end{figure}

\begin{figure} \epsscale{0.9} \plotone{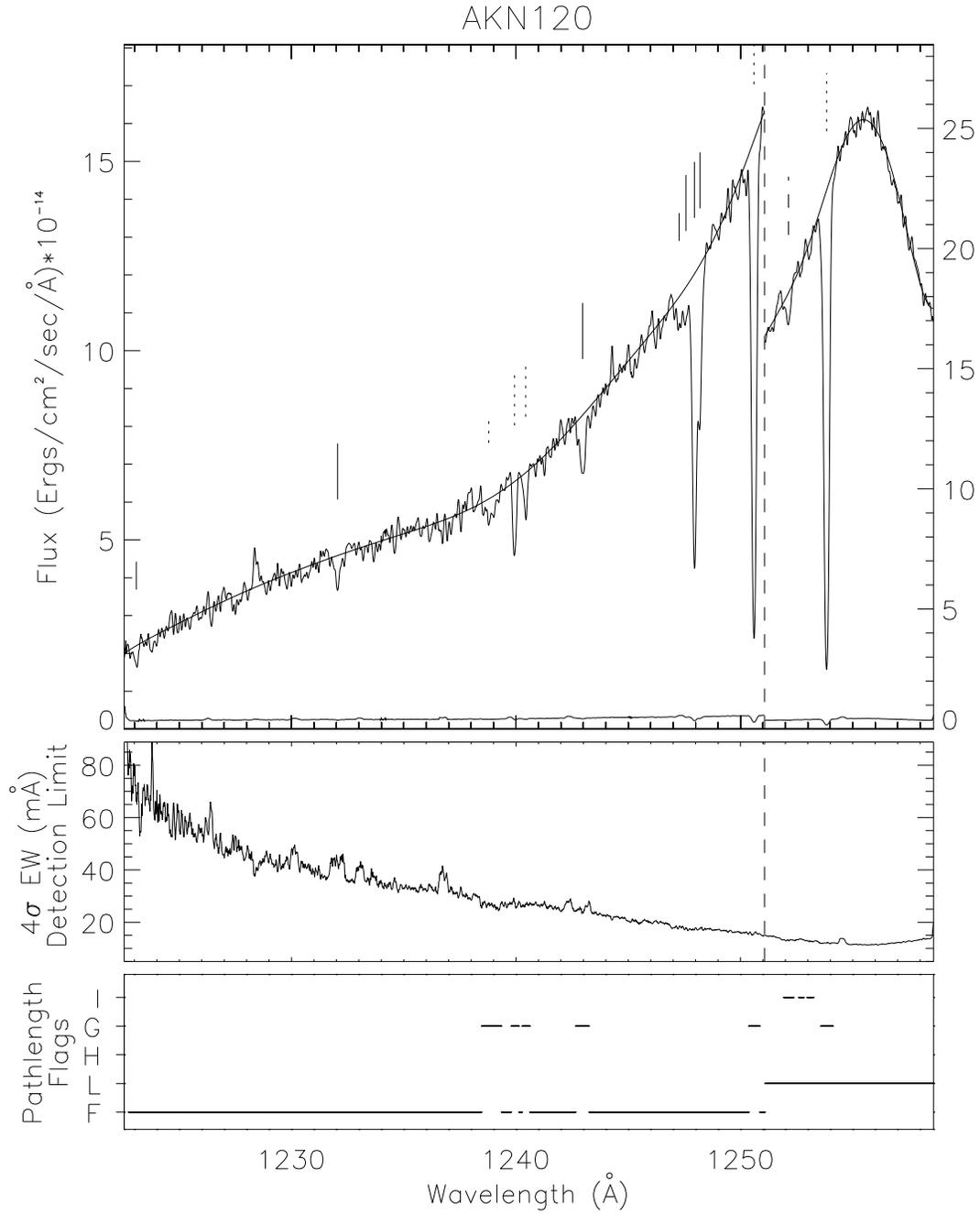}
\caption{\label{Akn120}GHRS/G160M Arakelian~120 spectrum.
 The dashed vertical line indicates the proximity limit as explained in 
 \S~\ref{sec:IIA}.  In the upper panel,  the left axis corresponds to flux blueward of the proximity limit, 
 while the right axis corresponds to flux redward of this convenient breakpoint.}
\end{figure}

\begin{figure} \epsscale{0.9} \plotone{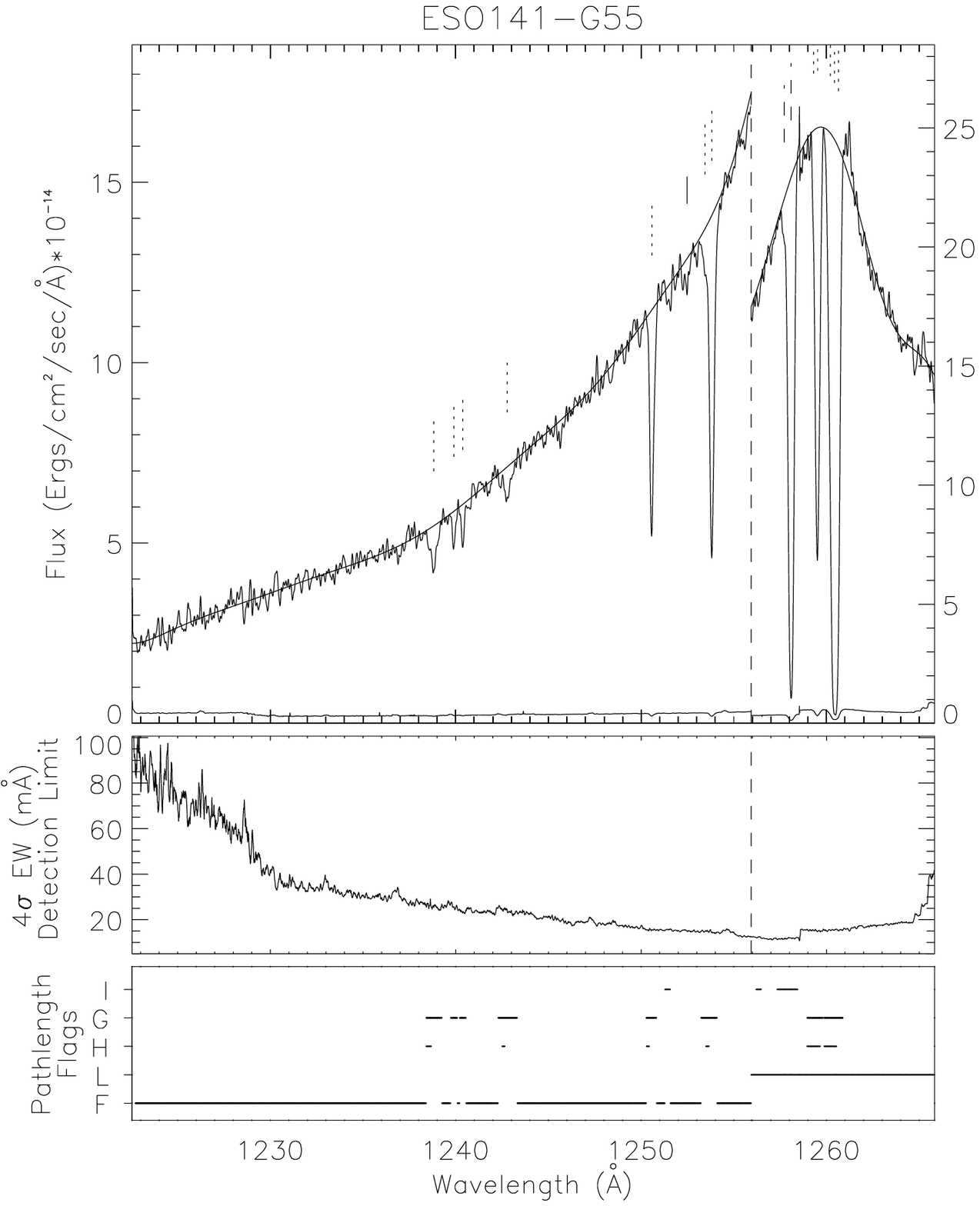}
\caption{\label{ESO141-G55}GHRS/G160M \ESO\ spectrum, sensitivity limits, and pathlength flags.}
\end{figure}

\begin{figure} 	\epsscale{0.9} \plotone{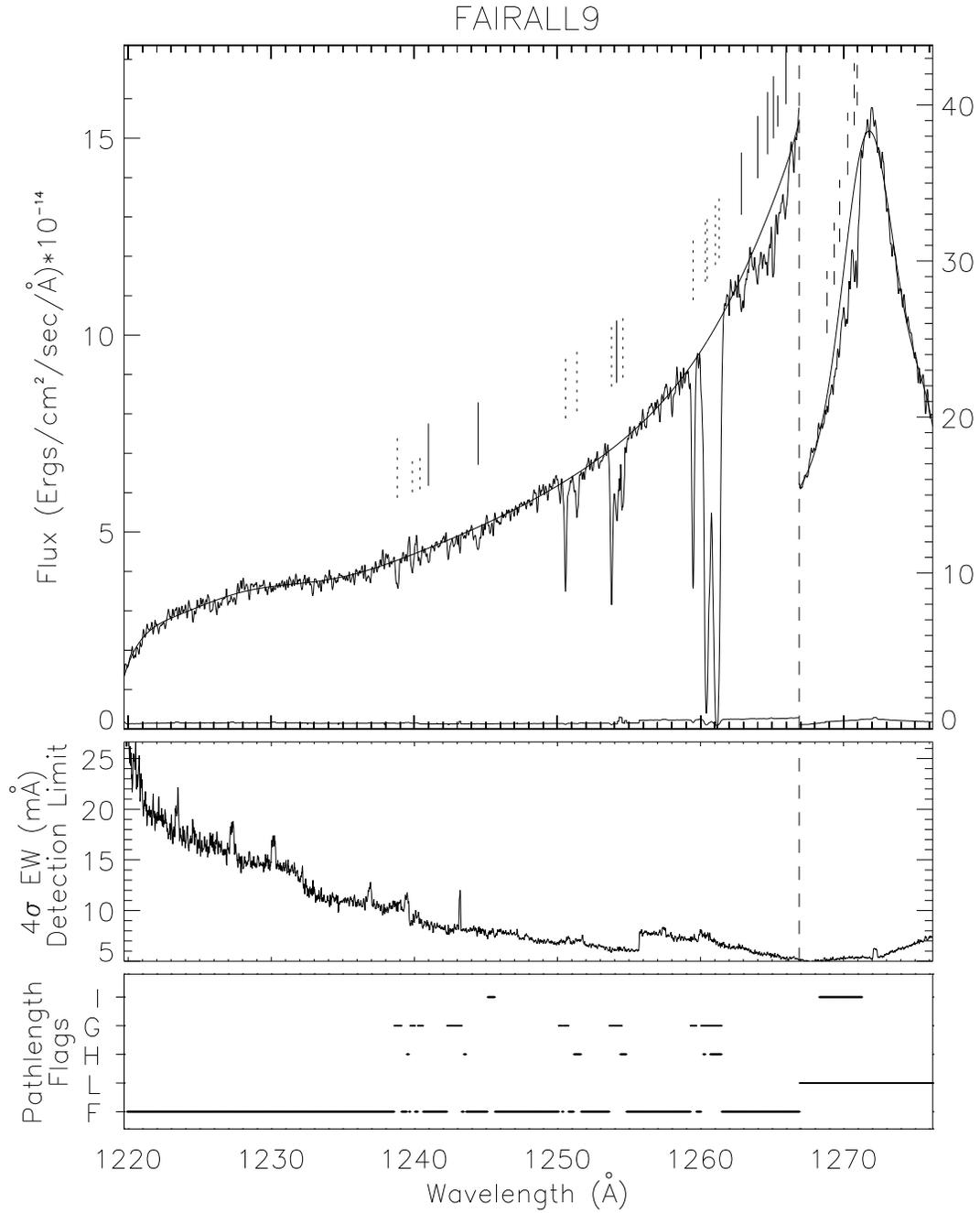}
\caption{\label{FAIRALL9}GHRS/G160M Fairall~9 spectrum, sensitivity limits, and pathlength flags.}
\end{figure}

\begin{figure} 	\epsscale{0.9} \plotone{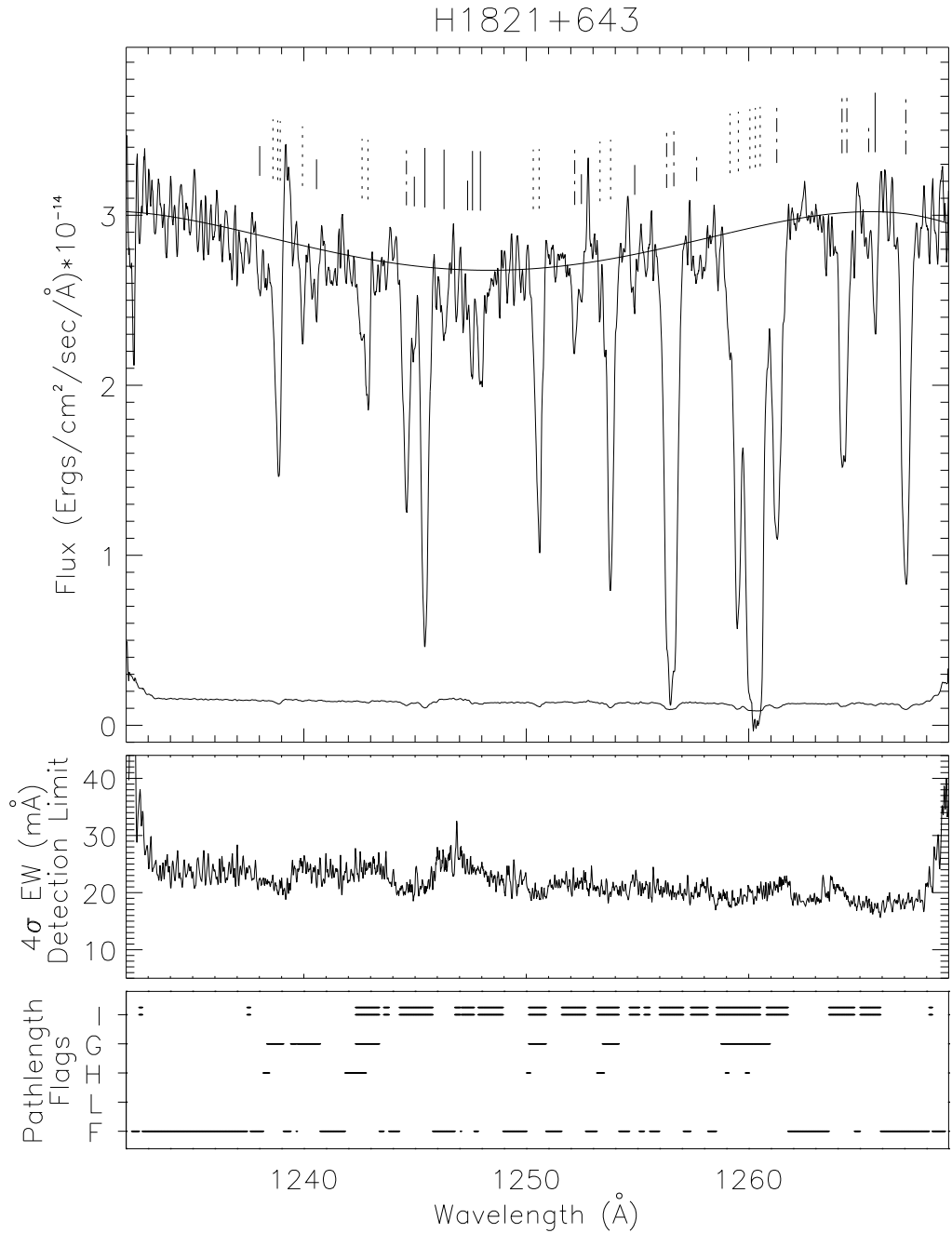}
\caption{\label{H1821+643}GHRS/G160M H~1821+643 spectrum. 
Features indicated in the upper panel by a dot-dashed line are determined to be intergalactic
non-\lya lines. The double lines in the ``{\bf I}'' row of the Pathlength Flags denotes pathlength excluded due to
these intergalactic non-\lya lines at higher \cz.}
\end{figure}

\begin{figure} \epsscale{0.9} \plotone{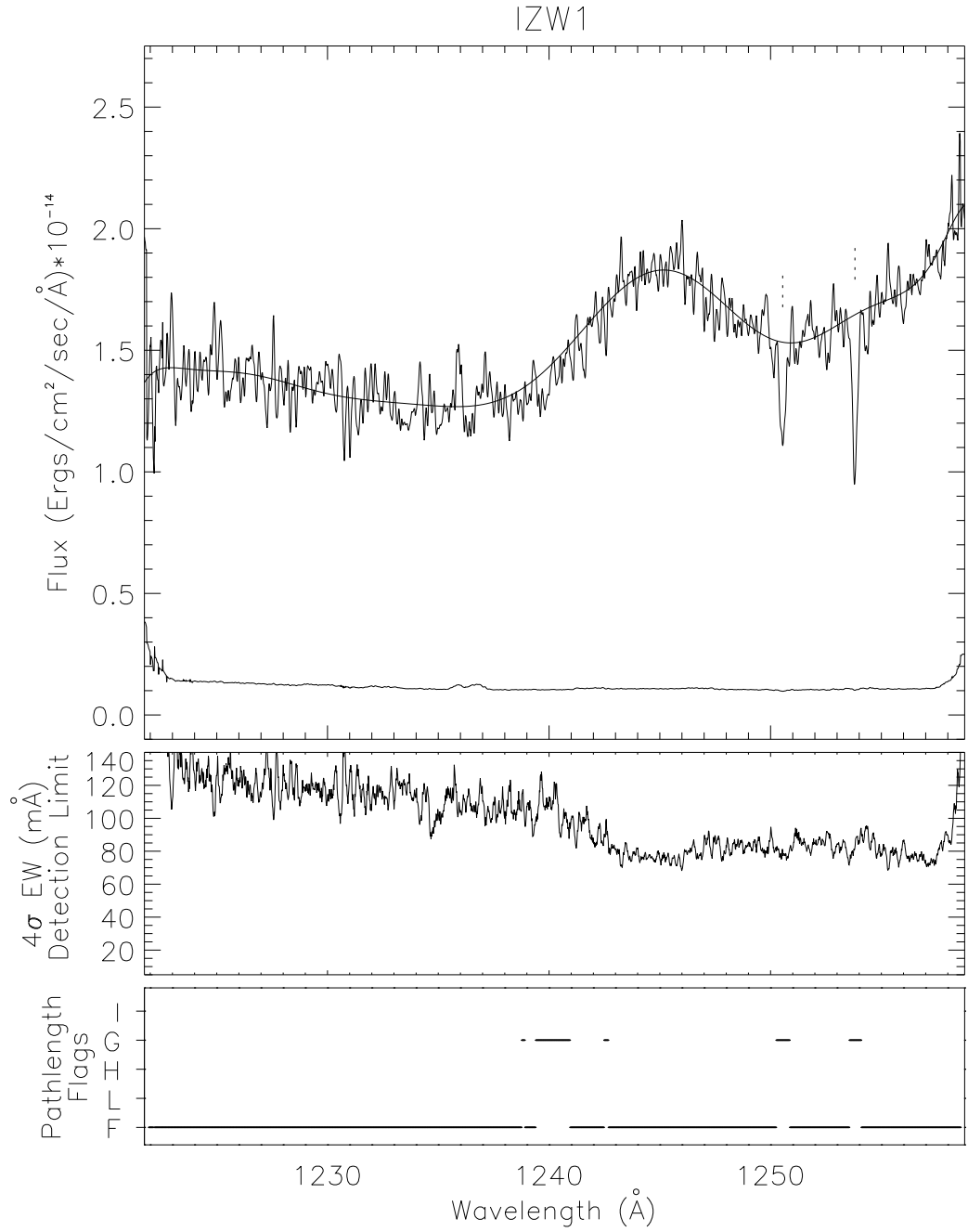}
\caption{\label{IZW1}GHRS/G160M I~ZW~1 spectrum, sensitivity limits, and pathlength flags.}
\end{figure}

\begin{figure} \epsscale{0.9} \plotone{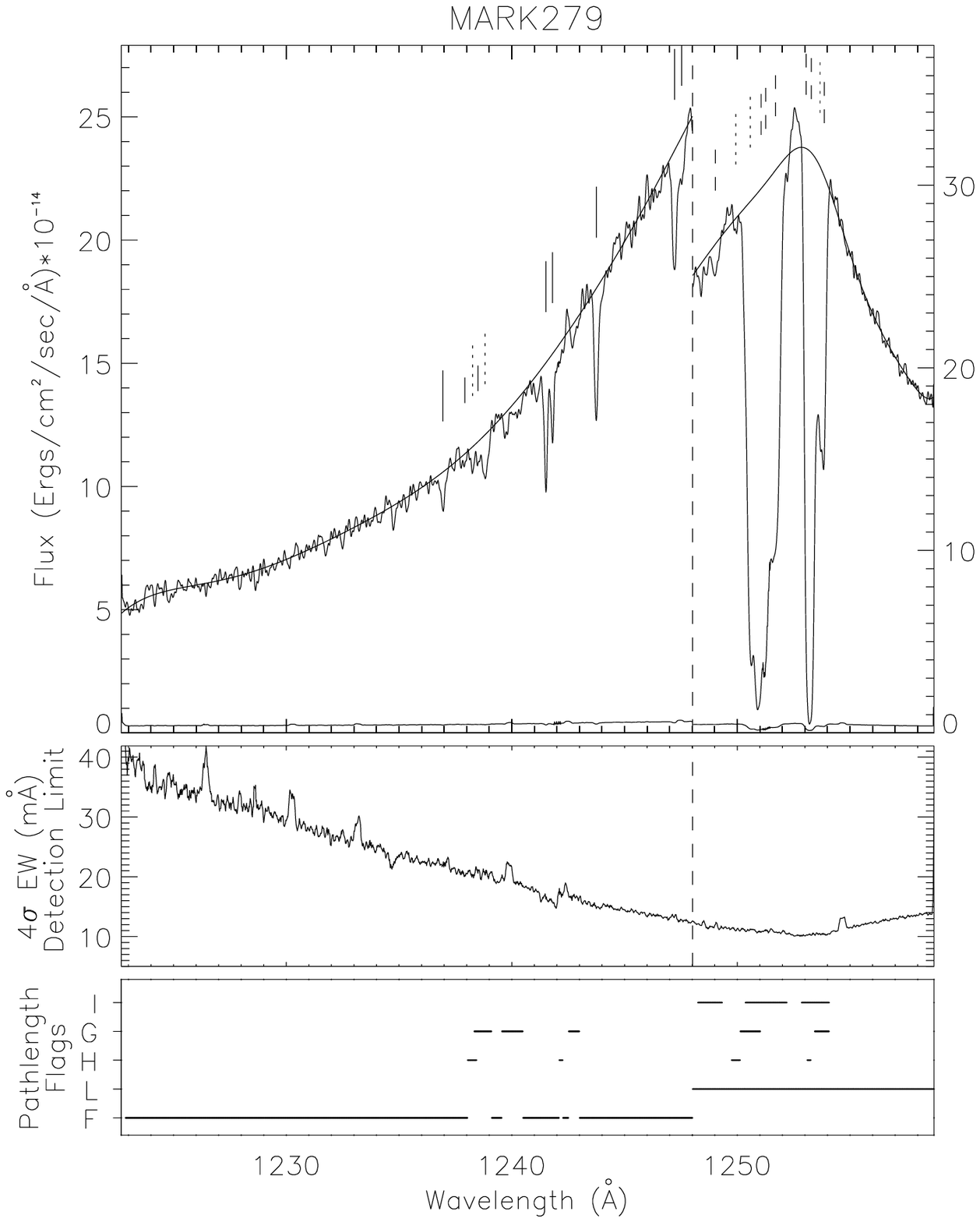}
\caption{\label{MARK279}GHRS/G160M Markarian~279 spectrum, sensitivity limits, and pathlength flags.}
\end{figure}

\begin{figure} \epsscale{0.9} \plotone{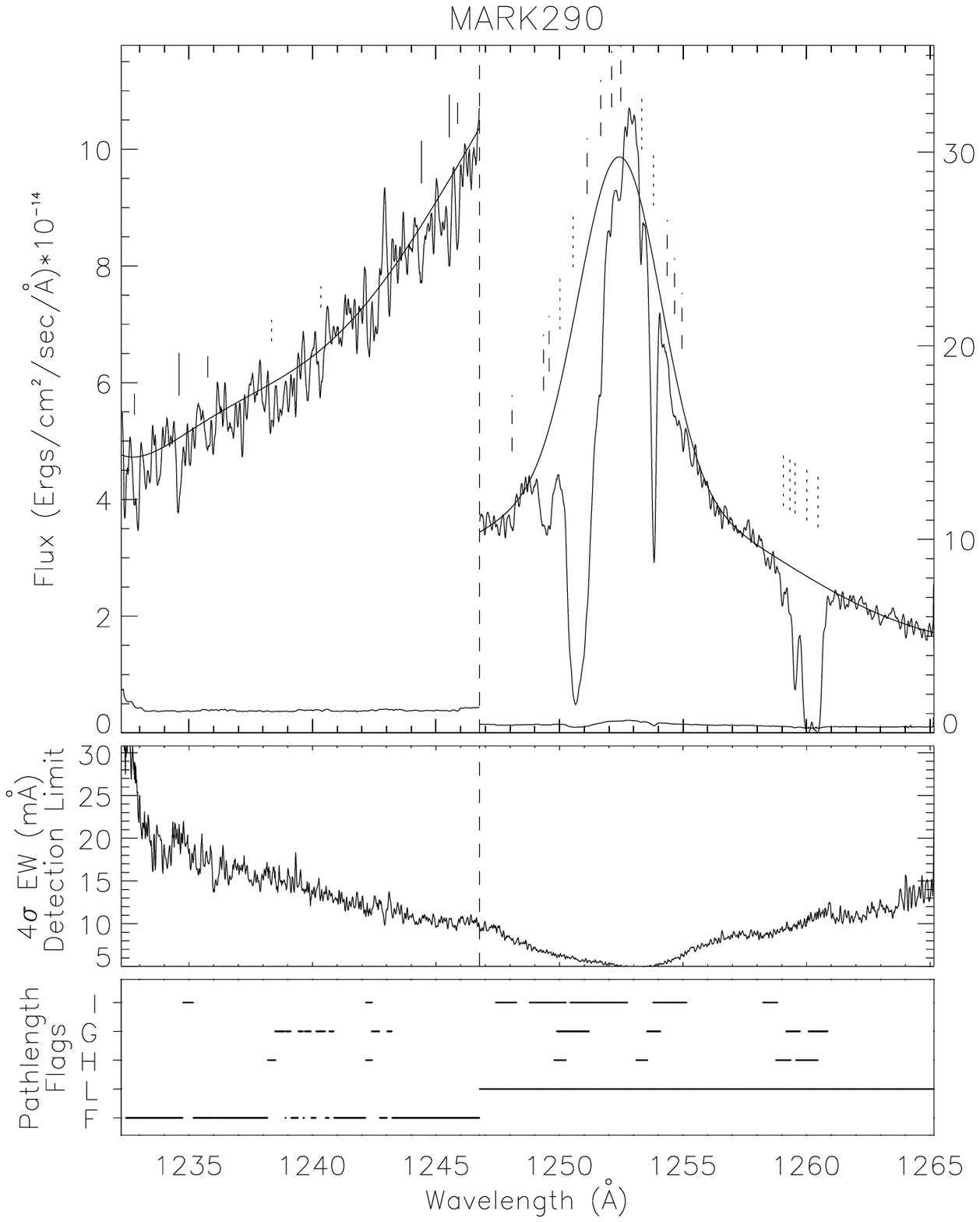}
\caption{\label{MARK290}GHRS/G160M Markarian~290 spectrum, sensitivity limits, and pathlength flags.}
\end{figure}

\begin{figure} \epsscale{0.9} \plotone{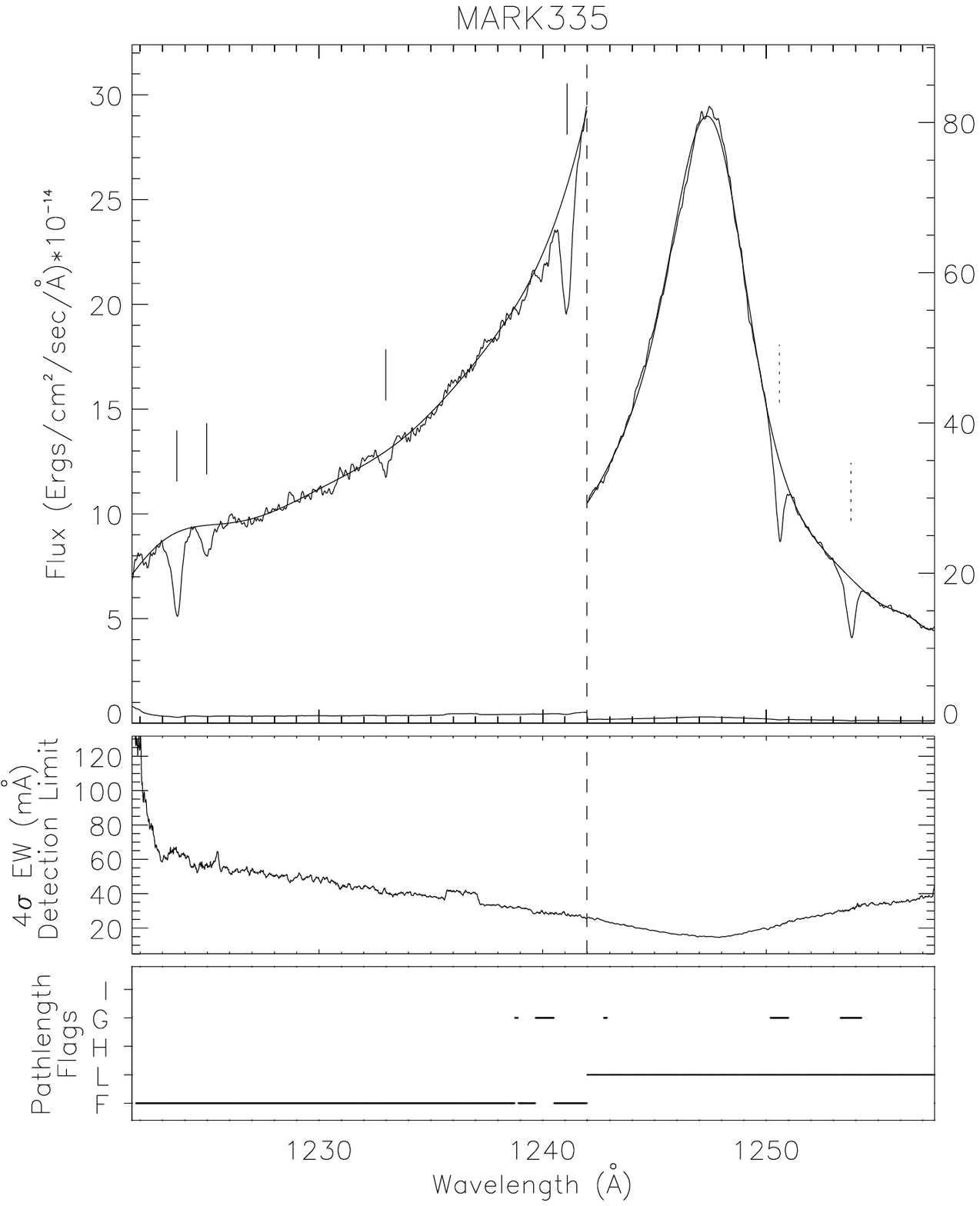}
\caption{\label{MARK335}GHRS/G160M Markarian~335 spectrum, sensitivity limits, and pathlength flags.}
\end{figure}

\begin{figure} \epsscale{0.9}	\plotone{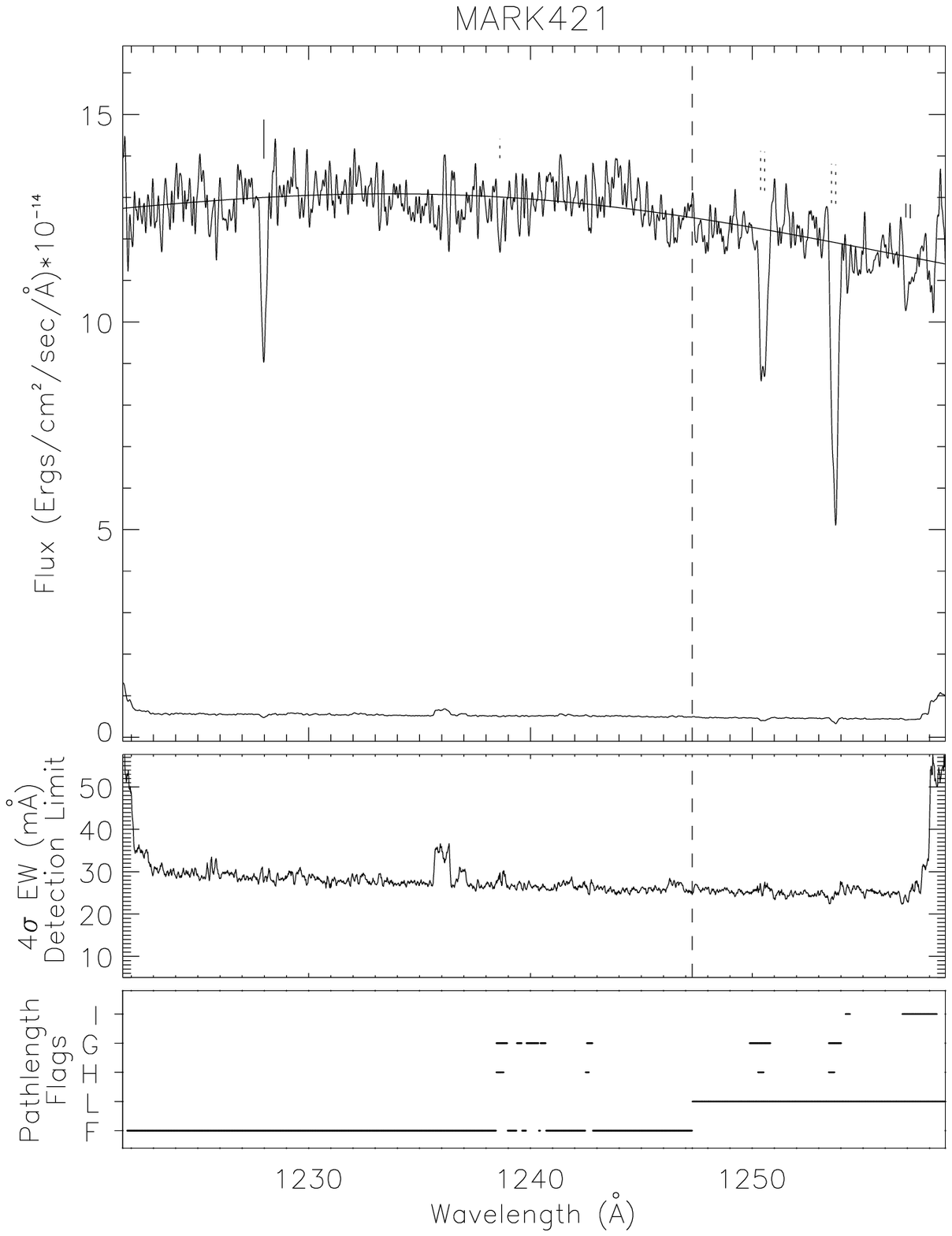}
\caption{\label{MARK421}GHRS/G160M Markarian~421 spectrum, sensitivity limits, and pathlength flags.}
\end{figure}

\begin{figure} \epsscale{0.9}	\plotone{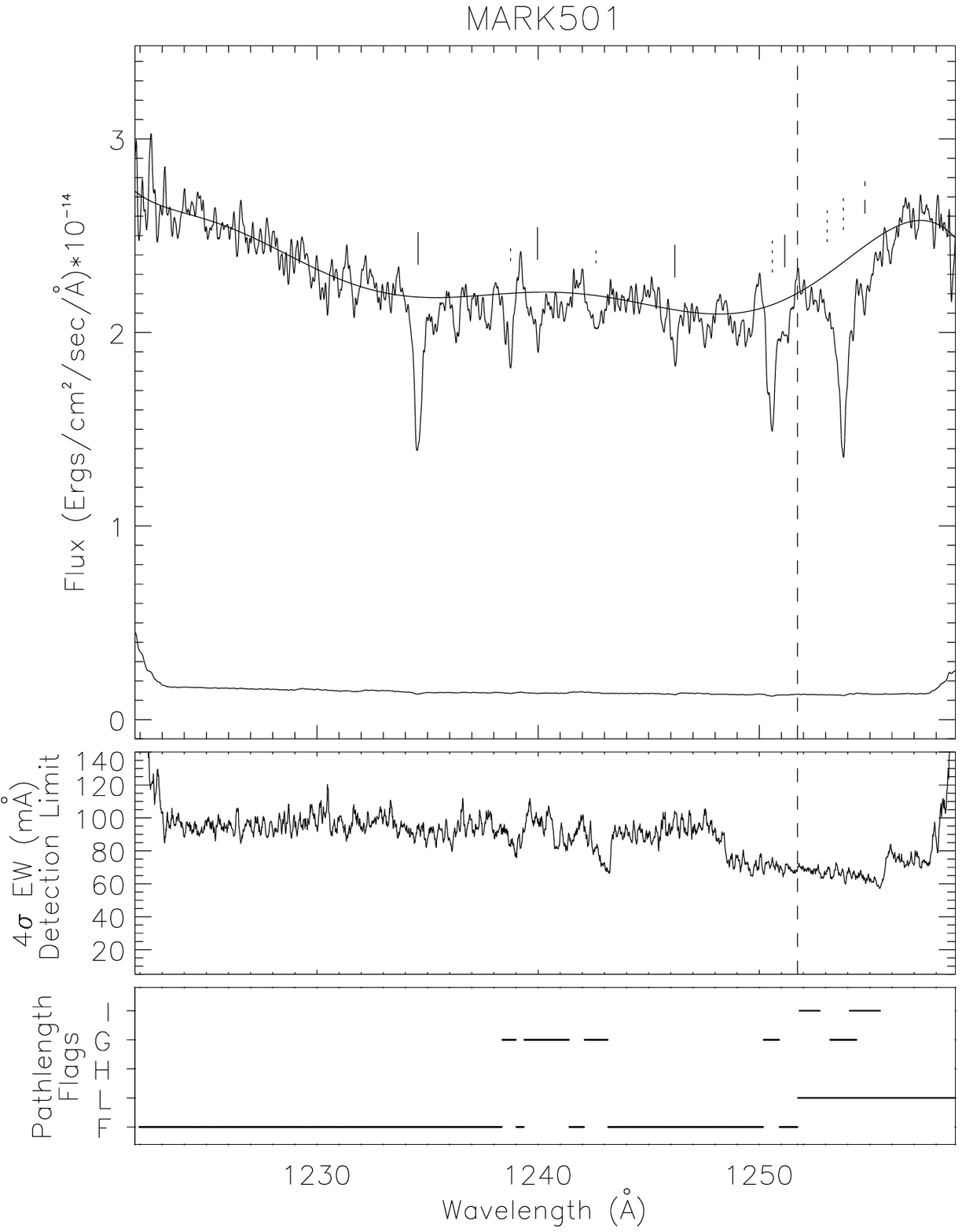}
\caption{\label{MARK501}GHRS/G160M Markarian~501 spectrum, sensitivity limits, and pathlength flags.}
\end{figure}

\begin{figure} \epsscale{0.9} \plotone{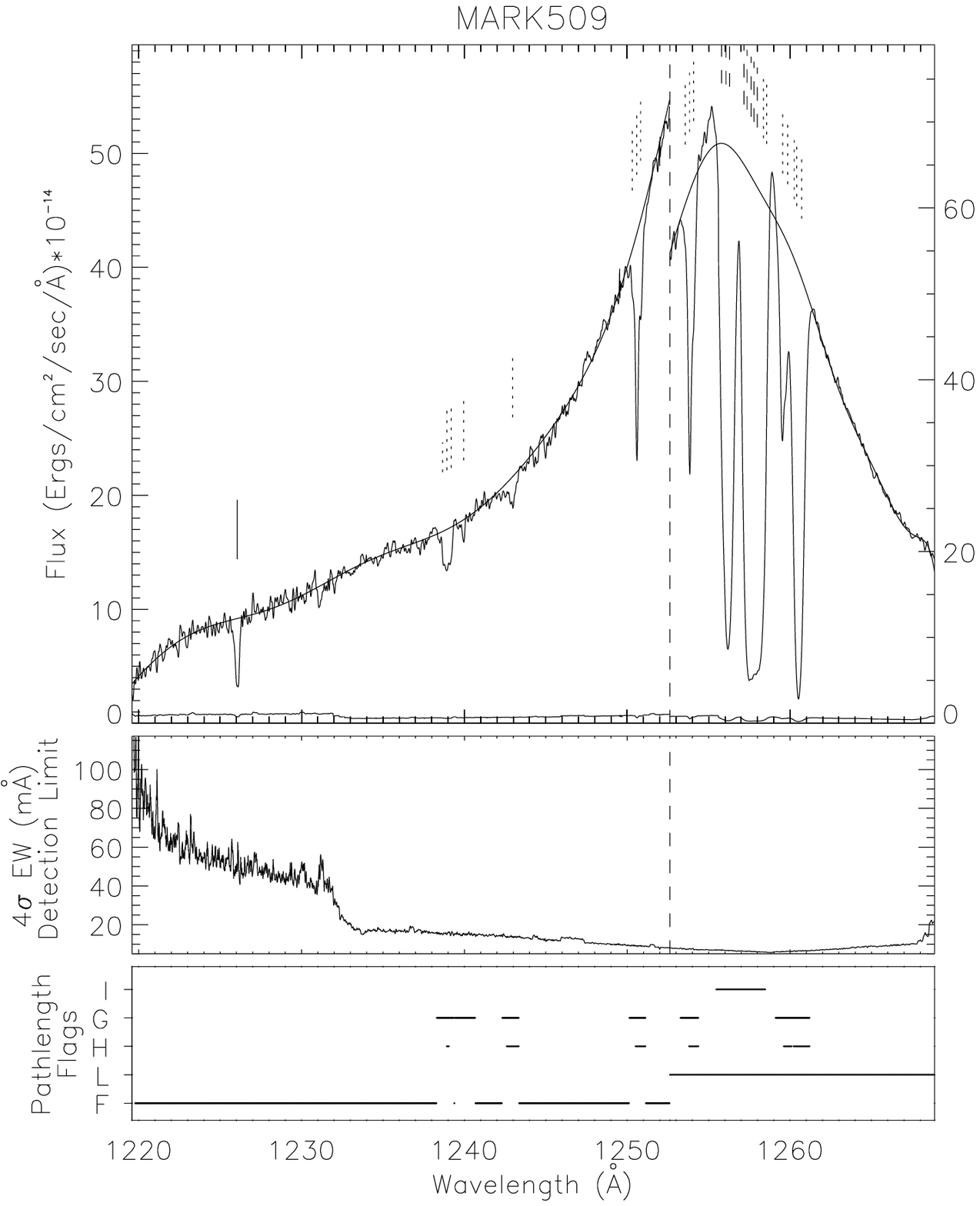}
\caption{\label{MARK509}GHRS/G160M Markarian~509 spectrum, sensitivity limits, and pathlength flags.}
\end{figure}

\begin{figure} \epsscale{0.9}	\plotone{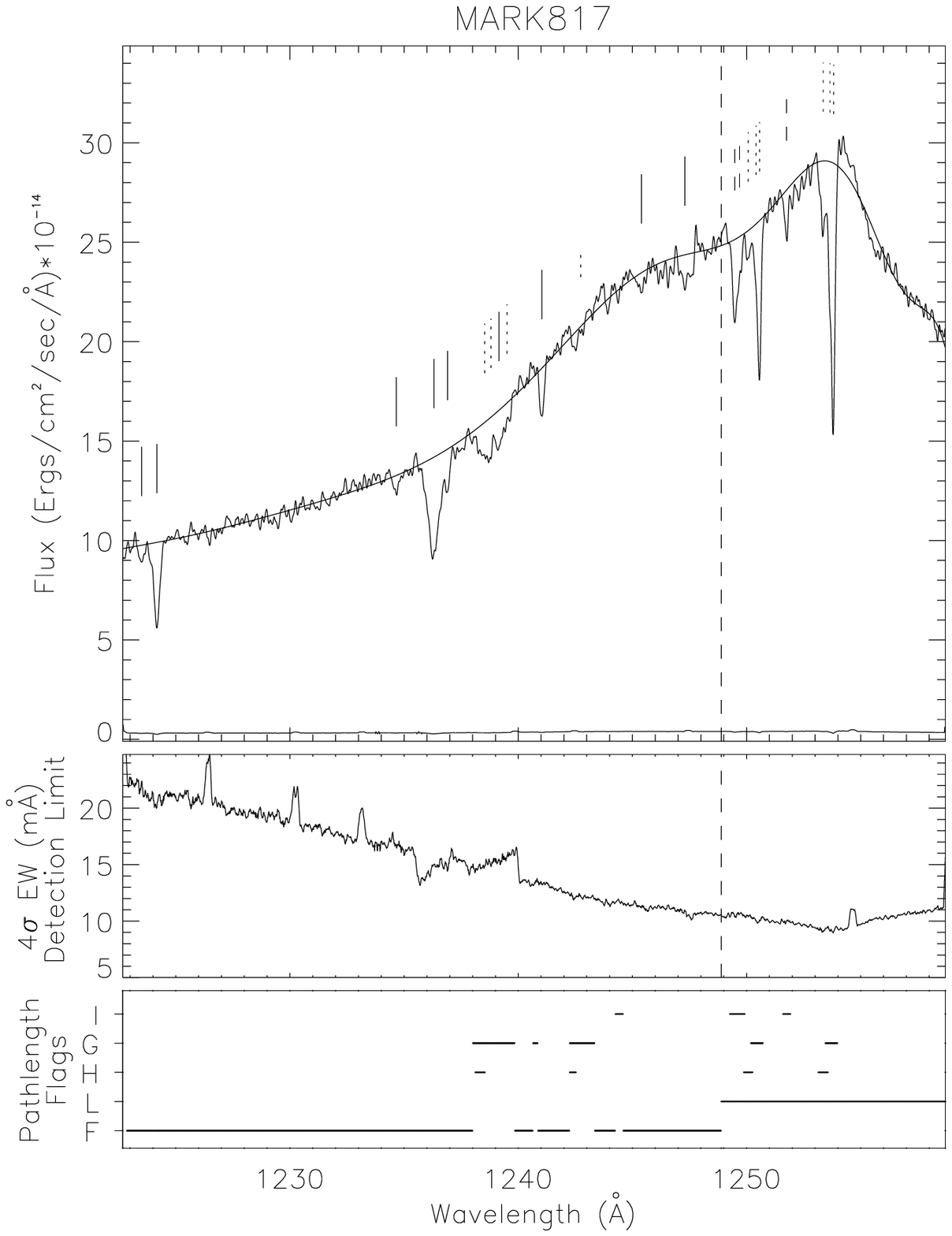}
	\caption{\label{MARK817}GHRS/G160M Markarian~817 spectrum, sensitivity limits, and pathlength flags.} 
\end{figure}

\begin{figure} \epsscale{0.9} \plotone{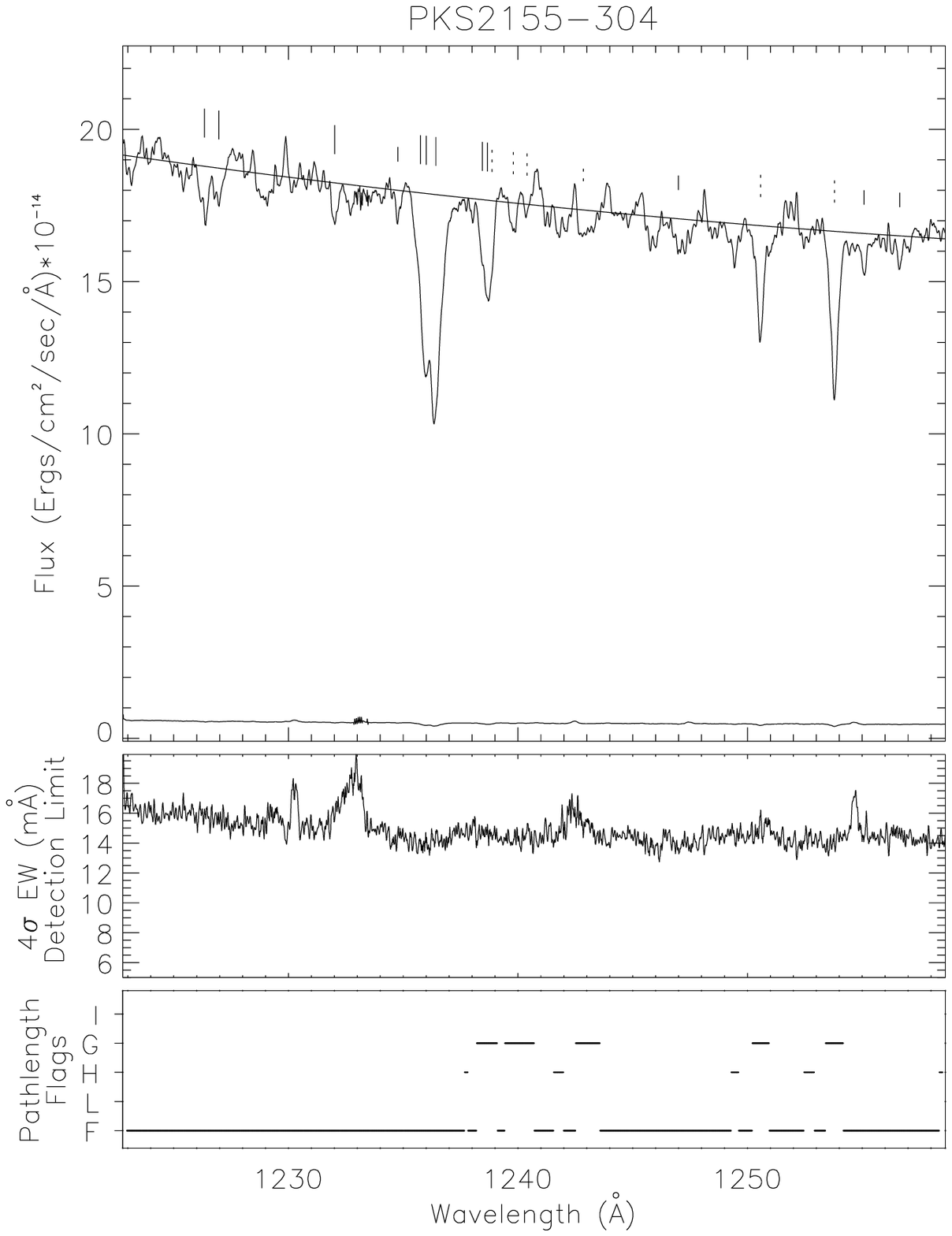}
	\caption{\label{PKS2155_PRE}Pre-COSTAR GHRS/G160M \PKS\ spectrum, sensitivity limits, and pathlength flags.}
\end{figure}

\begin{figure} \epsscale{0.9} \plotone{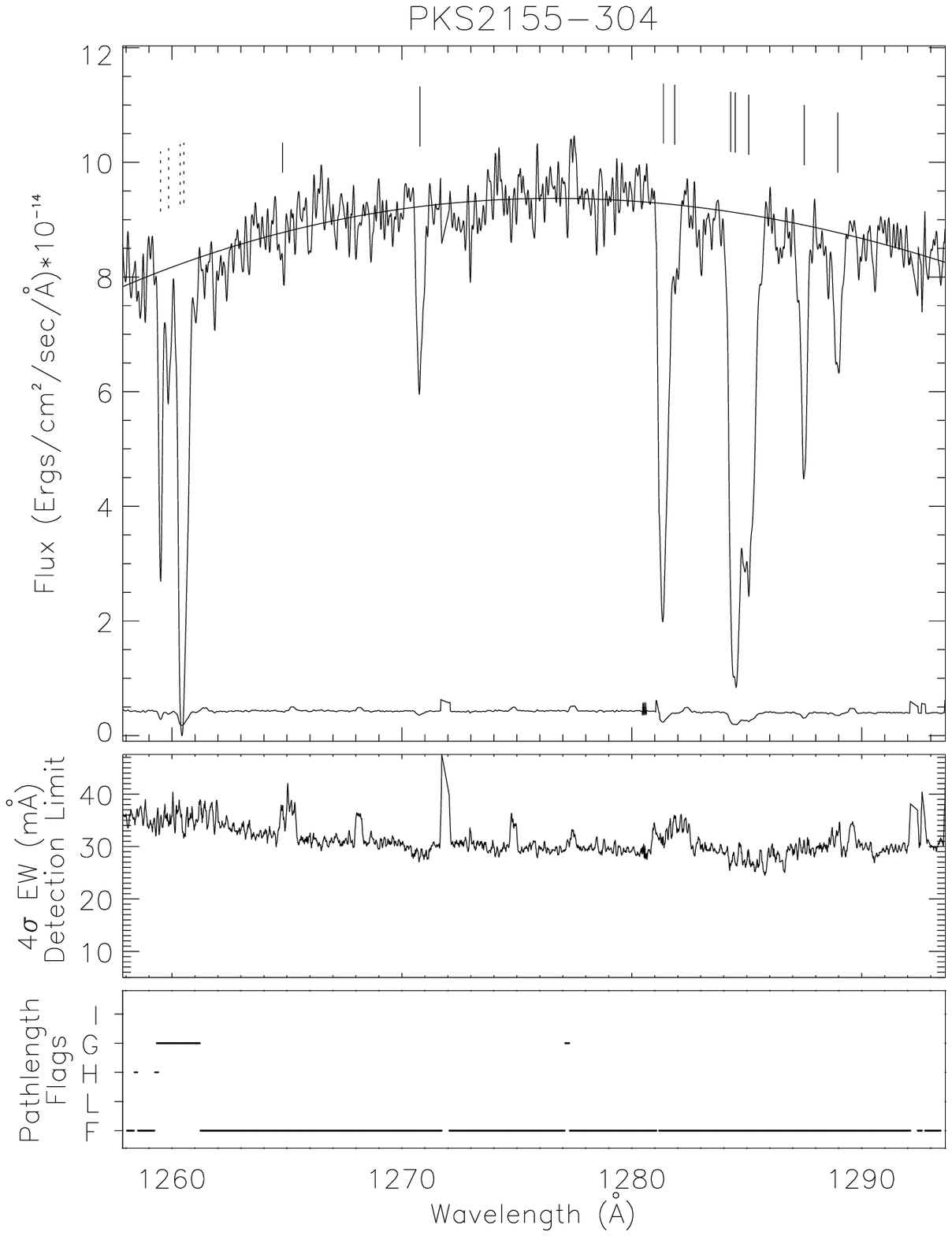}
\caption{\label{PKS2155_POST}Post-COSTAR GHRS/G160M \PKS\ spectrum, sensitivity limits, and pathlength flags.}
\end{figure}

\begin{figure} \epsscale{0.9} \plotone{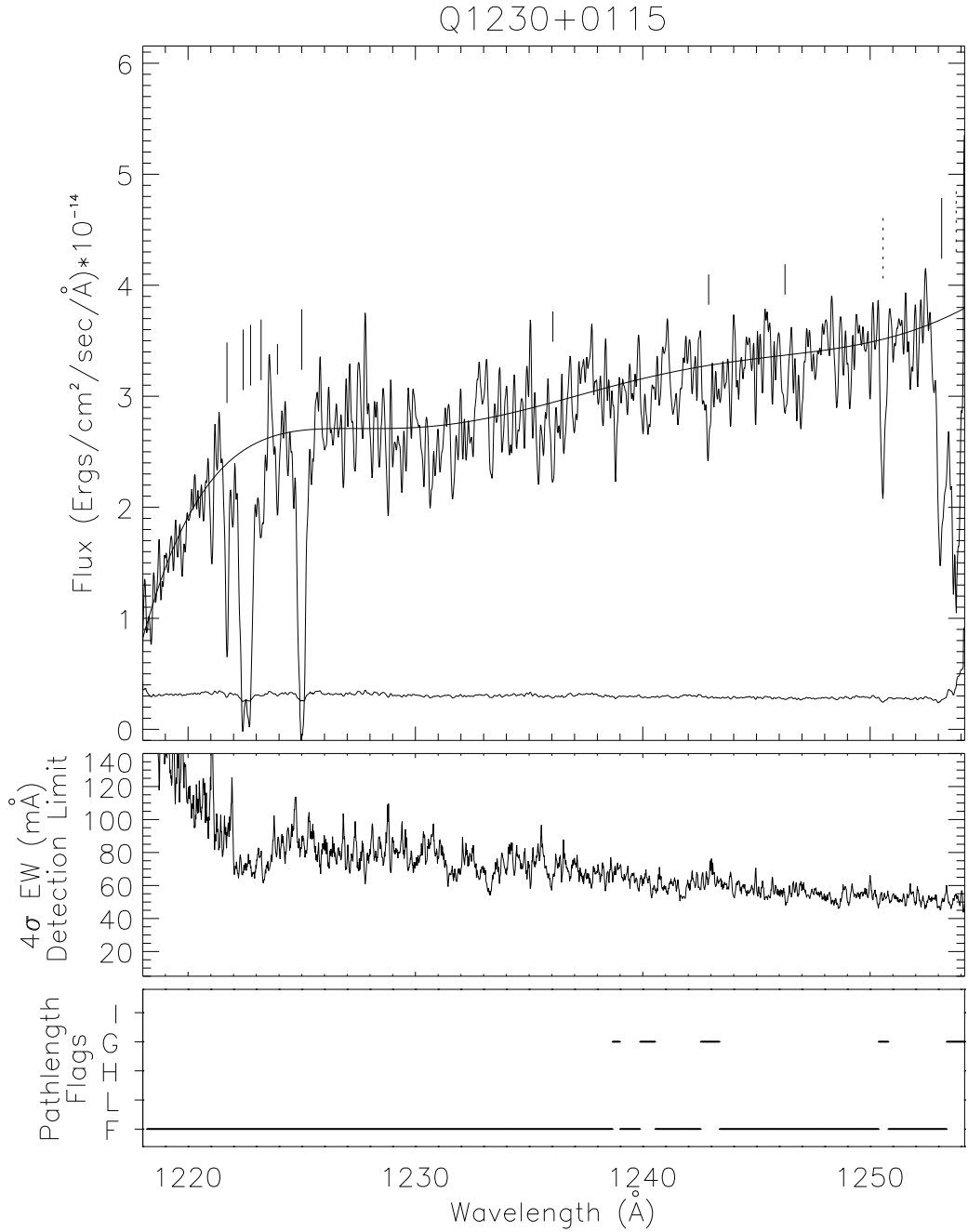}
 \caption{\label{Q1230+0115}GHRS/G160M Q~1230+0115 spectrum, sensitivity limits, and pathlength flags.}
\end{figure}

The GHRS/G160M spectra, error vectors, and pathlength accountings are presented in 
Figures~\ref{3C273}--\ref{Q1230+0115} for all targets.
The upper panel of each figure presents the composite HST/GHRS/G160M spectrum and the
1$\sigma$ error values, which are plotted below the flux vector.
As discussed in \S~\ref{sec:IIA}, our list of intergalactic
\lya absorbers is restricted to those more than 1,200\kms blueward of the target \lya emission.
This limit is indicated by the dashed vertical line in the upper and middle panels of the figures.  All non-Galactic features 
redward of this line are classified as intrinsic absorbers. All otherwise unidentified absorption features blueward
are taken to be intergalactic \lya absorption features, and are indicated by a solid vertical line above the feature. 
\lya absorbers with significance level (SL)~$ > 4\sigma$ (definite absorbers) are plotted with a longer vertical line than the
\tent\ absorbers. Galactic and high-velocity cloud (HVC) absorption lines are similarly represented with a dotted
line, intrinsic absorption lines with a dashed line, and intervening higher-\z\ absorption lines with a dot-dashed line. 
The solid line is the global continuum fit as described in \S~\ref{sec:global}. 

Note specifically that in several cases where \lya emission occurs in the observed band, the portion of the spectrum redward of the proximity limit
(vertical dashed line) is excluded from the region best fitted by the continuum. Thus, any intrinsic or Galactic absorptions
redward of the ``proximity limit'' may not have their EWs most accurately represented. A different continuum fitting 
procedure will be used to address these specific features in later publications.
The middle panel shows the
4$\sigma$ rest-frame EW (\Wno) sensitivity detection limit, per resolution element, as a function of
wavelength for each target. As with the upper panel, we include in our statistics only the portion of the spectrum that
lies $ >1,200$\kms blueward of the target \lya emission. Finally, the bottom panel of each figure summarizes the
available pathlength for inclusion in our sample and line statistics. The row marked {\bf `I'} indicates the portion of
the spectrum removed due to specific features intrinsic to the target or non-\lya 	lines associated with known
intervening systems at higher redshift not associated with the target AGN. The pathlength removed due to higher
redshift intervening absorbers is indicated by a double line to differentiate it from that removed due to intrinsic
absorption features.  The other rows are marked as follows: {\bf `G'} indicates pathlength attributed to non-HVC
Galactic features such as
\SIItriplet, {\bf `H'} indicates pathlength attributed to HVCs, {\bf `L'} indicates the
portion of the spectrum that is redward of \proxno, and  {\bf `F'} indicates those regions of the
spectrum that are available for the detection of intergalactic \lya absorbers. The {\bf `F'} row also corrects for our
inability to correctly detect and model features  near the edge of our wavelength coverage; we remove 10 pixels on the edges
of our wavelength coverage. Only for H~1821+643 does the intrinsic/intervening
{\bf `I'} accounting remove any significant pathlength not already removed by the {\bf `L'} row.

For a few objects, the  \prox ``proximity limit'' lies within the observed 
spectral band and is indicated by  a dashed vertical line. 
For targets with strong intrinsic \lya emission (Akn~120, \ESO, Fairall 9, Mrk~279, 
Mrk~290, Mrk~335, and Mrk~509),  fluxes are scaled differently
above and below this convenient breakpoint. For these spectra, the left axis corresponds 
to flux blueward of the proximity limit,  while the right axis corresponds to flux redward of this  
limit.

The line lists for each individual spectrum are presented sequentially in Table~\ref{linelist_all}.
The first column of Table~\ref{linelist_all} indicates the LSR-adjusted
wavelength and wavelength uncertainty for each feature. The wavelength 
scale of each spectrum was corrected to the LSR using the Galactic \hone and
\ion{S}{2} lines (1250, 1253, and 1259\ang, if available). The second column lists the non-relativistic velocity (\cz\ in\kmsno)
relative to the LSR. For \lya absorbers judged to be intrinsic to the AGN, velocities are listed relative to the AGN
narrow-line region (see \S~\ref{sec:IIA}). For intervening non-\lya absorbers, such as those found in H~1821+643, this column
reports the redshift (\z) of the absorber.  We also provide velocity or redshift uncertainties, based upon the  
total wavelength uncertainties (see \S~\ref{sec:LSR}). The third  column provides the single-component Doppler widths (\bb\ in\kmsno), and uncertainties,  for each feature as
estimated from the Gaussian width (\WG).  As explained previously, the Doppler widths have been corrected for the spectral
resolution of the GHRS and  the single-component Gaussian fits were restricted to the range $12<\bobs<100$\kmsno. 
The fourth column lists the rest-frame EW (\W in\mang)  and its uncertainty for each absorption
feature. This uncertainty includes both the statistical error of
the $\chi^2$ fit and our conservative estimate of the systematic error in the continuum placement (see \S~\ref{sec:global}).
The fifth column indicates the significance level (SL in \signo) of each feature.  
The SL is calculated by integrating the S/N (per RE) of the best fit Gaussian for each feature.
The final two columns present the identification (Id) and alternative identification (Alt Id) if applicable for each feature. 
Absorption lines that are determined not to be intergalactic \lya  have their identifications prefaced by {\bf g:}
(Galactic), {\bf h:} (HVC), {\bf i:} (intrinsic), or {\bf z:} (intervening).  Alternate line identifications that lie between
0.2--0.4\Ang from the expected location are indicated as speculative by the inclusion of a {\bf ?}
following the identification.
%3C273
% Akn 120
% ESO 141-G55
% Fairall 9
 % H1821+643
% I Zw 1
% Mark 279
% Mark 290
% Mark 335
\clearpage
 % Mark 421
% Mark 501
\clearpage
 % Mark 509
% Mark 817
% PKS 2155-304
% Q1230+0115
\begin{deluxetable}{lcccrcc}
\tablecolumns{7}
\tablecaption{HST/GHRS Absorption Features\label{linelist_all}}
\tablewidth{0pt}
\tablehead{
\colhead{Wavelength} &
\colhead{Velocity} &
\colhead{\bb} &
\colhead{\Wno} &
\colhead{SL} &
\colhead{Id} &
\colhead{Alt Id}\\
\colhead{(\AA)} &
\colhead{(\nokmsno)} &
\colhead{(\nokmsno)} &
\colhead{(m\AA)} &
\colhead{($\sigma$)} &
\colhead{ } &\colhead{ }
}
\startdata
\cutinhead{3C~273}
1219.786 $\pm$  0.024&  1015 $\pm$     6&  69 $\pm$   5& 369 $\pm$  36& 34.8&Ly$\alpha$&\nodata \\
1222.100 $\pm$  0.023&  1586 $\pm$     6&  72 $\pm$   4& 373 $\pm$  30& 42.0&Ly$\alpha$&\nodata \\
1224.587 $\pm$  0.150&  2199 $\pm$    37&  57 $\pm$  55&  29 $\pm$  35&  3.5&Ly$\alpha$&\nodata \\
1224.954 $\pm$  0.029&  2290 $\pm$     7&  54 $\pm$  33&  35 $\pm$  30&  4.2&Ly$\alpha$&\nodata \\
1234.704 $\pm$  0.029&  4694 $\pm$     7&  69 $\pm$  61&  25 $\pm$  29&  3.5&Ly$\alpha$&\nodata \\
1238.814 $\pm$  0.030&    -2 $\pm$     7&  87 $\pm$   8& 158 $\pm$  21& 27.2&g:NV1238.8&\nodata \\
1239.991 $\pm$  0.138&    16 $\pm$    33&  67 $\pm$  47&  36 $\pm$  34&  6.1&g:MgII1239.9&Ly$\alpha$ \\
1242.783 $\pm$  0.043&    -5 $\pm$    10&  76 $\pm$  15&  81 $\pm$  21& 13.5&g:NV1242.8&\nodata \\
\enddata
\end{deluxetable}
\normalsize

\normalsize
\clearpage
\section{CfA Pie Diagrams and Nearest Neighbor Distances}\label{sec:ApB}
\begin{figure}  \plotone{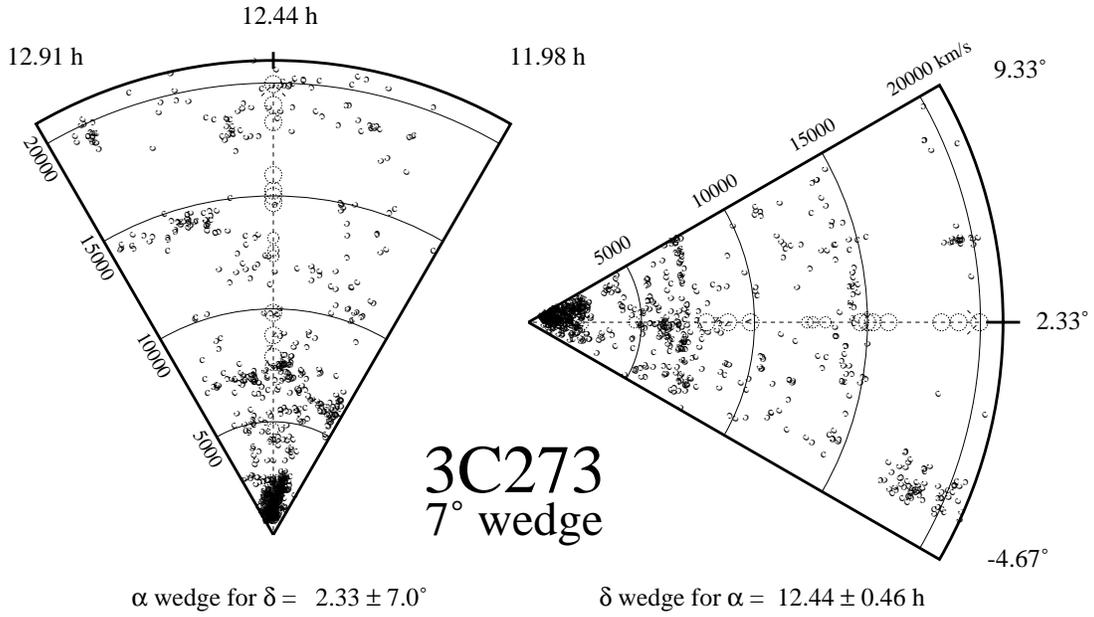}
\caption{Right ascension ($\alpha$) and declination ($\delta$) pie diagram for the 3C~273 sightline.\label{PIE_3C273}}
\end{figure}

\begin{figure}  \plotone{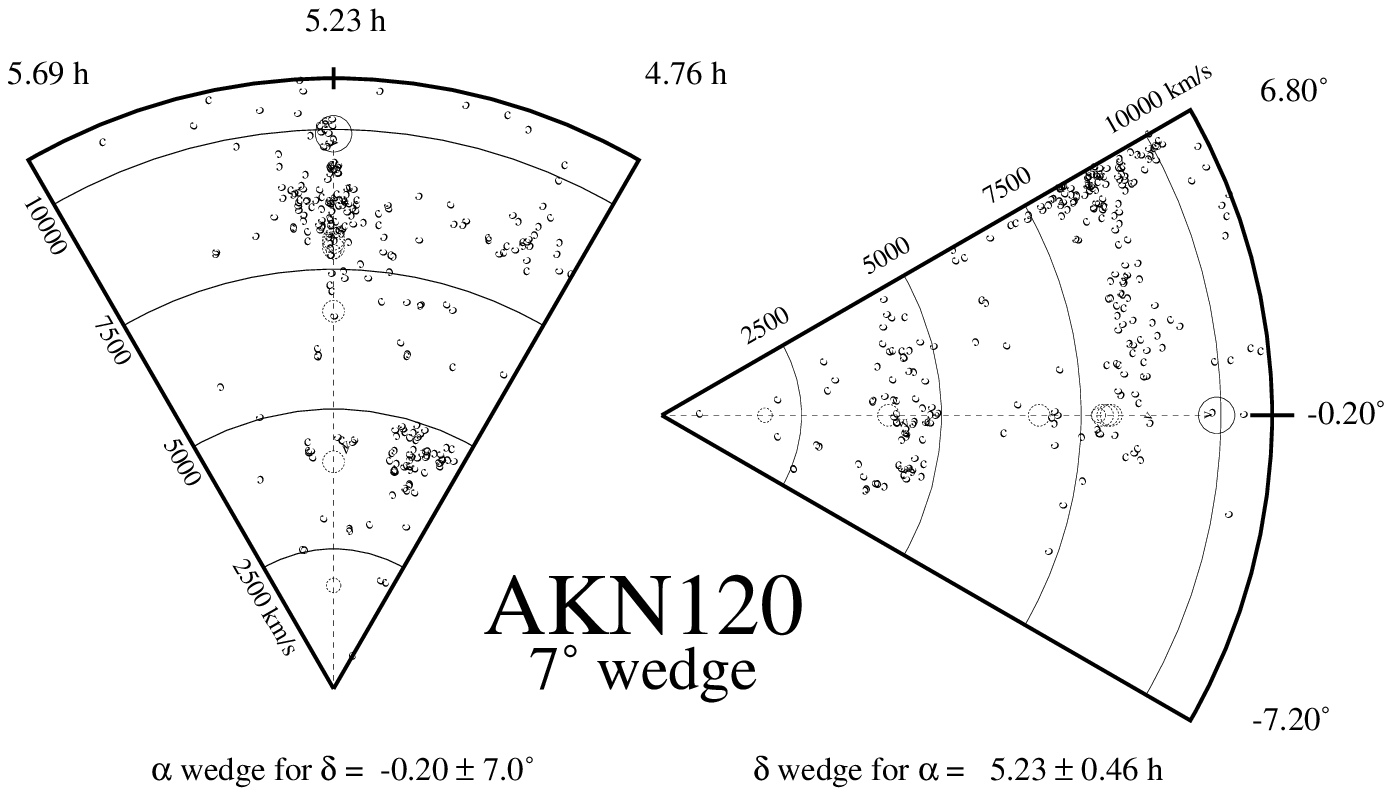}
\caption{Right ascension ($\alpha$) and declination ($\delta$) pie diagram for the Arakelian~120 sightline.\label{PIE_Akn120}}
\end{figure}

\begin{figure} \plotone{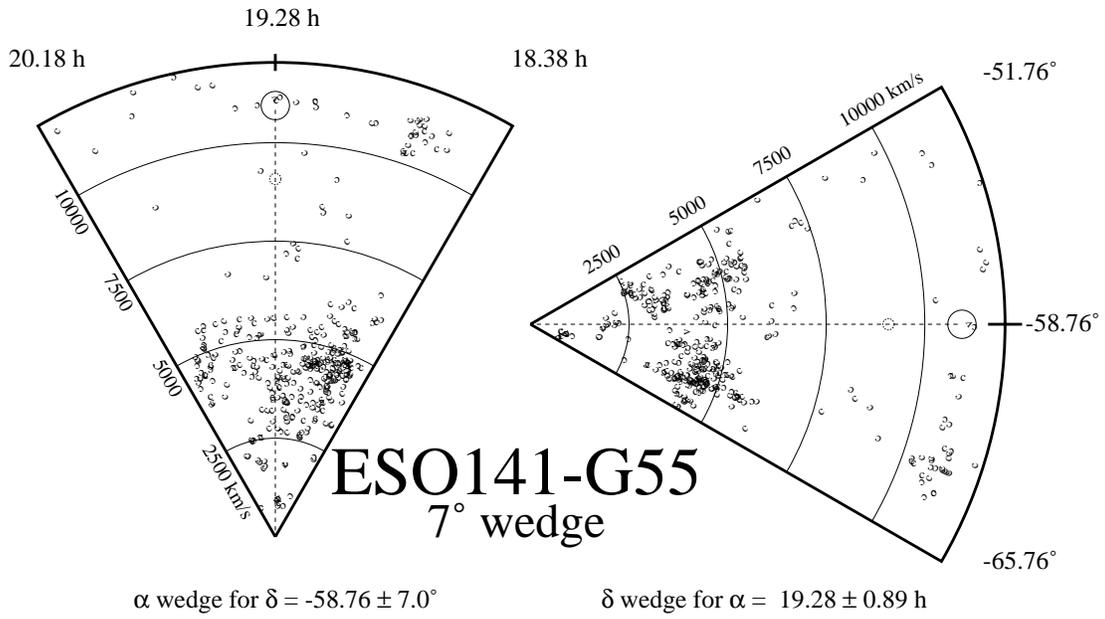}
\caption{Right ascension ($\alpha$) and declination ($\delta$) pie diagram for the ESO~141-G55 sightline.\label{PIE_ESO141-G55}}
\end{figure}

\begin{figure} \plotone{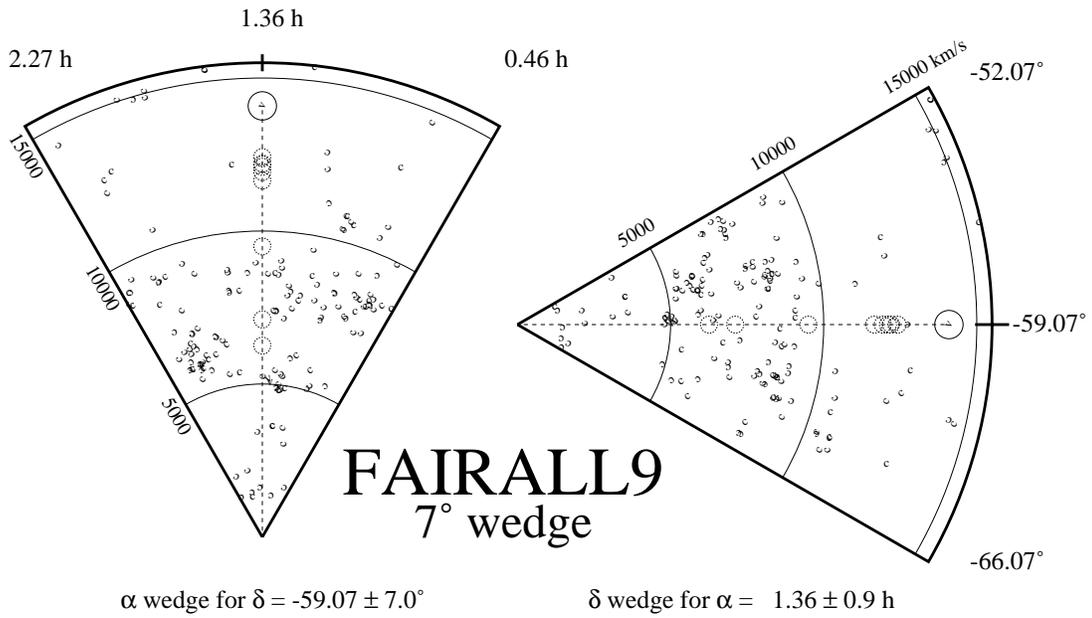}
\caption{Right ascension ($\alpha$)  and declination ($\delta$) pie diagram for the FAIRALL~9 sightline.\label{PIE_FAIRALL9}}
\end{figure}

\begin{figure} \plotone{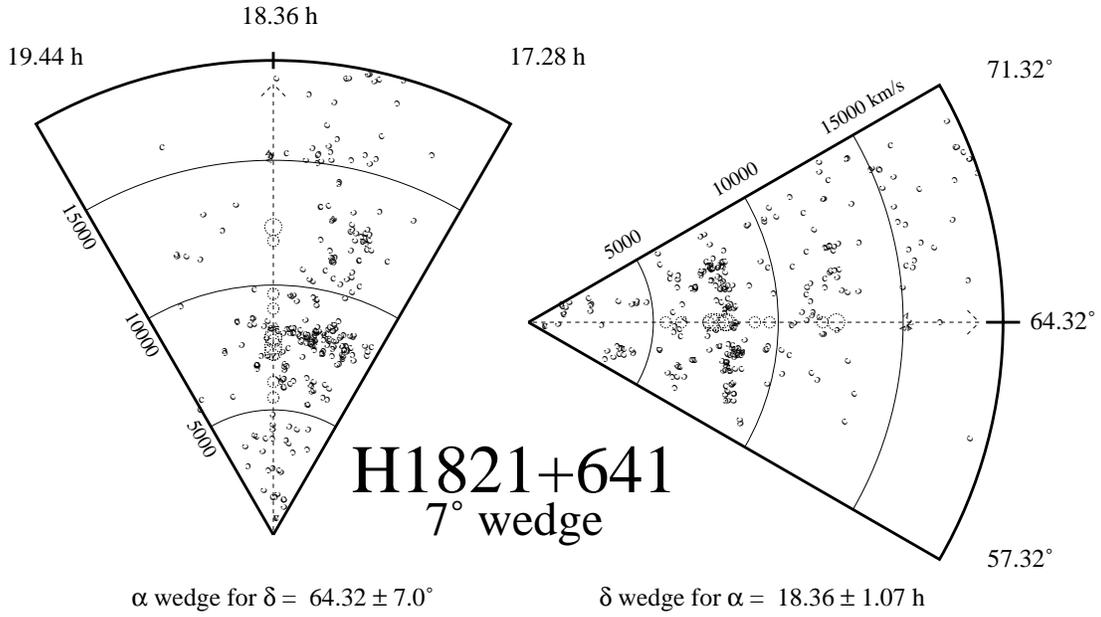}
\caption{Right ascension ($\alpha$) and declination ($\delta$) pie diagram for the H~1821+643 sightline.\label{PIE_H1821+643}}
\end{figure}

\begin{figure} \plotone{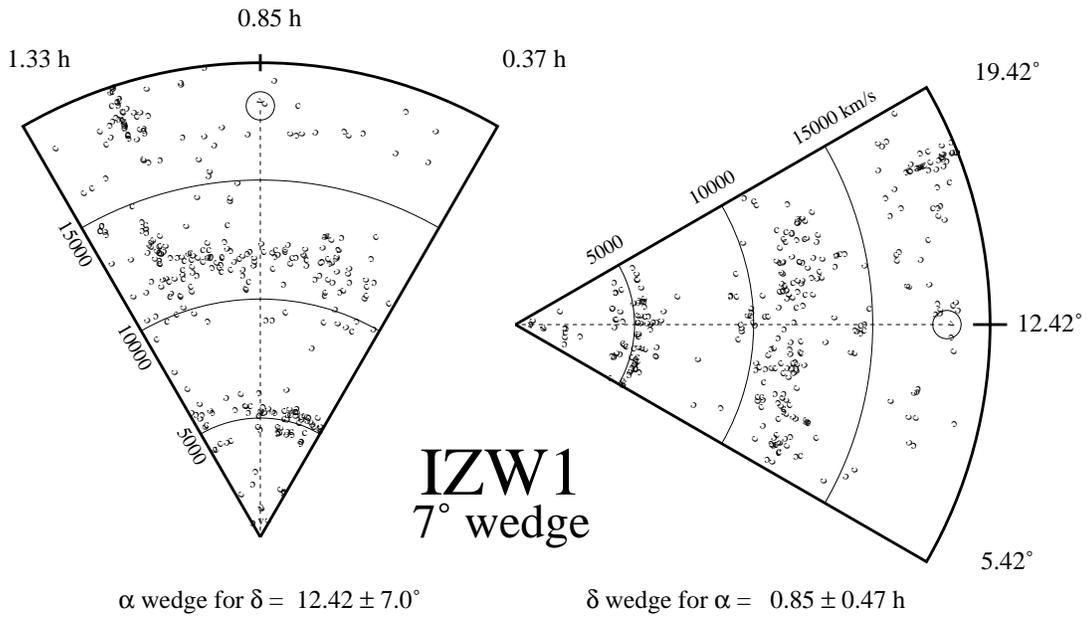}
\caption{Right ascension ($\alpha$) and declination ($\delta$) pie diagram for the I~ZW~1 sightline.\label{PIE_IZW1}}
\end{figure}

\begin{figure} \plotone{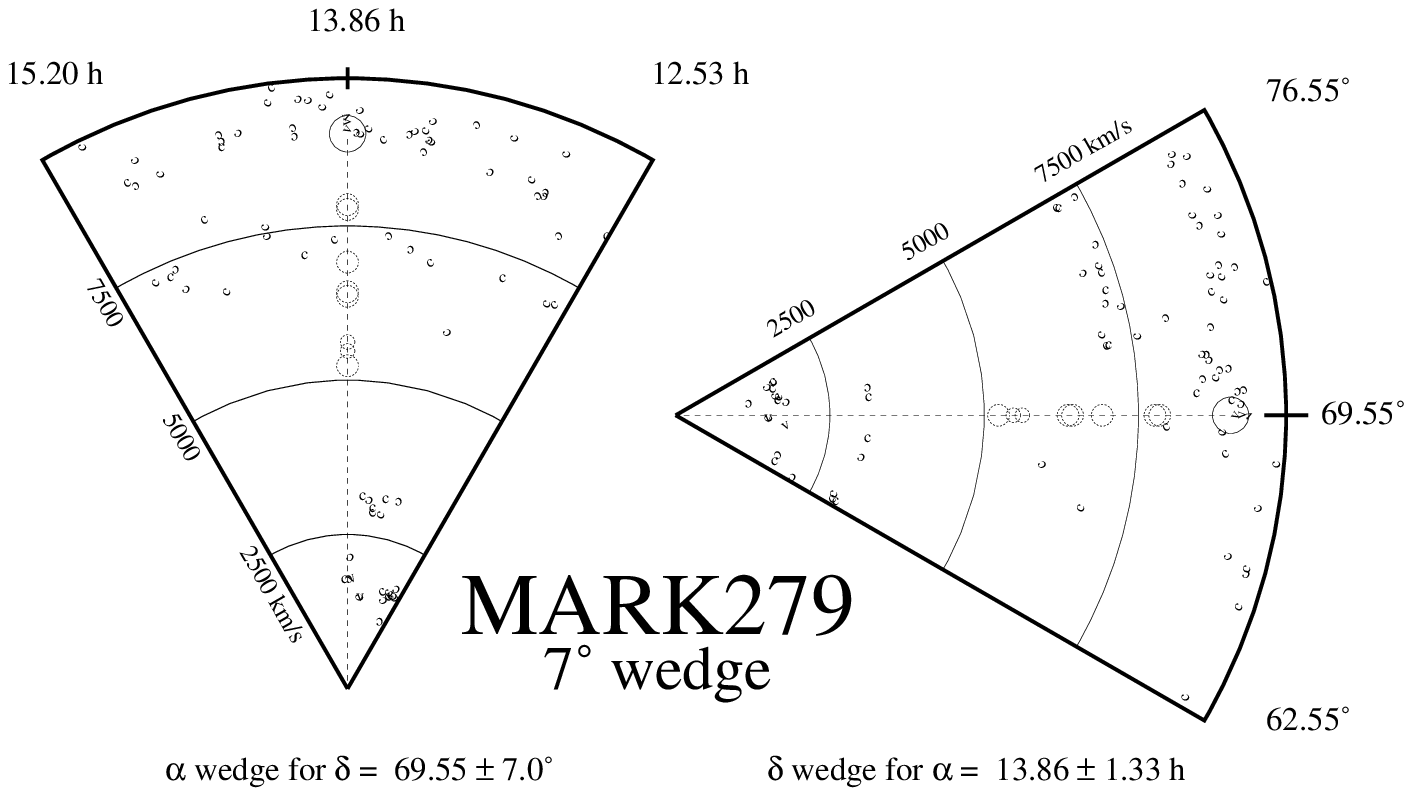}
\caption{Right ascension ($\alpha$) and declination ($\delta$) pie diagram for the Markarian~279 sightline.\label{PIE_MARK279}}
\end{figure}

\begin{figure} \plotone{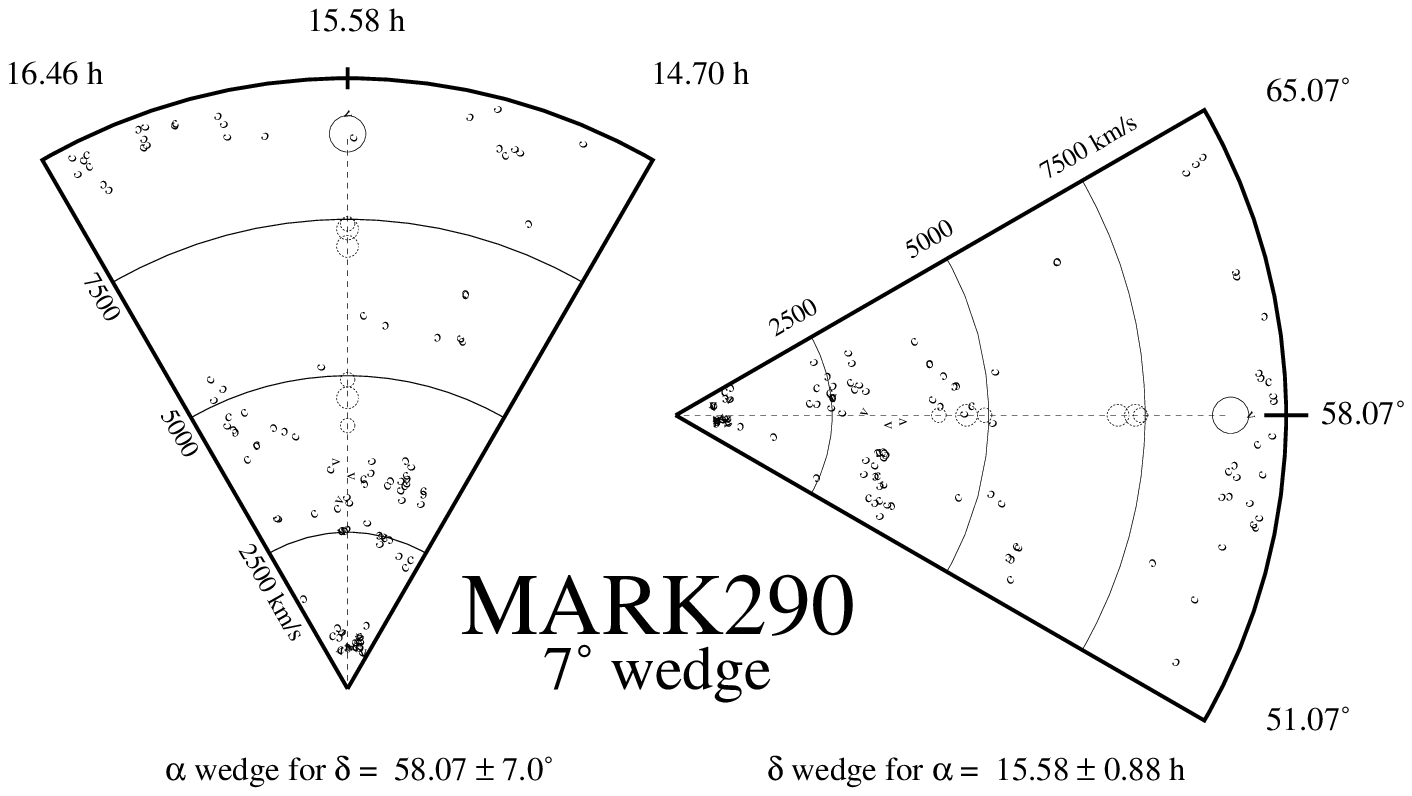}
\caption{Right ascension ($\alpha$) and declination ($\delta$) pie diagram for the Markarian~290 sightline.\label{PIE_MARK290}}
\end{figure}

\begin{figure} \plotone{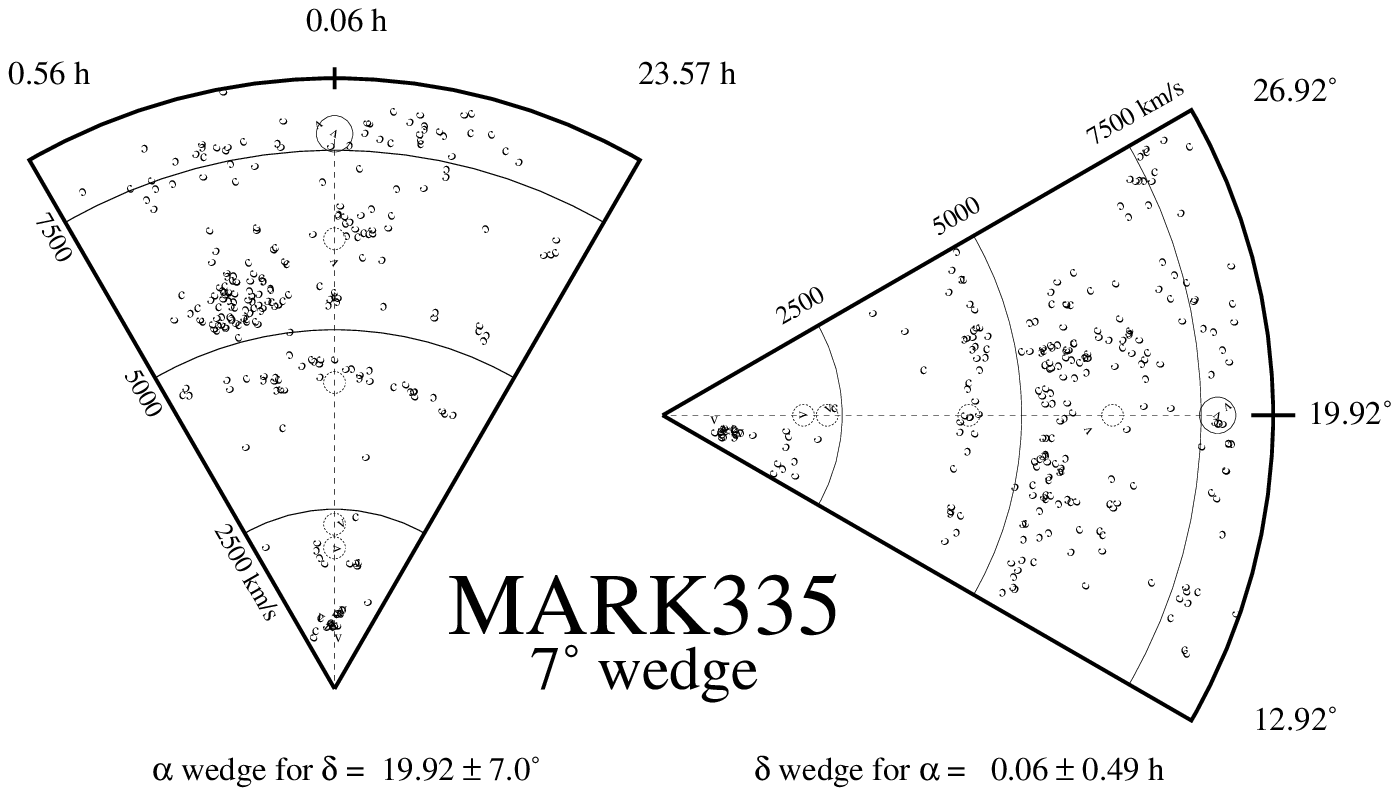}
\caption{Right ascension ($\alpha$) and declination ($\delta$) pie diagram for the Markarian~335 sightline.\label{PIE_MARK335}}
\end{figure}

\begin{figure} \plotone{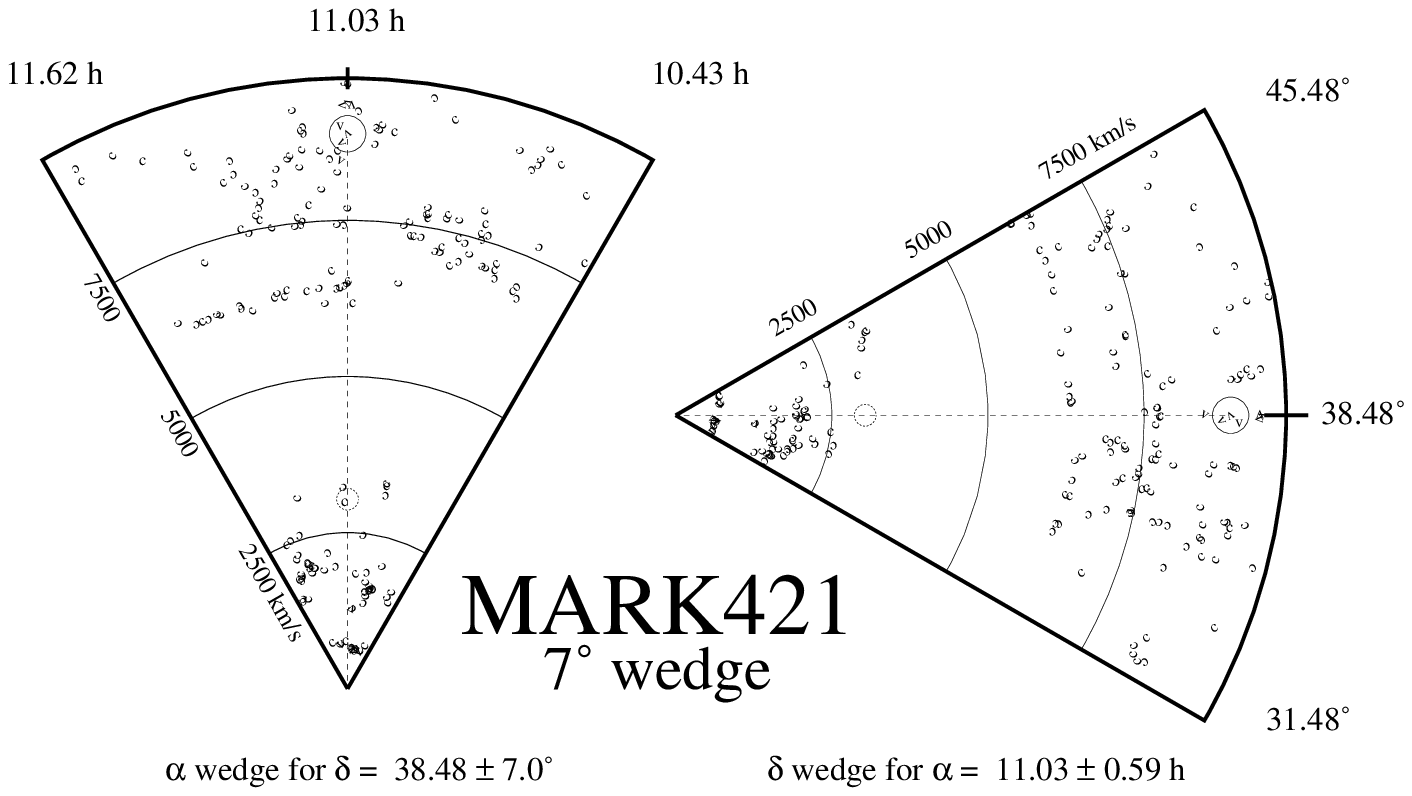}
\caption{Right ascension ($\alpha$) and declination ($\delta$) pie diagram for the Markarian~421 sightline.\label{PIE_MARK421}}
\end{figure}

\begin{figure} \plotone{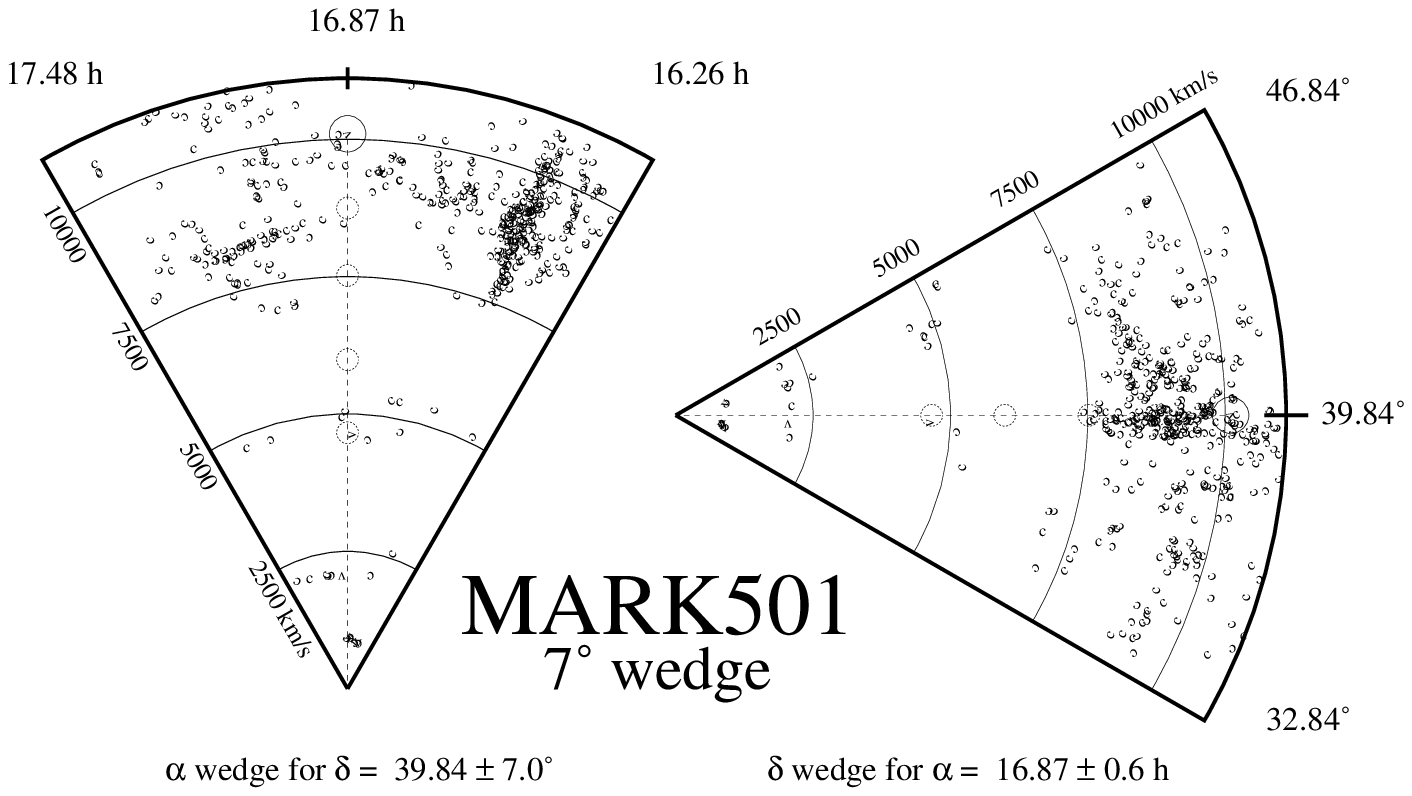}
\caption{Right ascension ($\alpha$) and declination ($\delta$) pie diagram for the Markarian~501 sightline.\label{PIE_MARK501}}
 \end{figure}

\begin{figure} \plotone{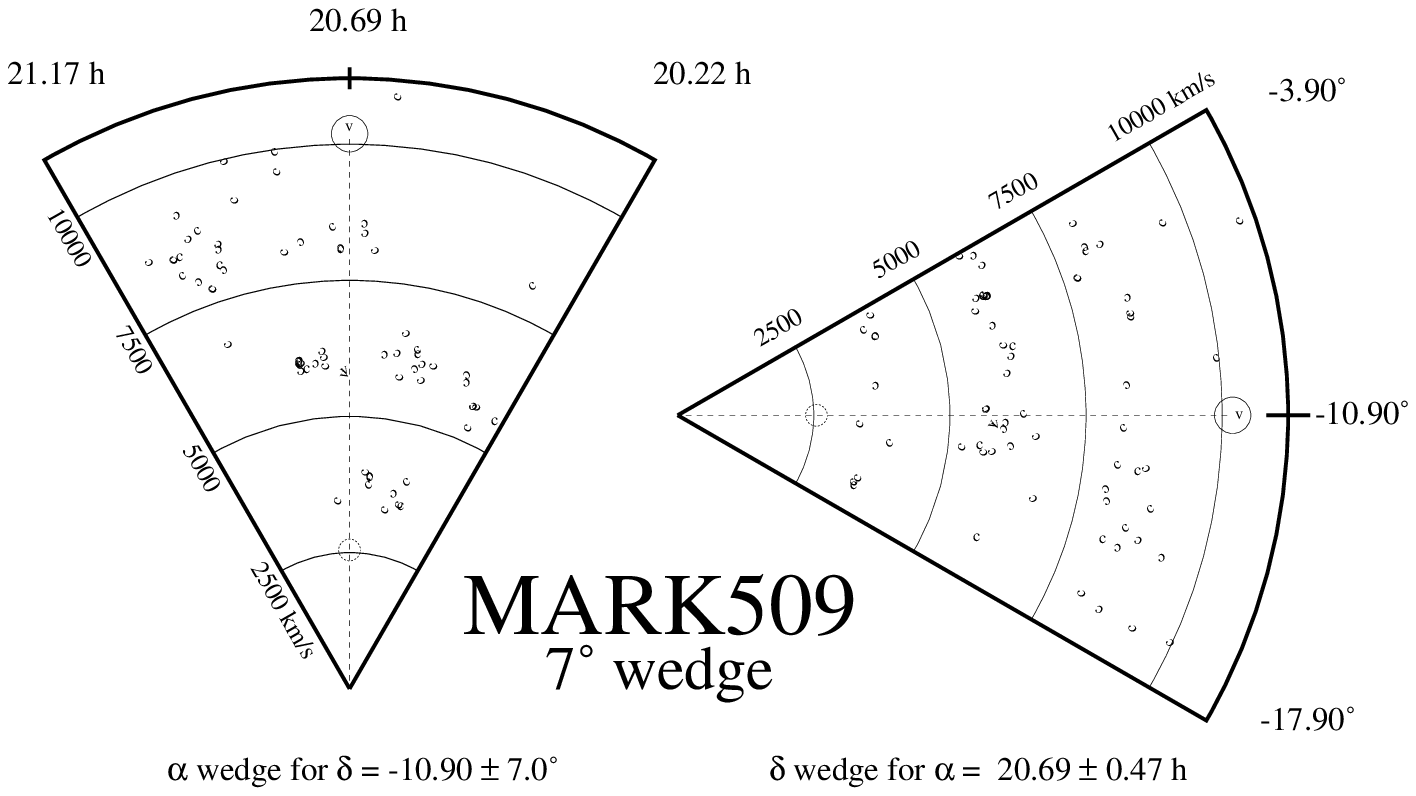}
\caption{Right ascension ($\alpha$) and declination ($\delta$) pie diagram for the Markarian~509 sightline.\label{PIE_MARK509}}
\end{figure}

\begin{figure} \plotone{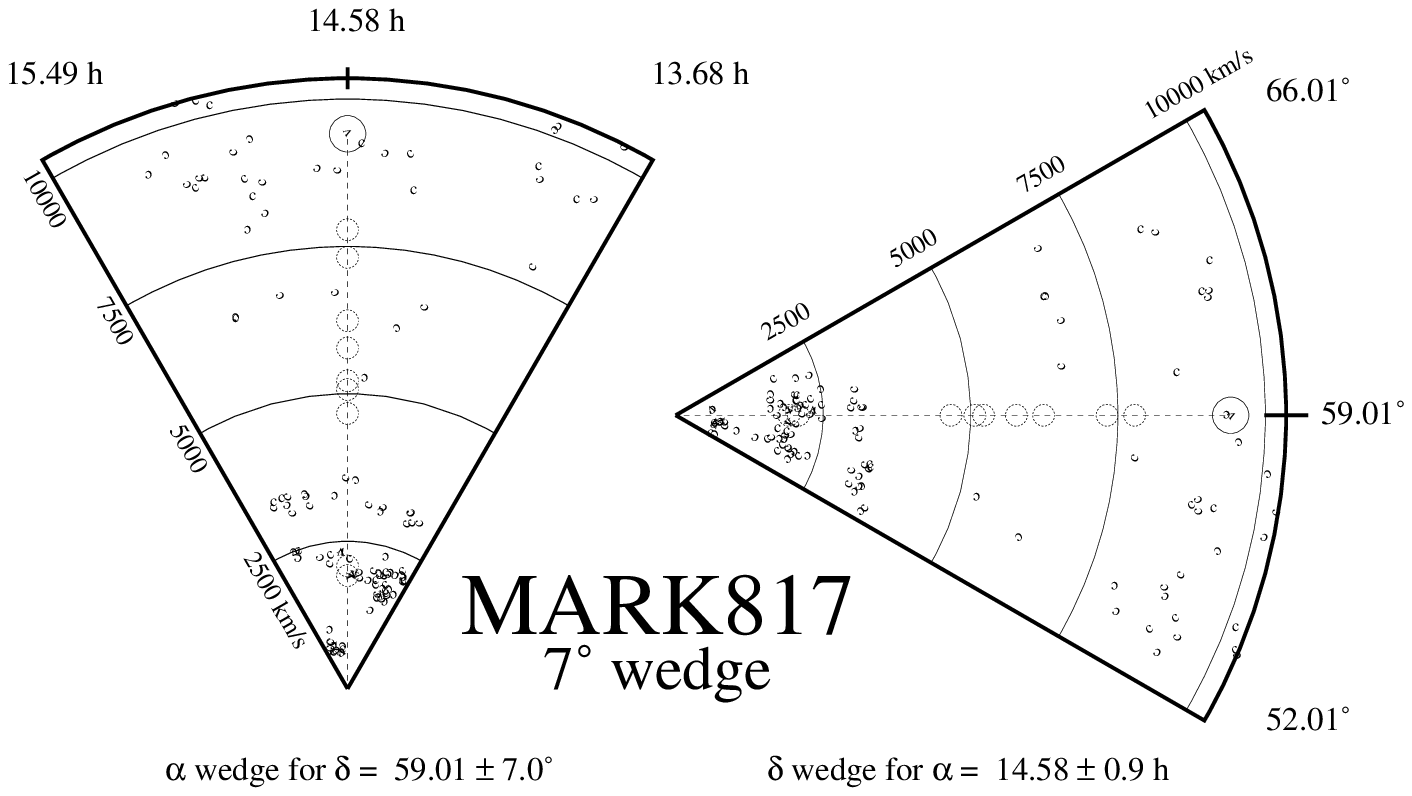}
\caption{Right ascension ($\alpha$) and declination ($\delta$) pie diagram for the Markarian~817 sightline.\label{PIE_MARK817}}
\end{figure}

\begin{figure}  \plotone{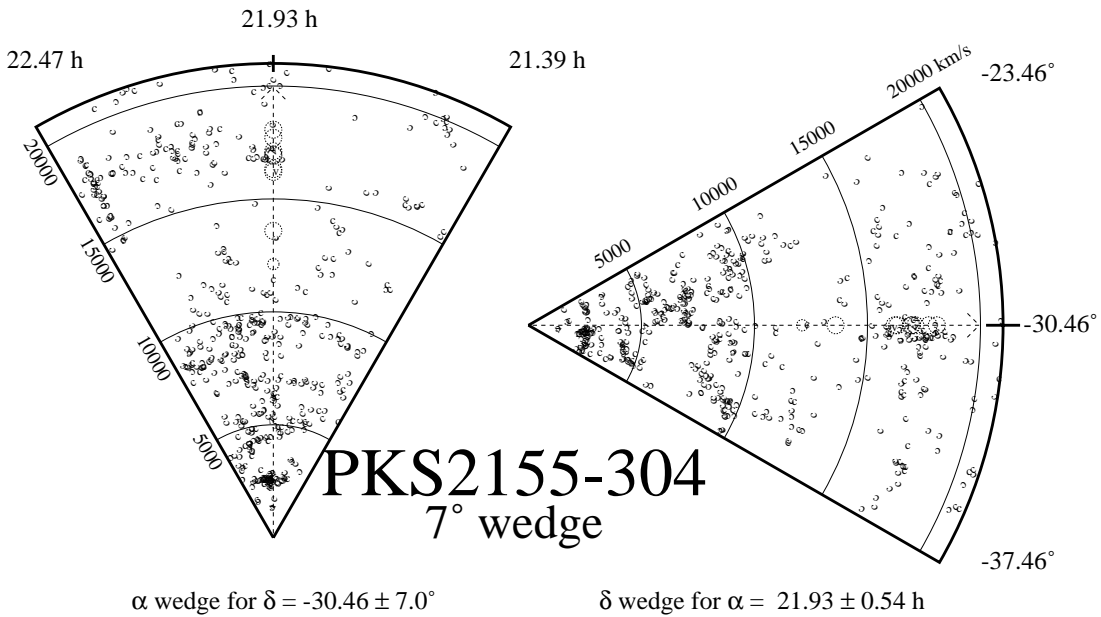}
\caption {Right ascension ($\alpha$) and
declination ($\delta$) pie diagram for the PKS2155-304 sightline. The arrow indicates that \PKS\ lies beyond the extent of the pie diagram at \z = 0.1165.\label{PIE_PKS2155-304}}
\end{figure}

\begin{figure} \plotone{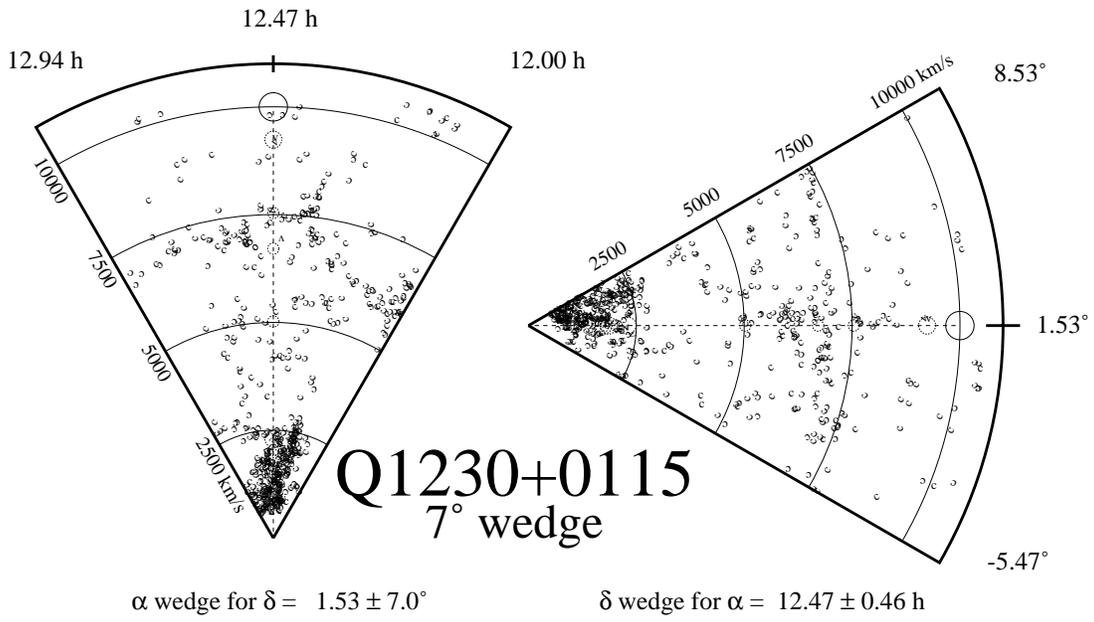}
\caption {Right ascension ($\alpha$) and declination ($\delta$) pie diagram for the Q~1230+0115 sightline. \label{PIE_Q1230+0115}}
\end{figure}

In this Appendix we present $\pm 7$\degr\ ``pie diagrams'' of our sightlines, and Table~\ref{nearest_master} 
which describes the three nearest known galaxies to each of our \lya absorbers.
In Figures~\ref{PIE_3C273}--\ref{PIE_Q1230+0115}, the coordinate along the slice of a pie diagram is the heliocentric recession velocity
in\kmsno. The dashed line in each diagram marks the sightline. Sightlines terminating with dashed arrows indicate 
that the target lies beyond the extent of the pie diagram (3C~273, H~1821+643, and \PKS). 
Objects whose positions are included by the pie diagram are marked with a large circle indicating the
 AGN target. The smaller circles indicate the positions along the sight line of the \lya absorbers;
the middle-sized circles are the definite (\real) detections, and the smallest circles are the possible (\tent)
detections. Each symbol ``c'' is an individual galaxy in the merged CfA redshift catalog; the orientation of the
``c'' for each galaxy is conserved between the wedges. The ``v''
symbols indicate CfA catalog galaxies within 2\degr\ of the sight line and are oriented consistently 
between the $\alpha$ and $\delta$ diagrams. Galaxies at \cz~\lt~500\kms have been deleted from
these plots.

 Table~\ref{nearest_master} includes the following information by column: 
(1) the absorber type where D=Definite (\real) or P=Possible (\tent);
(2) the heliocentric recession velocity of the absorber (non-relativistic \cz) converted from LSR using the 
standard value for the local standard of rest (see \S~\ref{sec:pie});
(3) the nearest galaxy name as given in the CfA catalog \citep{CFA}; 
(4) the absorber-galaxy distance perpendicular to the line-of-sight (LOS) in \hsfi Mpc; 
(5) the non-relativistic heliocentric recession velocity (\cz, in \kmsno) of the 
nearest galaxy as given in the CfA catalog; 
(6) the absorber-galaxy distance (in \hsfi Mpc) along the LOS, assuming a retarded Hubble
flow ($\pm 300$\kmsno); 
(7) the Euclidean 3D absorber-galaxy distance (in \hsfi Mpc), assuming a retarded Hubble flow 
($\pm 300$\kmsno; see \S~\ref{sec:pie} and \citet{Stocke95}); and
(8) the galaxy blue magnitude as given in the CfA Catalog.

As discussed in Paper~III, the completeness of the CfA catalog varies with position on the sky.
Therefore, direct conclusions about galaxy-absorber connections and ``void'' absorbers should
not be derived from Table~\ref{nearest_master}. However, if one adopts the very liberal
definition of an ``association'' as any absorber within 500\hsfi kpc of a CfA galaxy, and a
``void'' absorber as any galaxy whose nearest neighboring galaxy is $ > 6$\hsfi Mpc away, our
\real\ sample contains 20 galaxy-absorber ``associations'' and 21 galaxy ``void'' \lya absorbers
(using the total absorber-galaxy distances of column 7 in Table~\ref{nearest_master}).
Our \tent\ sample contains an additional two absorber-galaxy ``associations'' and four ``void'' absorbers.
% Akn 120
% ESO 141-G55
% Fairall 9
\clearpage
% H1821+643
% I Zw 1
% Mark 279
% Mark 290
\clearpage
% Mark 335
% Mark 421
\clearpage
% Mark 501
% Mark 509
% Mark 817
% PKS 2155-304
% Q1230+0115
\clearpage
\renewcommand{\arraystretch}{0.985}
\begin{deluxetable}{cclcccccl} 
\tablecolumns{8}
\tablewidth{0pt}
\tablecaption{Nearest three galaxies to each \lya absorber. \label{nearest_master} }
\tablehead{
\colhead{Abs\tablenotemark{a}}&
\colhead{Abs Vel}&
\colhead{Nearest}&
\colhead{$D_{\perp}$\tablenotemark{b}}&
\colhead{Gal Vel}&
\colhead{$D_{cz}$\tablenotemark{c}}&
\colhead{$D_{\rm tot}$\tablenotemark{d}}&
\colhead{$m_B$}\\
\colhead{Type}&
\colhead{(\nokmsno)}&
\colhead{Galaxy}&
\colhead{(\hsfi Mpc)}&
\colhead{(\nokmsno)}&
\colhead{(\hsfi Mpc)}&
\colhead{(\hsfi Mpc)}&
\colhead{} \\
\colhead{(1)}&
\colhead{(2)}&
\colhead{(3)}&
\colhead{(4)}&
\colhead{(5)}&
\colhead{(6)}&
\colhead{(7)}&
\colhead{(8)}  
}
\startdata
\cutinhead{3C~273}
D &    1012 &  12285+0157  &    0.170 &   1105 &    0.000  &     0.170  &  15.6  \\
D &    1012 &  A1225+0152  &    0.182 &   1298 &    0.000  &     0.182  &  16.3  \\
D &    1012 &  A1224+0146  &    0.232 &   1275 &    0.000  &     0.232  &  \nodata  \\
\tableline
D &    1583 &  A1225+0152  &    0.182 &   1298 &    0.000  &     0.182  &  16.3  \\
D &    1583 &  12277+0254  &    0.260 &   1635 &    0.000  &     0.260  &  15.3  \\
D &    1583 &  A1224+0146  &    0.232 &   1275 &    0.121  &     0.261  &  \nodata  \\
\tableline
P &    2196 &  A1225+0258  &    0.396 &   2329 &    0.000  &     0.396  &  16.3  \\
P &    2196 &  12270+0107  &    0.677 &   2234 &    0.000  &     0.677  &  17.0  \\
\enddata
\tablenotetext{a}{Type of absorber, D=Definite (\real) or P=Possible (\tent).}
\tablenotetext{b}{Galaxy distance, perpendicular to line-of-sight (LOS).}
\tablenotetext{c}{Absorber-galaxy distance along LOS, assuming retarded Hubble
flow ($\pm$300\kmsno).}
\tablenotetext{d}{Total 3D Euclidean absorber-galaxy distance, assuming retarded Hubble flow.}
\end{deluxetable}
\normalsize

\clearpage
%
% References
%
\bibliographystyle{apj}

\end{document}